\documentclass[aps, prd, twocolumn, superscriptaddress,nofootinbib,10pt]{revtex4}
\usepackage{graphicx}
\usepackage{dcolumn}
\usepackage{amssymb,amsmath,amsthm,mathrsfs}
\usepackage{epstopdf}
\usepackage{lipsum}
\usepackage{hyperref}
\usepackage{empheq}
\usepackage{color}
\usepackage{multirow}
\usepackage{floatpag}
\usepackage{mdframed}
\usepackage{afterpage}
\usepackage[export]{adjustbox}
\usepackage{bm}
\usepackage{footmisc}
\usepackage{enumitem}
  
\allowdisplaybreaks

\newcommand{\be}{\begin{equation}}
\newcommand{\ee}{\end{equation}}
 \newcommand{\bea}{\begin{eqnarray}}
\newcommand{\eea}{\end{eqnarray}}

\newcommand{\CL}{{\tt ${\mathcal C}$osmo${\mathcal L}$attice}~}
\newcommand{\CLns}{{\tt ${\mathcal C}$osmo${\mathcal L}$attice}}

\begin{document}

\title{Characterizing the post-inflationary reheating history,\\ Part I: single daughter field with quadratic-quadratic interaction}

\newcommand{\addressIFIC}{Instituto de F\'isica Corpuscular (IFIC), Universitat de Val\`{e}ncia-CSIC, E-46980, Valencia, Spain.}
\newcommand{\addressUNIBAS}{Department of Physics, University of Basel, Klingelbergstr. 82, CH-4056 Basel, Switzerland.}

\author{Stefan Antusch}
\affiliation{\addressUNIBAS}
\author{Daniel G. Figueroa}
\affiliation{\addressIFIC} 
\author{Kenneth Marschall}
\affiliation{\addressUNIBAS}
\author{Francisco Torrenti}
\affiliation{\addressUNIBAS}

\date{\today}

\begin{abstract}
We study the evolution of the energy distribution and equation of state of the Universe from the end of inflation until the onset of either radiation domination (RD) or a transient period of matter domination (MD). We use both analytical techniques and lattice simulations. We consider two-field models where the inflaton $\Phi$ has a monomial potential after inflation $V(\Phi) \propto |\Phi - v|^p$ ($p\geq2$), and is coupled to a daughter field $X$ through a quadratic-quadratic interaction $g^2\Phi^2 X^2$. We consider two situations, depending on whether the potential has a minimum at $i)$ $v = 0$, or $ii)$ $v > 0$. In the scenario $i)$, the final energy transferred to $X$ is independent of $g^2$ and entirely determined by $p$: it is negligible for $p < 4$, and of order $\sim 50\%$ for $p \geq 4$. The system goes to MD at late times for $p = 2$, while it goes to RD for $p > 2$. In the later case, we can calculate exactly the number of e-folds until RD as a function of $g^2$, and hence predict accurately inflationary observables like the scalar tilt $n_s$ and the tensor-to-scalar ratio $r$. In the scenario $ii)$, the energy is always transferred completely to $X$ for $p>2$, as long as its effective mass $m_X^2 = g^2(\Phi-v)^2$ is not negligible. For $p=2$, the final ratio between the energy densities of $X$ and $\Phi$ depends strongly on $g^2$. For all $p \ge 2$, the system always goes to MD at late times. 
\end{abstract}


\maketitle

\section{Introduction}

A phase of accelerated expansion in the early universe, {\it Inflation}~\cite{Starobinsky:1980te,Guth:1980zm,Linde:1981mu,Albrecht:1982wi}, provides a natural solution to the horizon and flatness problems of the hot big bang framework, as well as a mechanism to generate the right spectrum of primordial perturbations. Constraining cosmological observables, such as the scalar tilt $n_s$ or the tensor-to-scalar ratio $r$, provides insight into the inflationary epoch. Many inflationary models have been actually proposed, but not all of them are compatible with cosmological observations~\cite{Martin:2013tda,Planck:2018jri}. For example, recent analysis of the B-mode polarization of the Cosmic Microwave Background (CMB) set an upper bound on the inflationary Hubble scale as $H_{\rm inf} \lesssim 4.7\times10^{13}$ GeV~\cite{BICEP:2021xfz}, putting more pressure, when not directly ruling out, the parameter space of scenarios previously compatible with the data.

Inflation must be followed by a period of \textit{reheating}, during which the Universe ultimately has to reach a radiation dominated (RD) thermal state, at least before the onset of Big Bang Nucleosynthesis (BBN) at a temperature of $T_{\rm BBN}\sim 10^{-3}\text{ GeV}$ \cite{Kawasaki:1999na,Kawasaki:2000en,Hannestad:2004px,Hasegawa:2019jsa}. See~\cite{Bassett:2005xm,Allahverdi:2010xz,Amin:2014eta,Lozanov:2019jxc,Allahverdi:2020bys} for reviews on reheating. The first stage of reheating is often driven by a period of \textit{preheating}, characterized by a strong non-perturbative, out-of-equilibrium excitation of field fluctuations, typically resulting in exponentially growing particle number densities~\cite{Traschen:1990sw,Kofman:1994rk,Shtanov:1994ce,Kaiser:1995fb,Khlebnikov:1996mc,Prokopec:1996rr,Kaiser:1997mp,Kofman:1997yn,Greene:1997fu,Khlebnikov:1996zt,Kaiser:1997hg}. During preheating and later stages of reheating, the evolution of the universe is characterized by a time-dependent equation of state, determined by the (averaged) energy content distribution among the relevant degrees of freedom. In many models there is an intermediate time when the energy distribution and the equation of state of the universe stabilize, at least temporarily, and a stationary stage is developed~\cite{Micha:2002ey,Micha:2004bv,Antusch:2020iyq}. Understanding the reheating phase since the end of inflation till the onset of RD is one of the key challenges for making accurate predictions for CMB observables~\cite{Dai:2014jja,Martin:2014nya,Munoz:2014eqa,Gong:2015qha,Cook:2015vqa}. Helping to shed light on this understanding is one of the main goals of this paper.

Many analytical studies have been performed to investigate the phase of preheating, giving valuable insight into the initial linear regime of exponential field excitation after inflation, see e.g.~\cite{Kofman:1994rk,Kofman:1997yn,Greene:1997fu} for the case of parametric resonance with monomial inflaton potentials. The late time regime, on the other hand, is governed by non-linear dynamics. Thus, to investigate the dynamics of the (p)reheating phase in its full extent, it is necessary to resort to lattice field theory techniques (for a review see~\cite{Figueroa:2020rrl}). Post-inflationary dynamics have been studied with the help of such techniques in the past, in particular for models with monomial inflaton potentials. For potentials steeper than quadratic, the inflaton fragments due to self-resonance even in the absence of couplings to daughter field species, with the effective equation of state of the Universe transitioning towards RD $\sim 10$ e-folds after the end of inflation~\cite{Lozanov:2016hid,Lozanov:2017hjm}. For a quadratic potential, however, the inflaton can only fragment via gravitational effects in a much longer time scale~\cite{Musoke:2019ima}. 

The inclusion of interactions to other field species can change the post-inflationary dynamics significantly. The evolution of the energy distribution and equation of state during preheating was studied in the case of quadratic-quadratic inflaton-daughter field interactions in \cite{Podolsky:2005bw}, and more extensively in \cite{Maity:2018qhi} (see also \cite{Saha:2020bis} for a non-lattice study). The case of trilinear interactions was also considered in \cite{Dufaux:2006ee}, and a case with higher order effective couplings in \cite{Antusch:2015vna}. A fitting study for quadratic and quartic potentials was also carried out in~\cite{Figueroa:2016wxr}, where the dynamics was characterized as a function of the coupling strength. Furthermore, lattice studies of multifield models with non minimal couplings to gravity, transformed into non-minimal kinetic terms in the Einstein frame, have been carried out in \cite{Nguyen:2019kbm,vandeVis:2020qcp}, see also~\cite{DeCross:2015uza, DeCross:2016fdz, DeCross:2016cbs} for semi-analytical studies. Other works on preheating with non-minimal kinetic terms  include~\cite{Child:2013ria} in the context of DBI inflation and based on lattice simulations, and \cite{Krajewski:2018moi} and~\cite{Iarygina:2018kee} in the context of $\alpha$-attractor scenarios, based on lattice simulations and in semi-analytical computations, respectively.

Preheating studies have often focused on the early stages of the field dynamics, rather than on the very late time evolution of the system, when the energy densities and equation of state attain a stationary regime. In a recent letter~\cite{Antusch:2020iyq}, using lattice simulations, we captured for the first time the very late-time dynamics of the inflaton-daughter field system for a class of scenarios with monomial inflaton potentials around the origin during preheating, $V \propto |\phi|^p$ (with arbitrary power-law index $p \geq 2$). There we studied the evolution of the energy distribution and equation of state of the universe as a function the model parameters. The simulations were carried out with the \CL package~\cite{Figueroa:2020rrl,Figueroa:2021yhd}, which implements various evolution techniques that allow to decrease the simulation time and to execute the code parallellized in multi-core systems. In particular, we took advantage of the higher-order \textit{Velocity-Verlet} evolution algorithms implemented in the package, which allowed us to simulate the late-time regime of the system while preserving energy conservation sufficiently well. Furthermore, we also showed that simulations of the given scenario in (2+1) dimensions\footnote{With ``simulations in (2+1) dimensions" we mean solving the three-dimensional field equations in a 2D slice, with the discrete spatial derivatives adjusted to the new situation: for example, the $\mathcal{O} (\Delta x^2)$ Laplacian is sourced by four surrounding lattice points instead of six. We refer the reader to \cite{Felder:2000hq} for a discussion on this technique. \label{fn2Dsims}} can mimic the (3+1)-dimensional dynamics well, which allows to significantly increase the final simulated time.

The aim of the present paper is to complement and expand the results of~\cite{Antusch:2020iyq}. Regarding the setup for the inflaton-daughter field system, we consider now a singlet real scalar inflaton field $\Phi$ and model its potential by a monomial shape around a minimum $v$ after inflation, $V(|\phi|) \propto |\phi|^p$ ($p\geq2$) with $\phi \equiv \Phi - v $. As purely monomial potentials $V(\phi) \propto |\phi|^p$ are strongly disfavored for inflation\footnote{Monomial potentials during inflation are not ruled-out however if an appropriate non-minimal coupling between the inflaton and the Ricci scalar is also present~\cite{Tsujikawa:2013ila}.}, we rather use consistent inflaton potentials that flatten out towards large field values by developing a plateau, as inspired from $\alpha$-attractor models~\cite{Kallosh:2013hoa}. Only after inflation ends, the inflaton reaches the monomial shape of its potential $V(\phi) \propto |\phi|^p$ around $v$. We consider that the minimum of the inflaton potential can be both at $v = 0$ (as in~\cite{Antusch:2020iyq}), as well as at a non-vanishing values $v \neq 0$ (going beyond~\cite{Antusch:2020iyq}). In the case of $v=0$, we refer to the inflaton potential as \textit{centred}, while we call it \textit{displaced} potential otherwise for $v \neq 0$.
For the interaction between the inflaton $\Phi$ and the scalar daughter field $X$, we consider\footnote{We would like to emphasize that other interaction terms might play a relevant role as well, and that it would be interesting to extend our study in this direction in the future. However, any such consideration goes beyond the current scope of the present paper, where we limit ourselves to quadratic-quadratic interactions.} the ubiquitous quadratic-quadratic coupling $g^2\Phi^2 X^2$. It is the lowest order coupling allowed in case of a $Z_2$ symmetry, and it emerges, for example, as the leading term from gauge interactions of the form $(D_{\mu} \Phi)^{\dagger} (D_{\mu} \Phi)$~\cite{Figueroa:2015rqa}, with $D_{\mu}$ being a gauge covariant derivative. Furthermore, it is a marginal operator that does not introduce any new scale, and even if it was absent in the tree-level Lagragian, it is typically generated in the presence of other interactions from quantum effects~\cite{Gross:2015bea}. 

Using very long lattice simulations, we present a detailed study of the post-inflationary evolution of the different energy components and the equation of state as a function of the power-law coefficient $p$, the inflaton-daughter coupling $g^2$ and the inflaton potential minimum $v$. In particular, we characterize the late-time dynamics of the system, as well as the values that the different energy ratios and the equation of state take at the late stationary regime. We also present a detailed analytical formulation of the preheating process based on the linearized field equations, with arbitrary power-law coefficient $p$ of the monomial inflaton potential. Although this analysis can only be applied to the early dynamics of the system, it e.g.\ allows to obtain an estimation of the \textit{backreaction time} when the dominant process is either \textit{parametric resonance of the daughter field} or \textit{self-resonance of the inflaton}. Furthermore, the analytical results are also useful to interpret some of the features we observe in the lattice simulations. Finally, we also present a detailed calculation of the impact of reheating on the inflationary CMB observables. For $v = 0$ and $p > 2$, using our lattice results we can calculate exactly, as a function of $g^2$, the number of e-folds between the end of inflation and the onset of RD. This allows us to predict accurately inflationary observables like the scalar tilt $n_s$ and the tensor-to-scalar ratio $r$. This paper will be followed by a \textit{Part II}, in which we will treat the case of multiple daughter fields.

The paper is structured as follows: in Section \ref{Sec:Inflation} we describe the properties of our chosen inflationary potential and the post-inflationary inflaton oscillations. In Section \ref{Sec:LinearAn} we present a detailed linearized analysis of preheating, including the resonances of the inflaton and daughter field, considering both the effect of a vanishing ($v = 0$) and non-vanishing ($v \neq 0$) inflaton potential minimum. In Section \ref{Sec:LatResults} we present the results from our lattice simulations with \textit{centred} ($v=0$) and \textit{displaced} ($v\neq0$) potentials. In Section \ref{Sec:CMB} we study the impact of our results on CMB observables. Finally, in Section \ref{Sec:Conclusion} we discuss our results and conclude.

\section{Inflaton potential and oscillations} \label{Sec:Inflation}

We consider a singlet inflaton $\Phi$ with potential $V(\Phi)$ that exhibits a single minimum at some scale $\Phi = v$. For simplicity we consider the potential to be symmetric around the minimum, so it can be written as a function of $|\Delta \Phi | \equiv |\Phi - v|$. We will consider the potential to be a monomial $V \propto |\Phi - v|^p$ around the origin, with $p > 1$. In the absence of inflaton couplings to other fields, the dynamics of the system does not depend on $v$ and the field amplitude can be shifted as $\Phi \rightarrow \phi + v$, such that the minimum of the potential is at $\phi = 0$. For the remainder of this section we discuss the initial phase after inflation, where inflaton-daughter couplings are neglected and we only consider the inflaton potential $V(\phi)$. We will restore the interaction of the inflaton with a daughter field in Section~\ref{Sec:LinearAn}.

A paradigmatic example of monomial potentials is given by the chaotic inflation scenario, characterised by a function as
\be V (\phi) = \frac{1}{p} \lambda \mu^{4-p}   |\phi |^p \ , \label{eq:powlaw-pot}\ee
with $\lambda$ a dimensionless parameter, $p > 1$ a real exponent, and $\mu$ an energy scale. For $p=4$ and $p = 2$, Eq.~(\ref{eq:powlaw-pot}) reproduces the usual expressions $V(\phi) = (1 /4) \lambda \phi^4$ and $V(\phi) = (1/2) m^2 \phi^2$ (with $m^2 \equiv \lambda \mu^2$) respectively. As the potential depends on the absolute value of $\phi$, the existence of a minimum at $\phi = 0$ is ensured, even for odd or fractional values of $p$. 

\begin{figure}
    \centering
    \includegraphics[width=0.45\textwidth]{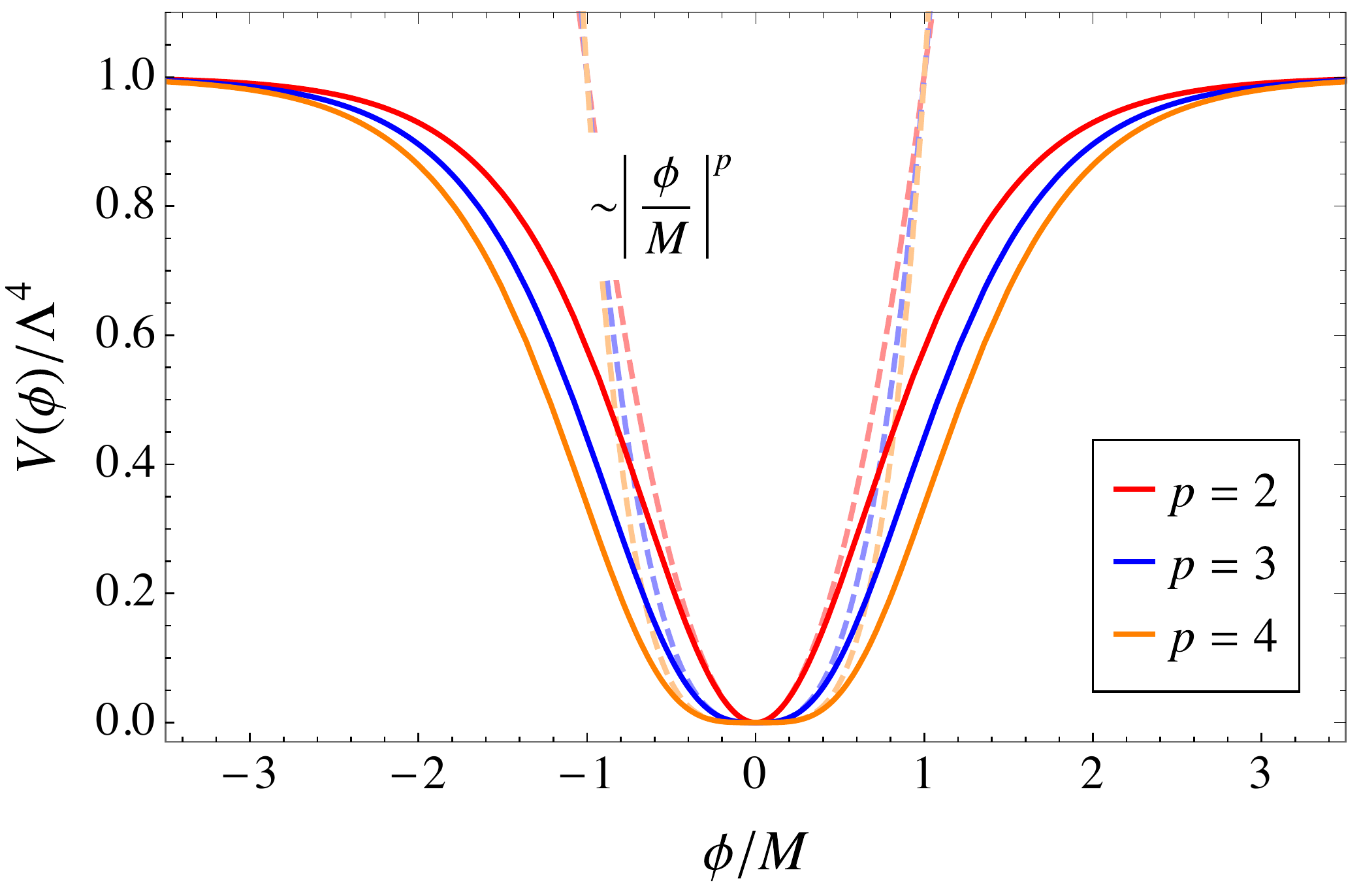}  \\ 
    \caption{Inflationary potential (\ref{eq:inflaton-potential}) for several values of $p$. The dashed lines show the monomial function (\ref{eq:powlaw-pot}) that approximates the potential well for small values of $\phi$.} \label{fig:infpotcentred}
\end{figure}

A monomial shape as in Eq.~(\ref{eq:powlaw-pot}) at all field amplitudes is excluded by CMB observations~\cite{Planck:2018jri,BICEP:2021xfz}. However, Eq.~(\ref{eq:powlaw-pot}) can still approximately describe the behaviour of a more general potential around its minimum, as long as it deviates from the monomial form at large field amplitudes, developing for instance a {\it plateau}. We parametrize such potentials in the following way,
\be V_{\rm t} (\phi) = \frac{1}{p} \Lambda^4{\rm tanh}^{p} \left( \frac{|\phi|}{M} \right) \ , \hspace{0.5cm} \Lambda \equiv \lambda^{\frac{1}{4}} \mu^{\frac{4-p}{4}} M^\frac{p}{4} \ . \label{eq:inflaton-potential} \ee
where $\Lambda$ and $M$ are parameters with dimensions of energy, appropriately chosen (give a value of $p$) to satisfy CMB constraints. The form of this potential is motivated by $\alpha$-attractor T-models of inflation \cite{Kallosh:2013hoa} and is depicted in Fig.~\ref{fig:infpotcentred}. The potential can be approximated by Eq.~(\ref{eq:powlaw-pot}) for small amplitudes $|\phi | \ll  M$, while it develops a plateau $V_{\rm t}(\phi) \rightarrow \Lambda^4 /p$ at large amplitudes $|\phi| \gg M$. The inflection point separating the positively and negatively curved regions of the potential is 
\be \phi_{\rm i} =  M {\rm arcsinh} \left( \sqrt{\frac{p-1}{2}} \right) \ , \label{eq:inflection-potential} \ee
which gives $\phi_{\rm i} / M \simeq \{0.66$, $0.89$, $1.03$, $1.15$, $1.26\}$ for $p = \{2$, $3$, $4$, $5$, $6\}$ respectively. Note that we get $\phi_{\rm i} \rightarrow \infty$ in the limit $M \rightarrow \infty$, as expected.

In these models, inflation takes place at large field values away from the minimum, where a slow-roll regime holds. The inflaton accelerates as it rolls towards the minimum, and eventually the slow-roll regime breaks. The end of inflation can be identified as the time when $\epsilon_H \equiv - \dot{H} /H^2 = 1$. For simplicity, we can also approximate the end of inflation by the condition $\epsilon_{_V}\equiv m_{\rm pl}^2 V_{,\phi}^2/(2V^2) = 1$, which allows to obtain analytical expressions for the inflaton amplitude at that time. In our model,
\be \epsilon_{_V} = \frac{2p^2m_{\rm pl}^2\rm{csch}^2\left(\frac{2|\phi|}{M}\right)}{M^2} \xrightarrow[M\to\infty]{} \frac{p^2m_{\rm pl}^2}{2|\phi|^2}\ ,  \label{eq:sr-epsilon} \ee
where we have also written the expression in the monomial limit $M \rightarrow \infty$. The field amplitude $\phi_*$ satisfying the condition $\epsilon_{_V}(\phi_*) \equiv1$, is given by
\be \phi_* = \frac{1}{2} M {\rm arcsinh} \left( \frac{\sqrt{2} p m_{\rm pl}}{M}\right)  \xrightarrow[M\to\infty]{} \frac{p m_{\rm pl}}{\sqrt{2}} \ . \label{eq:end-of-inflation} \ee

\begin{figure}
    \centering
    \includegraphics[width=0.45\textwidth]{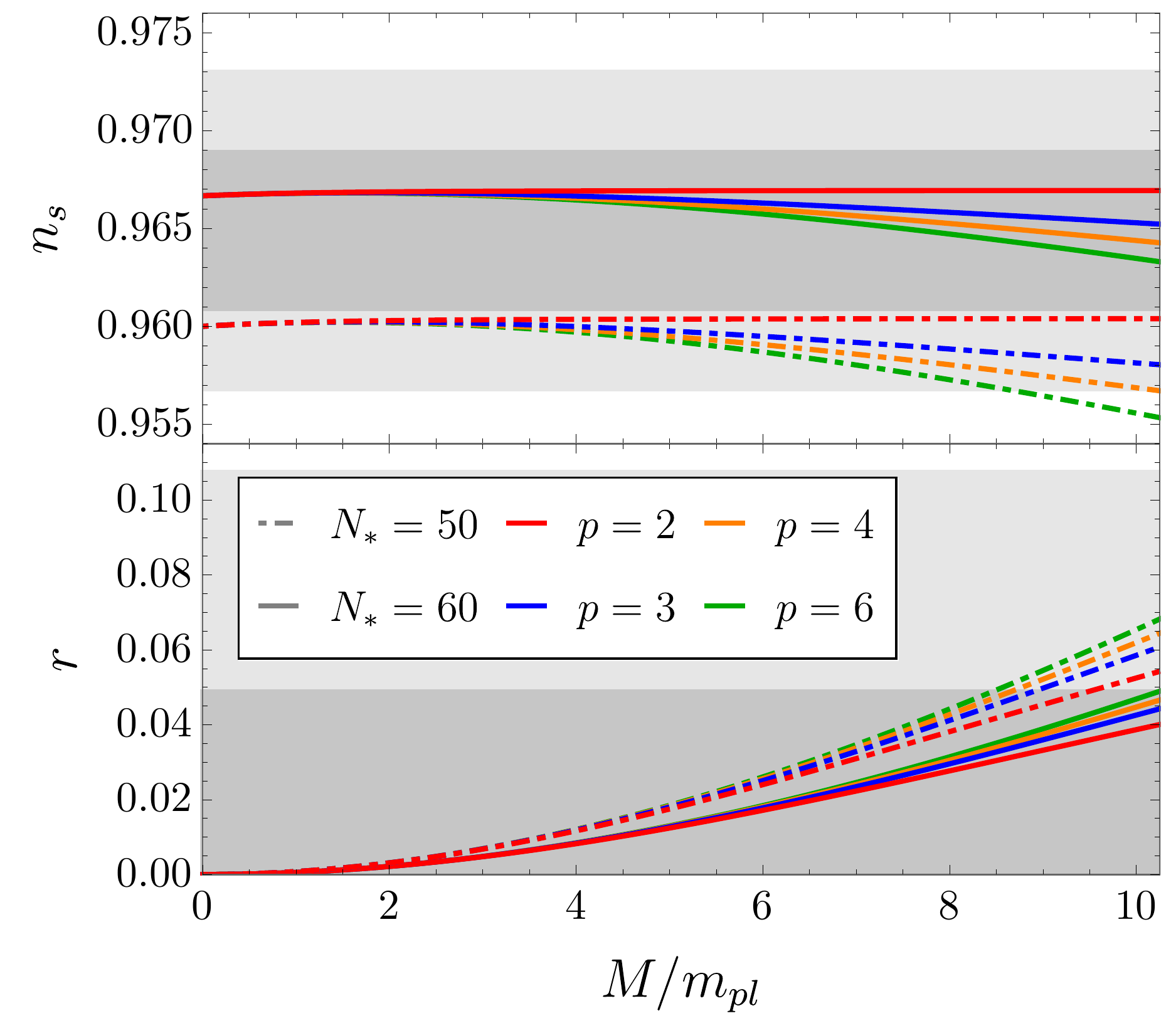}  \\ 
    \caption{Prediction for $n_s$ and $r$ as a function of $M$ for the inflationary potential (\ref{eq:inflaton-potential}), for different choices of $p$. The observational constraints for $n_s$ are depicted by the dark and light gray bands, which indicate the 68\% and 95\% C.L. intervals from Planck 2018 results \cite{Akrami:2018jri}.}
    \label{fig:nsr-M}
\end{figure}

\begin{figure}
    \centering
    \includegraphics[width=0.45\textwidth]{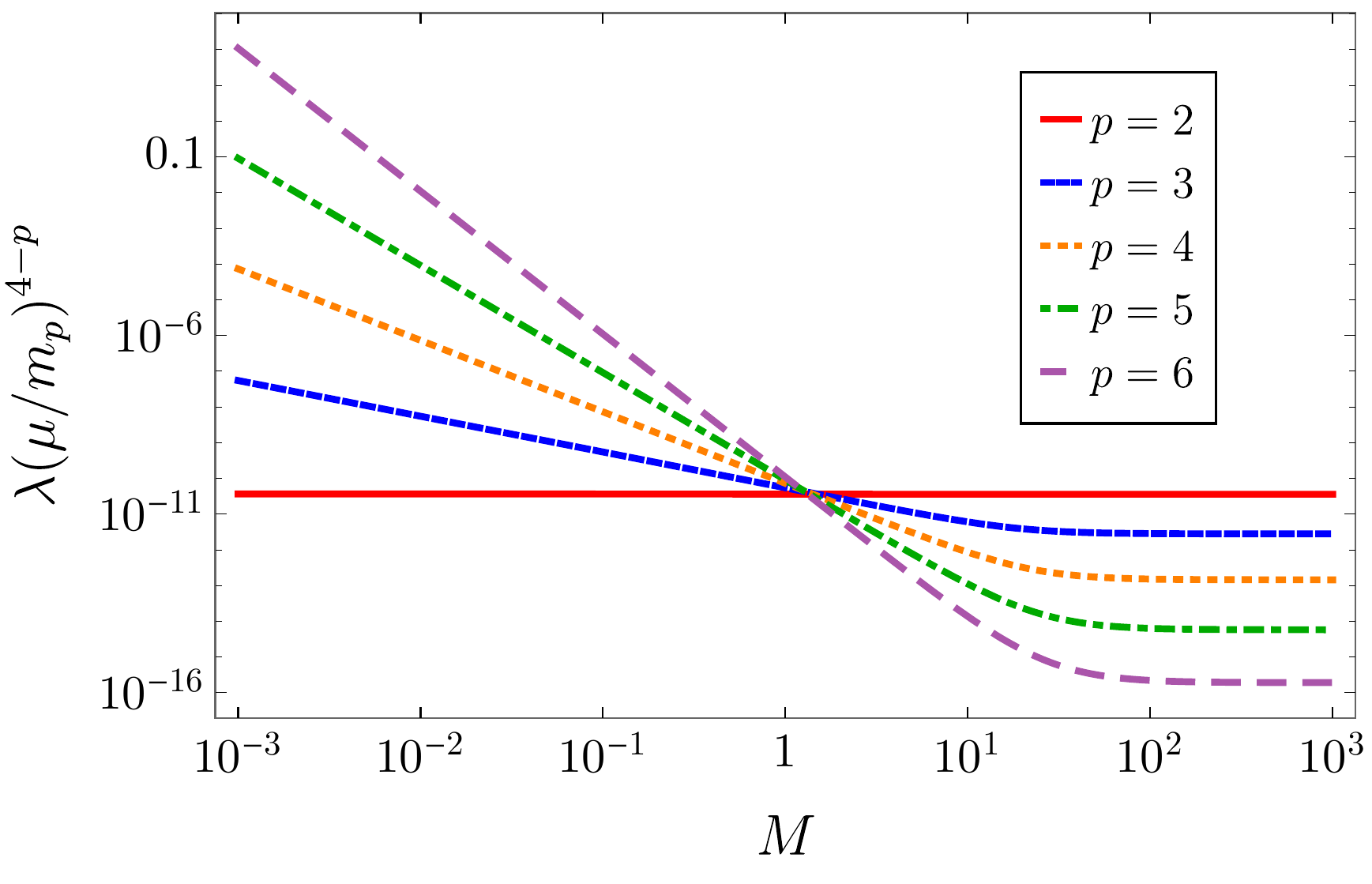}  \\ 
    \caption{Coefficient $\lambda (\mu / m_{\rm pl})^{4-p}$ appearing in the first term of the Taylor expansion of the potential (\ref{eq:powlaw-pot}) as a function of $M$, for different choices of $p$.}  \label{fig:mum}
\end{figure}

The parameters $\{p, \Lambda, M\}$ can be constrained by matching the model predictions of inflationary observables with current observational constraints. In particular, we want to explain the observed values of the amplitude $A_s$ and  spectral tilt $n_s$ of the scalar perturbations, while respecting at the same time the upper bound for the tensor-to-scalar ratio $r$. Predictions are obtained by evaluating $A_s$ and $n_s$ as a function of the potential $V$ and the slow-roll parameters $\epsilon_{_V}$ and $\eta_{_V} \equiv  m_{\rm pl}^2 V_{,\phi \phi} / V$, when the pivot scale $k_\mathrm{CMB} =0.05\, \rm Mpc^{-1}$ crosses the Hubble radius, approximately $\sim 50-60$ e-folds before the end of inflation. Current bounds for $A_s$ and $n_s$ are \cite{Planck:2018jri} 
\bea 
A_s &=& \frac{(V_k/m_{\rm pl}^4)}{24 \pi^2 \epsilon_{V_k}} \nonumber \simeq 2.1 \cdot 10^{-9}  \ ,\\ 
n_s &=& 1+2 \eta_{V_k} - 6 \epsilon_{V_k} = 0.9668 \pm 0.0037 \ , 
\eea 
where quantities with a subindex $k$ must be evaluated when $k_\mathrm{CMB}$ crosses the horizon.
Moreover, the upper bound of the tensor-to-scalar ratio is \cite{BICEP:2021xfz}
\be r_{0.05} = 16 \epsilon_{V_k} < 0.036 \ . \label{eq:Boundr005}\ee
In Fig.~\ref{fig:nsr-M} we show the predictions for $n_s$ and $r$ as a function of $M$ for our inflationary model, for $N_k = 50$ and $N_k=60$. The constraint for $r$ translates to the upper bound $M \lesssim 10 m_{\rm pl}$ in our potential. In particular, by fitting $A_s$, we can determine the following relation,
\be \Lambda^4 =  \frac{3\pi^2A_sM^2m_{\rm pl}^2}{N_k^2}f(p,M,N_k) \ , \label{eq:LabelNk}\ee
with $f = f(p,M,N_k)$ a complicated function that obeys $f \sim 1$ for $M \ll 1$. We provide the exact form of $f$ in Appendix \ref{App:TanhPotential}. For $M\simeq10 m_{\rm pl}$ we get $0.8<f<1.3$ for $p \in [2,6]$. Eq.~(\ref{eq:LabelNk}) can be used to obtain an expression for the coefficient $\lambda \mu^{4-p}$ in front of the monomial potential (\ref{eq:powlaw-pot}), which approximates the full potential (\ref{eq:inflaton-potential}) at small amplitudes. We show in Fig.~\ref{fig:mum} the dependence of this coefficient on $M$, for $p=2 - 6$. It is quite independent on $M$ for $p=2$, giving $m^2 \equiv \lambda \mu^2 \simeq 10^{-11}$ at all scales. In the monomial limit $M \rightarrow \infty$, it is given by
\be \lambda \mu ^{4-p} \equiv \frac{\Lambda^4}{M^p} \xrightarrow[M\to\infty]{} \frac{12\pi^2p^3A_s m_{\rm pl}^{4-p}}{(2pN_k+p^2/2)^{\frac{p+2}{2}}} \ . \label{eq:mu}\ee

The inflationary period terminates when $\phi \lesssim \phi_{*}$, and then the inflaton starts oscillating around the minimum of the potential. For $p \geq2$, we find that $\phi_{\rm i}>\phi_{*}$ for $M \geq M_c \equiv 1.633m_{\rm pl}$. By choosing $M > M_c$ we guarantee that the post-inflationary oscillations of $\phi$ are always contained in the positive-curvature region of the potential. In this case, the inflaton potential can be well approximated by the power-law expression (\ref{eq:powlaw-pot}) during the oscillatory regime. We assume this mass choice in the following. 
 
\subsection{Inflaton oscillations}\label{sec:2A}

The equation of motion (EOM) of an inflaton with potential (\ref{eq:powlaw-pot}) can be written as
\be \ddot \phi - \frac{1}{a^2} \nabla^2\phi + 3 H \dot \phi + \lambda \mu^{4-p} \,|\phi|^{p-2} \phi = 0 \,, 
\label{eq:generic-eom} 
\ee
where $H \equiv {\dot{a}/a}$ is the Hubble rate. Here we consider the first oscillations after inflation, when the effects of interactions with other fields are negligible. Hence, we can also discard the laplacian term in the EOM, and consider $\phi$ as approximately homogeneous. Under these circumstances, the inflaton solution can be approximately parametrized as the product of a \textit{decaying amplitude function} ${\mathcal A}_\phi(t)$ and an \textit{oscillatory function} ${\mathcal F}(t)$ \cite{Turner:1983he}, as
\be \label{eq:ApproxVarPhi} \phi (t) \simeq  {\mathcal A}_\phi(t){\mathcal F}(t) \ , \hspace{0.4cm} {\mathcal A}_\phi(t) \equiv \phi_{\star} \left(\frac{t}{t_{\rm {\star}}}\right)^{-2/p} \ ,  \ee
where $\phi_{\star} \equiv \phi (t_{\star})$ is the field amplitude at a certain time scale $t = t_{\star}$ close to the end of inflation (the exact definition of $\phi_{\star}$ is given in Eq.~(\ref{eq:InitCond-Star}) below). The oscillation period of $F(t)$ changes with time for $p \neq 2$, and it is given by
\be \Omega_{\rm osc} \equiv \sqrt{V_{,\phi} \, \phi^{-1} } = \lambda^{\frac{1}{2}}  \mu^{\frac{4-p}{2}}  {\mathcal A}_\phi^{\frac{p-2}{2}} = \omega_{\star} \left(\frac{t}{t_{\star}}\right)^{\frac{2}{p} - 1} \ , \label{eq:OmegaOsc}\ee
where $\omega_{\star} \equiv  \lambda^{\frac{1}{2}}  \mu^{\frac{4-p}{2}} \phi_{\star}^{\frac{p-2}{2}}$ is the oscillation frequency at $t = t_{\star}$. Shortly after the end of inflation, the oscillations of the homogeneous inflaton give rise to the following \textit{effective} equation of state \cite{Turner:1983he},
\begin{eqnarray}
    \bar{w}_{\rm hom} \equiv \frac{\langle p_\phi \rangle}{ \langle \rho_\phi \rangle} = \frac{p-2}{p+2} \ , \label{eq:EoSoscillations}
\end{eqnarray}
where $\langle \rho_\phi \rangle$ and $\langle p_\phi \rangle$  are the \textit{oscillation-averaged} energy density and pressure of the inflaton respectively. For power-law coefficients $p \in [2,4]$, the equation of state lies within the range $\bar{w}_{\rm hom} \in[0,1/3]$, but a period with a \textit{stiff} equation of state $\bar{w}_{\hom} > 1/3$ can also be envisaged for $p >4$.  Correspondingly, the evolution of the scale factor during this regime can be approximately described as
\be a(t) \simeq  a_{\star} \left( 1 + \frac{3p}{2+p}   H_{\star} (t-t_{\star}) \right)^{\frac{2 + p}{3p}} \ , \label{eq:SfT}\ee
where $a_{\star} \equiv a(t_{\star})$ and $H_{\star} \equiv  H (t_{\star})$. For monomial potentials, after few oscillations it holds that $H_{\star} (t - t_{\star}) \gg 1$, and hence that $a(t)/a_{\star} \sim  (H_{\star}t)^{\frac{2}{3 (1+w)}} = (H_{\star}t)^{\frac{2+p}{3p}}$. We set $a_{\star}= 1$ for convenience from now on. 

Using Eq.~(\ref{eq:SfT}), we can predict the following scaling behaviors,
\be {\mathcal A}_\phi \propto a^{-\frac{6}{p+2}} \ , \hspace{0.4cm} \Omega_{\rm osc} \propto a^{\frac{3 (2-p)}{(p+2)}} \ . \label{eq:PhiOmegaScaling}\ee
In light of (\ref{eq:PhiOmegaScaling}), it is convenient to redefine the field and spacetime variables. First, we introduce an inflaton amplitude transformation as
\be \varphi \equiv a^{\frac{6}{p+2}} \frac{ \phi}{\phi_{\star}} \ , \label{eq:newvars3} \ee
such that the amplitude of the oscillations of $\varphi$ remains constant in time. Second, we define new spacetime variables $(u, \vec{y})$ as 
\be t \rightarrow u \equiv \int_{t_{\star}}^t {\omega_{\star} \, a(t')^{\frac{3 (2-p)}{2+p}}dt'} \ , \hspace{0.4cm} \vec{x} \rightarrow \vec{y} \equiv \omega_{\star} \vec{x} \ . \label{eq:newvars1}\ee
In this way, the oscillation period of the function $ \varphi = \varphi (u)$ will be approximately constant in time.  We will collectively refer to the variables defined in Eqs.~(\ref{eq:newvars3})-(\ref{eq:newvars1}) as  \textit{natural  variables}. Derivatives with respect to natural time and space variables will be denoted as $' \equiv  d/ du$ and $\nabla_{\vec y} \equiv d / d \vec{y}$ respectively. 

In terms of natural variables, the homogeneous inflaton EOM is written as
\be \varphi'' + ( |\varphi |^{p-2}  + F(u) )\varphi = 0 \label{eq:inflaton-homeq} \ , \ee
where
$F(u)$ is a time-dependent functional of the scale factor and its derivatives,
\be F[a'/a,a''/a] \equiv \frac{6}{(p+2)}\left(\frac{(p-4)}{(p+2)} \left( \frac{a'}{a} \right)^2 - \frac{a''}{a}\right) \ . \label{eq:sfterm-eom} \  \ee
By substituting Eq.~(\ref{eq:SfT}) into (\ref{eq:newvars1}), it turns out that the scale factor in natural time evolves as $a \sim u^{\frac{p+2}{6}}$, so that $F(u) \propto u^{-2}$. Therefore, the term $\propto F(u)$ in Eq.~(\ref{eq:inflaton-homeq}) becomes negligible after a few inflaton oscillations, and we discard it in the following analysis. 

In order to solve the differential equation (\ref{eq:inflaton-homeq}), we need to decide at which time to set the initial conditions, i.e.~we need to define the time $t=t_{\star}$. By construction, this corresponds to $u=0$ in natural time, and to an initial field amplitude ${\varphi(u=0)=1}$. We then define $t_{\star}$ from a condition over the initial time-derivative, in particular requiring that $\varphi' (u=0) = 0$. In physical variables, this can be written as
\be \varphi' (u=0) = 0 \,\,\, \rightarrow \,\,\, \dot{\phi}_{\star} = - \frac{6 H_{\star}}{p+2}  \phi_{\star} \ . \label{eq:InitCond-Star}\ee
The field amplitude $\phi_{\star}$ can be obtained by solving the coupled inflaton and scale factor equations numerically, with initial conditions deep in the slow-roll regime, until the above condition holds. This gives
\bea 
\frac{ \phi_{\star} }{ m_{pl} } &\simeq & \{0.97, 1.34, 1.72, 2.09, 2.47\} , \label{eq:PhiHStar} \\
\frac{ H_{\star} }{ \omega_{\star} } &\simeq & \{0.49, 0.59, 0.69, 0.79, 0.89\} , \hspace{0.15cm} \text{for}  \hspace{0.15cm}  p=2, 3, 4, 5, 6,  \nonumber 
\eea    
where we also indicate the values of the Hubble parameter $H_{\star} = H (t_{\star})$ at this time. Note that we have $\phi_{\star} \approx \phi_*$, although for $p>2$ the oscillatory regime always begins shortly after the end of inflation.

In the approximation $F(u) \simeq 0$, the solution of Eq.~(\ref{eq:inflaton-homeq}) for $p=2$ is simply $\varphi = \cos (u)$, whereas for $p=4$, the solution is given instead by an elliptic cosine, the {\it Jacobi} function $\varphi = {\rm cn} (u, 1/2)$. In this second case, the solution can be nonetheless approximated well by the first term of its harmonic expansion, $\varphi  \simeq {\rm cos} (0.8472 u)$~\cite{Greene:1997fu}. Actually, we have checked that for any value of $p \geq 2$, the numerical solution for $\varphi$ can always be approximated by simple cosine functions with different oscillation periods. In particular, we can write
\be \varphi \simeq \cos ( \beta_{\bar{\varphi}} u) \ , \hspace{0.4cm} \beta_{\bar{\varphi}} \equiv \frac{ 2 \pi }{ \omega_* T_{\rm {\varphi}} } \ , \label{eq:Infl-sol} \ee
where $T_{\rm {\varphi}}$ is the oscillation period of the inflaton, which can be computed as follows,
\bea \omega_*T_{\rm {\varphi}} &=& 4\int_{u({\varphi}=0)}^{u({\varphi}=1)}du=4\int_{{\varphi} = 0 }^{{\varphi} = 1} \frac{1}{{\varphi}'}d {\varphi} \label{eq:Inflaton-period} 
\\
&=&\sqrt{8p}\int_{{\varphi}=0}^{{\varphi}=1} \frac{d {\varphi}}{\sqrt{1-|{\varphi}|^p}} 
=\sqrt{8p\pi}\frac{\Gamma[1+\frac{1}{p}]}{\Gamma[\frac{1}{2}+\frac{1}{p}]}\ ,\nonumber \eea
In the third equality we have used that the energy $E_{{\varphi}} \equiv \frac{1}{2} {\varphi}'^2 + \frac{1}{p} |{\varphi}|^p = \frac{1}{p}$ is conserved.
We obtain
\be \beta_{\bar\varphi} \simeq  \{1, 0.92, 0.85, 0.79, 0.75\} , \hspace{0.18cm} \text{for}  \hspace{0.2cm}  p=2, 3, 4, 5, 6.
\label{eq:BetaNumbers} \ee
  
We now derive an expression for the \textit{oscillation-averaged} energy density stored in the inflaton homogeneous condensate,
\bea \rho_{\bar \phi} &\equiv & \frac{1}{2} \langle\dot{{\phi}}^2\rangle  + \langle V ({\phi})\rangle  \nonumber \\*
&\simeq & \omega_{\star}^2 \phi_{\star}^2 a^{\frac{-6 p}{p+2}} \left( \frac{1}{2} \langle {\varphi}'^2 \rangle + \frac{1}{p} \langle | {\varphi}|^p \rangle \right) \nonumber \\*
&=& \frac{\omega_{\star}^2 \phi_{\star}^2}{p a^{\frac{6 p}{p+2}}} \ . \label{eq:en-homc}\eea
In the first equality we have extracted the scale factor from the oscillation averages, as it does not change significantly during one oscillation period. In the second equality we have used that $\langle \varphi'^2 \rangle = \langle \varphi^p \rangle = 2 /(p+2)$, as computed in Eqs.~(\ref{eq:dotVarphi2}) and (\ref{eq:Varphip}) of Appendix \ref{App:PeriodAverages}. As expected, the energy redshifts as non-relativistic matter for $p=2$, and as radiation for $p=4$.

Finally, let us remark that the inflaton solution (\ref{eq:Infl-sol}) has been obtained under the assumption in Eq.~(\ref{eq:inflaton-homeq}) that $F(u) = 0$ holds exactly at all times. This is however not really true during the initial oscillations of the inflaton. Thus, neither the oscillation-averaged equation of state nor the scale factor is given exactly by Eqs.~(\ref{eq:EoSoscillations}) and (\ref{eq:SfT}) during the initial oscillatory stage. We quantify next the discrepancy between our analytical approximations and the real solution, obtaining the latter by solving numerically the coupled equations of the field and scale factor evolution with initial conditions deep inside the slow-roll regime. In this regard, we introduce the following parametrizations,
\bea 
\bar{\varphi} (u) &\simeq& (1+\delta_1) \cos ( (1+\delta_2) \beta_{\bar{\varphi}} u )  \ , \label{eq:inf-corr} \\
a(u) &\simeq& (1- \delta_3) \left( 1 + \frac{6}{p+2}\frac{H_{\star}}{\omega_{\star}} u \right)^{\frac{p+2}{6}} \ , \label{eq:sf-corr}
\eea
where $\delta_a$ ($a=1,2,3$) are three correction factors that account for such discrepancy. Fitting Eqs.~(\ref{eq:inf-corr}) and (\ref{eq:sf-corr}) to the numerical solutions, we find reasonable agreement by fixing
\bea \delta_1 &=& \{0.00, 0.02, 0.04, 0.04, 0.05\}, \nonumber \\
\delta_2 &=& \{0.00, 0.02, 0.04, 0.05, 0.07\}, \nonumber \\
\delta_3 &=& \{0.12, 0.17, 0.23, 0.28, 0.34\},\nonumber \\
&& \hspace{2cm} \text{for} \hspace{0.2cm} p=\{2, 3, 4, 5, 6\}. \eea 
While the corrections for the inflaton solution are minor, the corrections for the scale factor are more sizeable and become more significant for larger values of $p$.

\section{Analytical analysis of post-inflationary dynamics}\label{Sec:LinearAn} 

Having discussed the post-inflationary oscillations of the inflaton homogeneous mode, we now proceed to include a quadratic-quadratic interaction with a massless scalar field $X$ and study the growth of the inflaton and daughter field fluctuations. For the analytical discussion we present now, we consider the following potential during preheating,
\bea V(\Phi, X) &\equiv& V (\Phi) + V_{\rm i} (\Phi, X) \nonumber \\ 
&=& \frac{1}{p} \lambda \mu^{4-p} |\Phi - v|^p + \frac{1}{2} g^2 \Phi^2 X^2 \ , \label{eq:FullPotential} \eea
where $g^2$ is a dimensionless coupling constant and
$\Phi$ is the inflaton. We will refer to the potential with $v=0$ as the \textit{centred} potential, and to the one with $v > 0$ as the \textit{displaced} potential. It is convenient to express, as in the previous section, the potential in terms of a \textit{shifted field} $\phi\equiv\Phi-v$, such that the potential is always centred around $\phi=0$. \textit{From now on we will only work with the shifted field} $\phi$ \textit{and refer to it as the inflaton}. The total potential then reads
\bea V (\phi, X) &=& V (\phi)+V (X)+V_{\rm i} (\phi,X) \nonumber \\ 
&=& \frac{1}{p} \lambda \mu^{4-p} |\phi|^p + \frac{1}{2}g^2v^2X^2 \nonumber \\
&+& \frac{1}{2}g^2(2v\phi+\phi^2)X^2\ .\label{eq:shifted-inflaton-potential1}\eea
We see that for $v \neq 0$ a trilinear interaction arises between the inflaton and the daughter field, and the latter also acquires a non-zero mass $m^2_X \equiv g^2 v^2$. In any case, the post-inflationary dynamics of the inflaton homogeneous mode during the initial linear regime does not depend on the choice of $v$ and behaves as described in Section \ref{sec:2A}. In particular, the inflaton still oscillates around the minimum of the potential with time-dependent frequency (\ref{eq:OmegaOsc}), which gives rise to the effective equation of state (\ref{eq:EoSoscillations}). Therefore, it is still convenient to work with \textit{natural} field variables ($\varphi$, $\chi$), 
\be \varphi \equiv \frac{1}{\phi_{\star}} a^{\frac{6}{p+2}} \phi \ , \hspace{0.4cm} \chi \equiv \frac{1}{\phi_{\star}} a^{\frac{6}{p+2}} X \ , \label{eq:newvars2} \ee
as well as with the \textit{natural} spacetime variables (\ref{eq:newvars1}).

In Section \ref{subsec:Analytics-centred} we present, for the centred potential, an analytical study of the two relevant resonant excitation processes after inflation: inflaton self-resonance and parametric resonance of the daughter field. In Section \ref{sec:Lin:DisplacedPotential} we extend our study of parametric resonance to the case of the displaced potential. Finally, in Section \ref{subsec:EnEos} we present useful expressions for the equation of state and energy components of the two-field system.  

\subsection{Analytical analysis of resonances: Centred potential} \label{subsec:Analytics-centred}

Let us first consider the centred potential case with $v=0$. The inflaton and daughter field equations in natural variables are
\bea
&& \hspace{-0.4cm}\varphi'' - a^{\frac{-(16 - 4p)}{2+p}} \nabla^2_{\vec{y}} \varphi +  ( |\varphi|^{p-2} + \tilde q \chi^2  + F(u) ) \varphi = 0 \ , \label{eq:fullEOMs1} \hspace{0.7cm}  \\
&& \hspace{-0.4cm}\chi''- a^{\frac{-(16 - 4p)}{2+p}} \nabla^2_{\vec{y}} \chi + ( \tilde q\varphi^2 + F(u) ) \chi = 0 \ , \label{eq:fullEOMs2}  
\eea
where $F(u) \sim u^{-2}$ is the function defined in Eq.~(\ref{eq:sfterm-eom}) (which becomes negligible after a few inflaton oscillations), and $\tilde q \equiv \tilde q (u)$ is the time-dependent \textit{effective resonance parameter}, defined as
\be\label{eq:ResParam}
\tilde q (u) \equiv q_{\star} a^{\frac{6 (p-4)}{p+2}} \ , \hspace{0.3cm} q_{\star} \equiv \frac{g^2\phi_{\star}^2}{ \omega_{\star}^2} \ , \ee
with $q_{\star}$ the (dimensionless) \textit{initial} resonance parameter. At the onset of oscillations we have $\tilde q = q_{\star}$, but as the Universe expands, $\tilde q$ changes in different ways depending on the value of $p$: it decreases for $p<4$, increases for $p>4$, and remains constant for $p=4$.

Let us now expand the fields at linear order as
\bea \varphi (\vec{y}, u) & \equiv & \bar{\varphi} (u) + \delta \varphi (\vec{y},u) \ , \\
\chi (\vec{y}, u) &\equiv &  \delta \chi (\vec{y},u) \ , \eea
where the bar denotes the inflaton homogeneous mode. We have set the initial homogeneous mode of the daughter field to zero, as it is made purely out of vacuum quantum fluctuations. Under the approximation $F = 0$ we have $\bar{\varphi} \simeq \cos ( \beta_{\bar \varphi} u)$, with $\beta_{\bar \varphi}$ given in 
Eq.~(\ref{eq:Inflaton-period}) for each value of $p$. On the other hand, the EOM in momentum space of the inflaton and daughter field fluctuations at linear order are
\bea\delta \varphi^{''}_k + \tilde\omega_{k,\varphi}^2 \delta \varphi_k = 0&& \hspace*{-0.55cm}\ , \hspace{0.3cm} \tilde\omega_{k,\varphi} \equiv \sqrt{ \tilde \kappa^2 (a) + (p-1) |\bar{\varphi}|^{p-2} }  \label{eq:linear1} \ , \hspace{0.6cm}  \\
\delta \chi^{''}_k +  \tilde\omega_{k,\chi}^2  \delta \chi_k = 0&& \hspace*{-0.55cm}  \ , \hspace{0.3cm}  \tilde\omega_{k,\chi} \equiv \sqrt{ \tilde \kappa^2 (a) + \tilde q (a) \bar{\varphi}^2} \ , \label{eq:linear2}\eea
where we define in natural units the \textit{resonance momenta} $\tilde \kappa$ as 
\be \tilde \kappa(a) \equiv  \kappa\, a^{\frac{-(8 - 2p)}{2+p}}\,,~ \kappa \equiv \frac{k}{\omega_{\star}}\,. \label{eq:resonant-momentum}  \ee
Note that $\tilde\kappa$ is neither the comoving momentum $\kappa \equiv k / \omega_{\star}$ nor the physical momentum $\kappa_{\rm ph} (a) \equiv {\kappa} / a$, although it coincides with the former for $p = 4$ and the latter for $p = 2$. Therefore, even if the resonance happens at constant scales in terms of $\tilde \kappa$, the range of excited physical/comoving momenta typically change as the universe expands.

As seen in the lineralized equations, the fluctuations of both fields have time-dependent effective masses induced by the oscillations of the inflaton field. In \textit{natural} (dimensionless) units,
\bea
m^2_{\varphi,\mathrm{eff}} &=& (p-1) |\bar{\varphi} |^{p-2} 
\label{effmass1} \ , \\
m^2_{\chi,\mathrm{eff}} &=& \tilde q (a) \bar{\varphi}^2 \label{effmass2} \ .
\eea
Each time the inflaton crosses the minimum of its potential, the effective masses vary non-adiabatically (for sufficiently large $\tilde q$), inducing a strong growth of the fluctuations of both fields. By means of a Floquet analysis we shall see that the field fluctuations have exponentially growing solutions for certain ranges of $\tilde \kappa$, with the effect controlled by $p$ in the case of the inflaton, and by $\{q_{\star}, p\}$ in the case of the daughter field. Due to this, the particle number of both fields grows as $n_k^\varphi \sim |\delta \varphi_k |^2 \propto e^{2\mu_k u}$ and $n_k^\chi \sim |\delta \chi_k |^2 \propto e^{2\nu_k u}$ respectively, where $\mu_k$ and $\nu_k$ are the \textit{Floquet indices} of each field. This gives rise to two distinct resonance effects that play a major role in the early preheating phase: \textit{self-resonance of the inflaton}, and \textit{parametric resonance of the daughter field}. In the following we describe these resonances in more detail.

\subsubsection{Self-resonance of the inflaton}\label{sec:3A}

\begin{figure}
    \centering
    \includegraphics[width=0.47\textwidth]{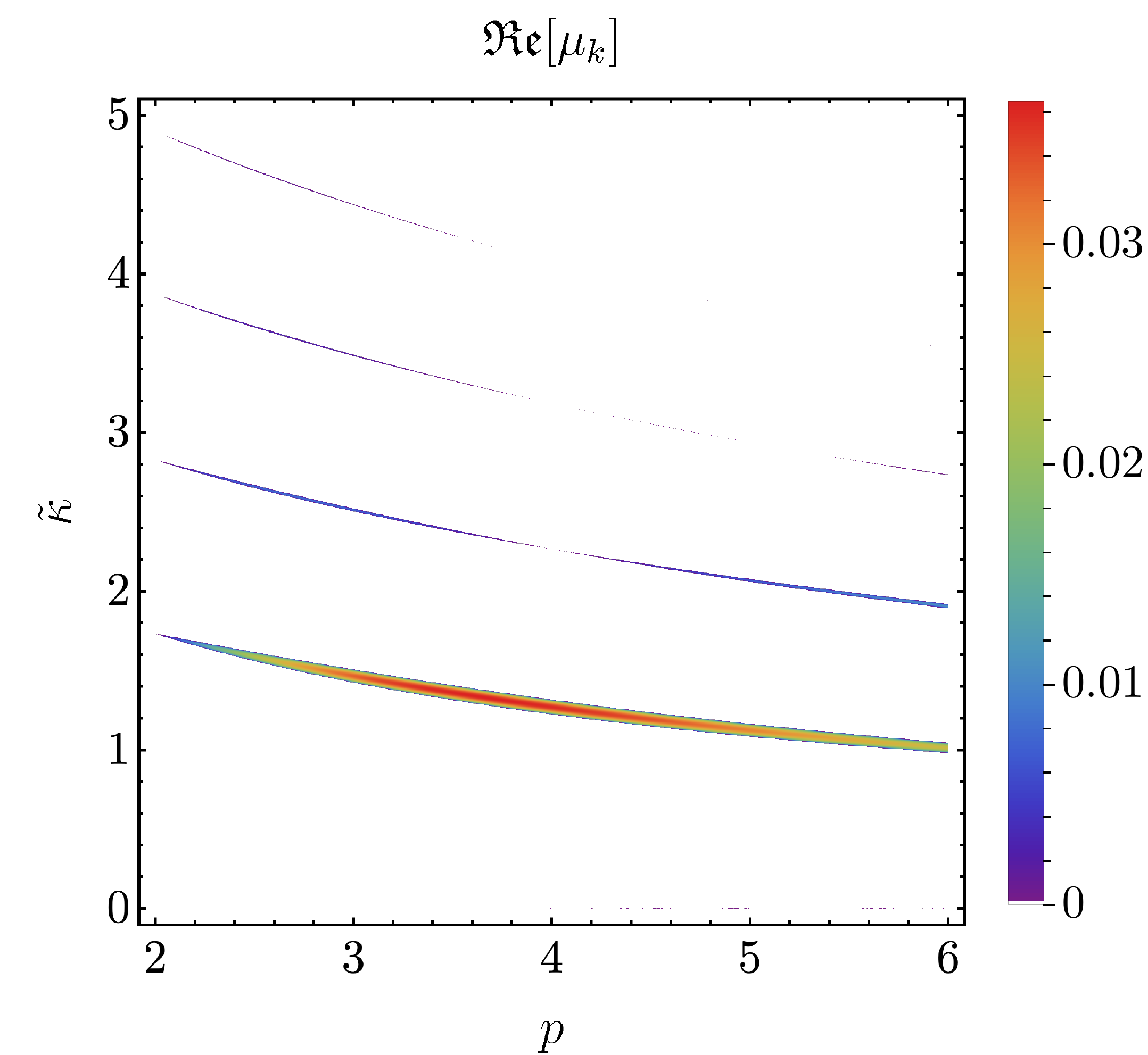}
    \caption{   
        Floquet chart for inflaton self-resonanace, which shows the real part of the inflaton Floquet index $\mu_k$ as a function of $p$ and $\tilde \kappa$, obtained from the solution of Eq.~(\ref{eq:linear1}) and (\ref{eq:inflaton-homeq}). White areas correspond to $\mathfrak{Re} [\mu_k] = 0$. 
        }
    \label{fig:FloquetDiagrams}
\end{figure}

For $p \neq 2$, the fluctuations of the inflaton have a time-dependent effective mass due to the oscillations of the homogeneous mode (see \ref{effmass1}). This leads to exponential growth of some modes in a process of \textit{self-resonance}. More precisely, Eq.~(\ref{eq:linear1}) admits solutions of the type $\varphi_k \sim e^{\mu_k u}$ with $\mu_k$ the Floquet index. Fig.~\ref{fig:FloquetDiagrams} shows the real part of this coefficient as a function of $\tilde \kappa$ and $p$. The band with the lowest momenta is the widest one and exhibits the largest average Floquet index for all values of $p$, so it dominates the self-resonance process. The maximum Floquet index in the entire parameter space is $\mu_k \simeq 0.036$, obtained for $p \simeq 3.6$. Remarkably, the expansion of the universe does not diminish the strength of the resonance for a fixed value of $\tilde \kappa$, as the corresponding Floquet index remains constant. 
This is a fundamental difference with respect to parametric resonance of daughter fields as we shall see.

After the first zero-crossing of the inflaton, we have $\rho_{\delta \phi} \ll \rho_{\bar{\phi}}$, where $\rho_{\delta \phi}$ and $\rho_{\bar{\phi}}$ are the energy of the inflaton fluctuations and homogeneous mode respectively, see Eq.~(\ref{eq:en-homc}). However, the energy stored in the fluctuations grows exponentially fast as the inflaton oscillates, and it eventually becomes large enough that nonlinear effects become relevant, leading to the decay of the inflaton homogeneous mode. We refer to this moment as the \textit{backreaction} time. An analytical estimation of this quantity was obtained in Ref.~\cite{Lozanov:2017hjm}, which correctly approximates the more exact result obtained from lattice simulations. Ref.~\cite{Lozanov:2017hjm} parametrization depends on a constant $\delta$, which is constrained with input from the lattice. In the following we propose an alternative estimation for the backreaction time that does not require such input.

We define the backreaction time as the instance when $\rho_{\delta \phi} = \rho_{\bar{\phi}}$. The expression of $\rho_{\delta \phi}$ can be written as follows,
\bea \rho_{\delta \phi} &=& \frac{1}{2 \pi^2 a^3} \int dk k^2 \Omega_{k,\phi} n_{k,\phi} \nonumber \\
&=& \frac{\omega_{\star}^4  }{2 \pi^2}  a^{\frac{12 (2 -p)}{p+2}} \int d {\tilde \kappa} \tilde \kappa^2 \tilde\omega_{k,\varphi}  n_{k,\phi} \ , \label{eq:DaughterIntegral}\eea
where we have defined the frequency of the mode as 
\bea
\Omega_{k,\phi} &\equiv& \sqrt{(k/a)^2 + (p-1) \lambda \mu^{4-p} |\bar \phi|^{p-2}}\nonumber  \\
&=&  \omega_{\star} a^{\frac{3 (2-p)}{p+2}} \sqrt{\tilde \kappa^2 + (p-1) | \bar \varphi|^{p-2} }\nonumber\\ 
&\equiv& \omega_{\star}  a^{\frac{3 (2-p)}{p+2}}  \tilde\omega_{k,\varphi}\,.
\eea

\begin{figure}
    \centering
    \includegraphics[width=0.48\textwidth]{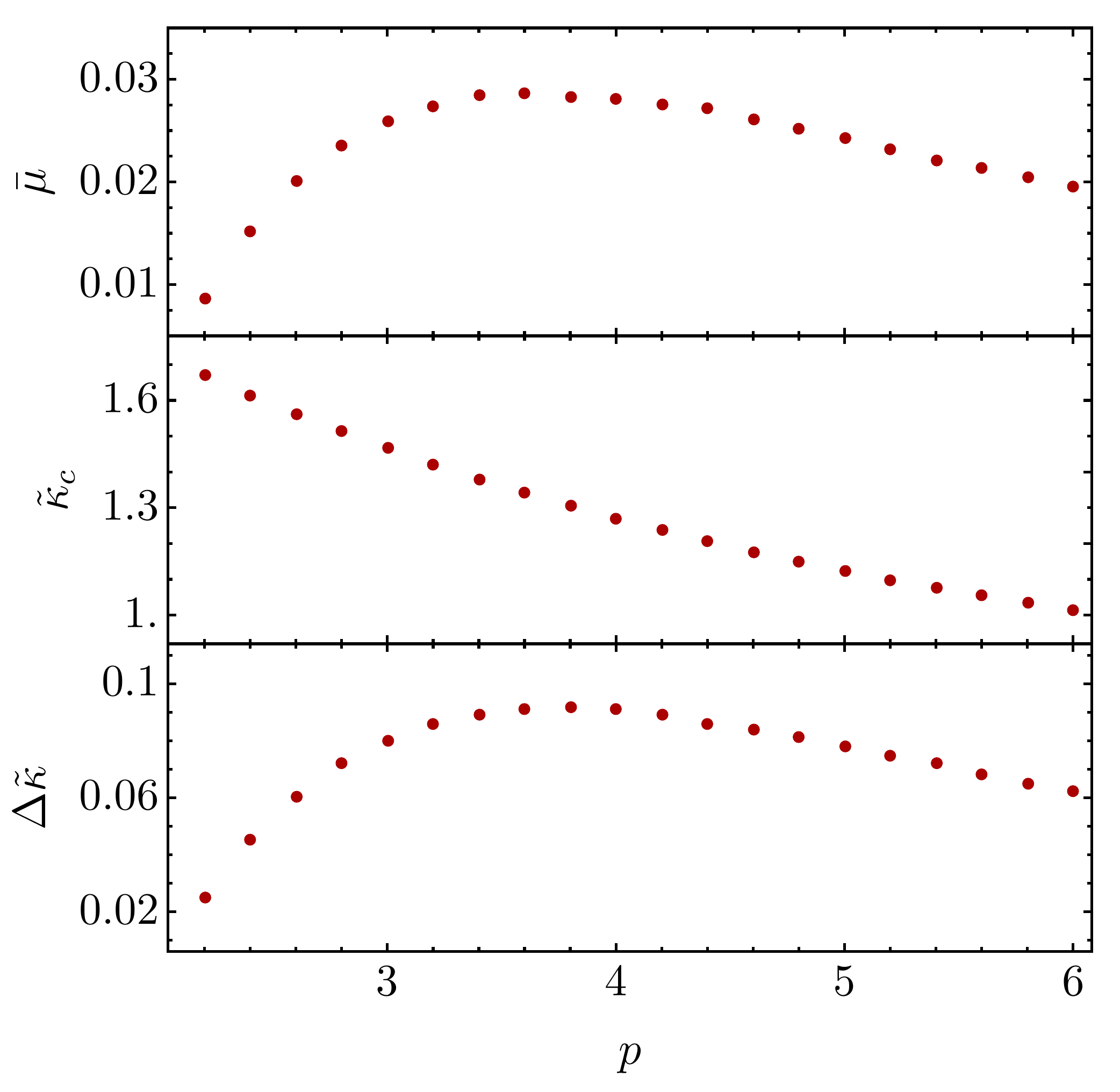}
    \caption{Parametrization of the main resonance band appearing in the Floquet chart of inflaton self-resonance. From top to bottom: average Floquet index $\bar{\mu}$, average momenta of the band $\tilde \kappa_{\rm c}$ and band width $\Delta \kappa$ as a function of $p$.}
    \label{fig:SelfResProperties}
\end{figure}

According to the Floquet diagram of Fig.~\ref{fig:FloquetDiagrams}, the most important contribution to the integral comes from the lowest-momenta resonance band. Therefore, we can roughly approximate the exponentially growing particle number of the excited modes as
\bea
n_{k,\phi} \simeq \begin{cases}
    e^{2 \bar{\mu}u}  \hspace{0.4cm}  &\text{for } \tilde \kappa_{-} < \tilde \kappa < \tilde \kappa_{+} \ , \\
    0 \hspace{0.7cm} &\text{otherwise}. \label{eq:Nk-selfres}
\end{cases}
\eea
where $\tilde \kappa_{-}$ and $\tilde \kappa_{+}$ are the minimum and maximum (resonance) momenta of the main resonance band, and we approximate $\mu_k$ by the average Floquet index within the band, denoted as $\bar{\mu}$. We can parametrize the band in a similar fashion as in Ref.~\cite{Lozanov:2017hjm}. In particular, we depict in Fig.~\ref{fig:SelfResProperties} the average Floquet index $\bar{\mu}$, the band width $\Delta \tilde \kappa \equiv \tilde \kappa_+ - \tilde \kappa_-$, and the central momentum of the band $\tilde {\kappa}_{\rm c} \equiv (\tilde \kappa_+ + \tilde \kappa_{-})/2$, all as a function of $p$.

\begin{figure}
    \centering
    \includegraphics[width=0.48\textwidth]{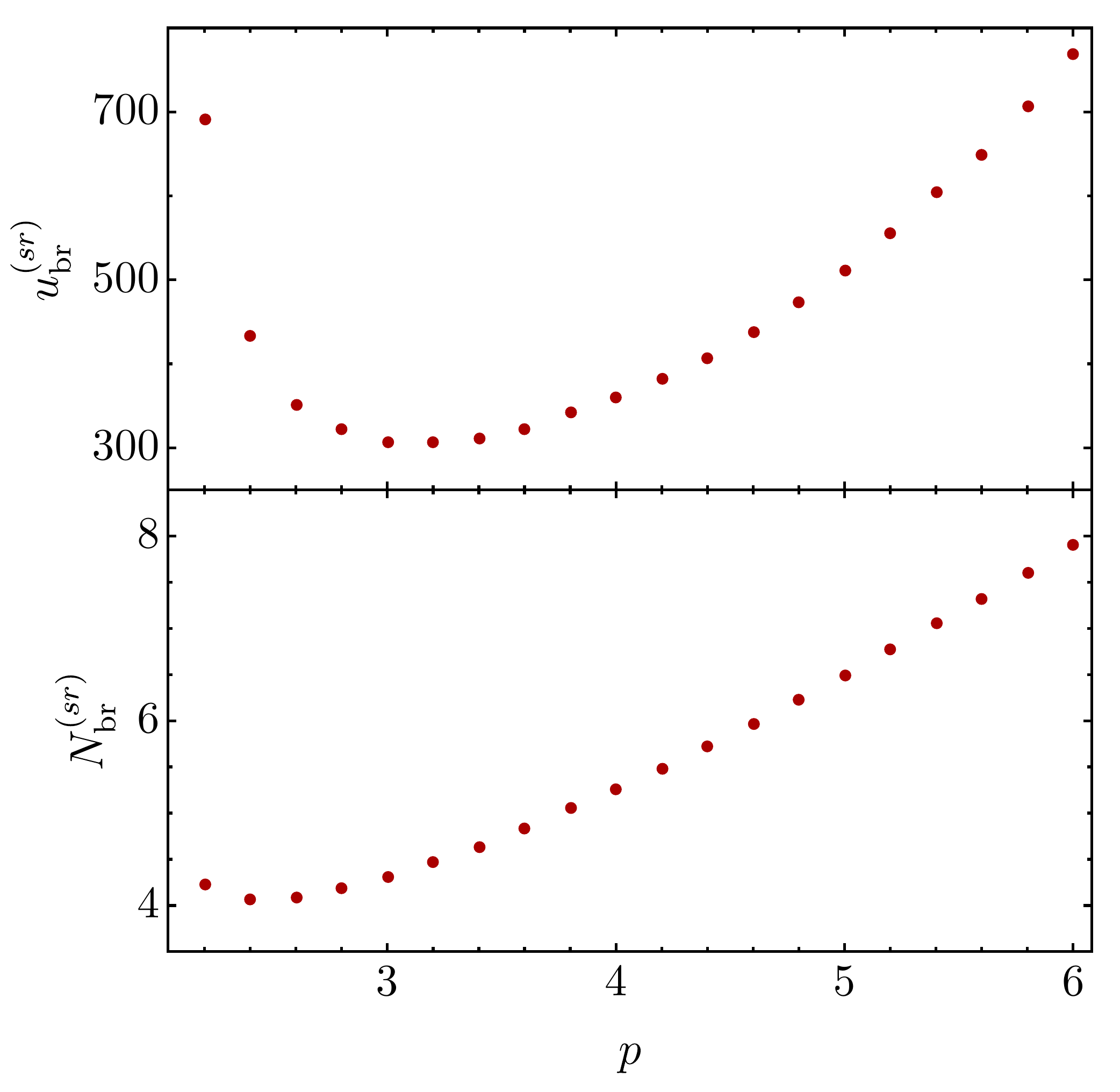}
    \caption{Estimate of the backreaction time for inflaton self-resonance as a function of $p$, in terms of natural time $u_{\rm br}$ (top) and post-inflationary number of e-folds $N_{\rm br}$ (bottom).} 
    \label{fig:PvszbrNbr}
\end{figure}

Plugging (\ref{eq:Nk-selfres}) into (\ref{eq:DaughterIntegral}) leads to
\bea
\label{eq:InfFluctEnergy}
\rho_{\delta \phi} &\simeq& \frac{\omega_{\star}^4  }{2 \pi^2}  a^{\frac{12 (2 -p)}{p+2}}  \int_{\tilde \kappa_-}^{\tilde \kappa_+} d {\tilde \kappa} \tilde \kappa^2 e^{2 \bar{\mu} u} \tilde\omega_{k,\varphi} \\
&\approx & \frac{\omega_{\star}^4  }{2 \pi^2}  a^{\frac{12 (2 -p)}{p+2}} \tilde \kappa_{\rm c}^2 \Delta \tilde \kappa   \sqrt{\tilde \kappa_{\rm c}^2 + (p-1) \langle | \bar{\varphi}|^{p-2} \rangle } e^{2 \bar{\mu} u} \ , \nonumber \eea
where we have approximated the inflaton effective mass by its oscillation-averaged expression (see Eq.~(\ref{eq:dotVarphipm2}) in Appendix \ref{App:PeriodAverages} for a computation of $\langle |\bar{\varphi}|^{p-2} \rangle$). Note that the field fluctuations are neither fully relativistic nor non-relativistic, as the typical excited (resonance) momenta are ${\tilde \kappa \sim (p-1) \langle | \bar{\varphi}|^{p-2} \rangle} \sim \mathcal{O} (1)$. By equating (\ref{eq:InfFluctEnergy}) to (\ref{eq:en-homc}), we can estimate the \textit{backreaction time} as the solution to the following equation,
\bea u_{\rm br} & \simeq & \frac{1}{2 \bar{\mu} (p)}  \times  \left[\frac{6 (p-4)}{p+2}  \log a(u_{\rm br}) + \log \left( \frac{\phi_{\star}^{4-p}}{\lambda \mu^{4-p}} \right) \right. \nonumber\\
&&\left. + \log  (2 \pi^2 / p) -  \log \Delta \tilde \kappa (p) -  2 \log  \tilde \kappa_{\rm c} (p) \right. \nonumber \\
 &&\left.  - \frac{1}{2} \log \left( \tilde \kappa_{\rm c}^2 (p) + (p-1) \frac{\Gamma[\frac{p-1}{p}]  \Gamma[\frac{p+2}{2p}]}{\Gamma[\frac{1}{p}] \Gamma[\frac{3p-2}{2p}]} \right)  \right] \label{eq:ubr-selfr},
\eea      
which is implicit for $p \neq 4$. We can solve numerically this expression for monomial potentials using Eqs.~(\ref{eq:mu}), (\ref{eq:PhiHStar}) and (\ref{eq:sf-corr}). The solution is shown in Fig.~\ref{fig:PvszbrNbr}, where we depict both $u_{\rm br}$ and the number of e-folds $N_{\rm br} \equiv \log (a(u_{\rm br}))$ as a function of $p$. We observe that the minimum backreaction time corresponds to $u_{\rm br} \simeq 300$, attained at $p \simeq 3.2$. If we either decrease or increase $p$, the backreaction time increases, reaching $u_{\rm br} \gtrsim 700$ for $p=2$ and $p=6$. The behaviour is different in terms of e-folds, as we obtain $N_{\rm br} = 4 - 4.5$ for $p=2.2-3$, while it grows (almost linearly) for larger values of $p$, reaching $N_{\rm br} \simeq 8$ for $p=6$.

\subsubsection{Parametric resonance of the daughter field} \label{sec:3B} 

The oscillations of the inflaton condensate can also trigger an exponential growth of the fluctuations of the daughter field. If $\tilde q > 1$, the effective frequency of the daughter field changes non-adiabatically as $\tilde\omega_{k,\chi}' / \tilde\omega_{k,\chi}^2 \gg 1$, so that each time the inflaton crosses zero, it excites daughter field fluctuations via a process of parametric resonance. Due to this, daughter field modes evolve as $\chi_k^2 \sim e^{2 \nu_k u}$ with  $\mathfrak{Re}[\nu_k] > 0$ for specific resonance momentum bands in $\tilde \kappa$, where $\nu_k$ is the corresponding Floquet index. Compared to the self-resonance case, the Floquet index $\nu_{k} \equiv \nu_{k} (\tilde q,p)$ depends now not only on $p$, but also on $\tilde q$, and hence on the scale factor.  Therefore, unlike the case of inflaton self-resonance, the strength of the resonance of $X$ for a fixed $\tilde \kappa$ changes as the universe expands.

\begin{figure*}
    \includegraphics[width=0.99\textwidth]{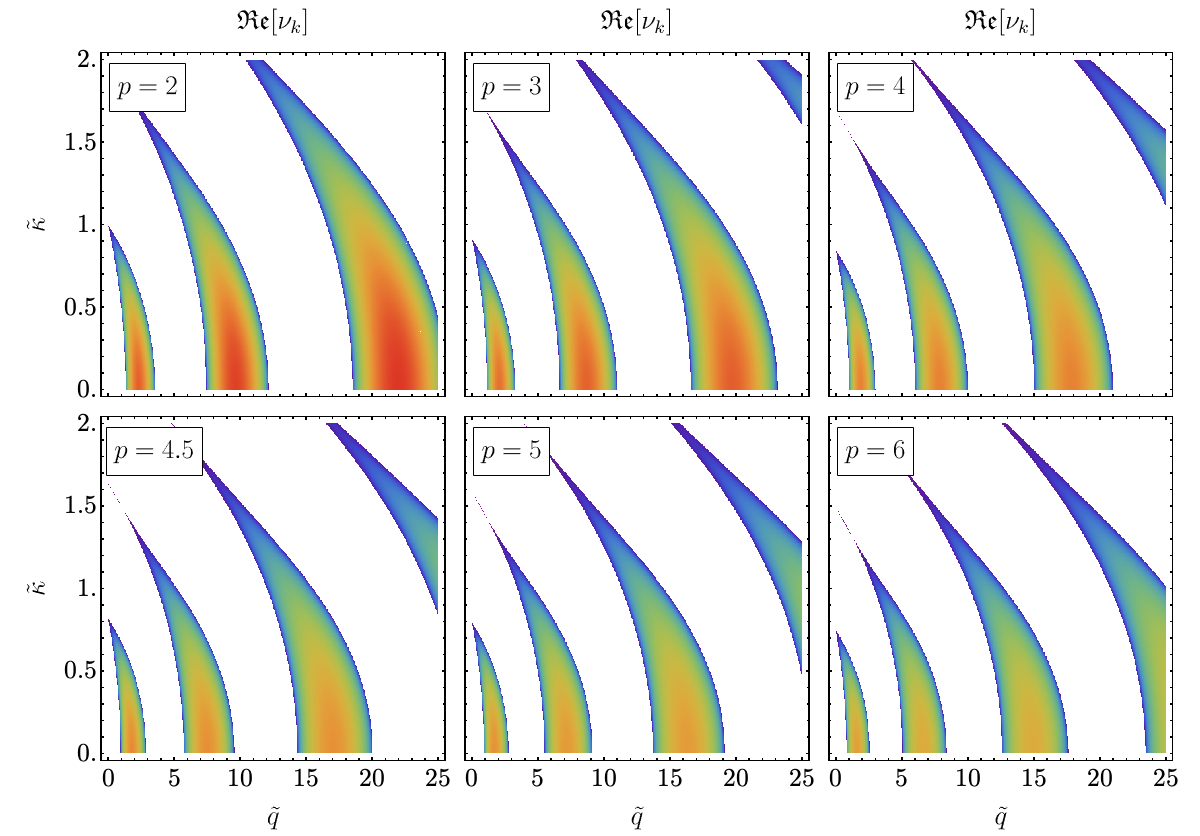}\vspace{0.1cm} \\
	\centering  
    \includegraphics[width=0.38\textwidth]{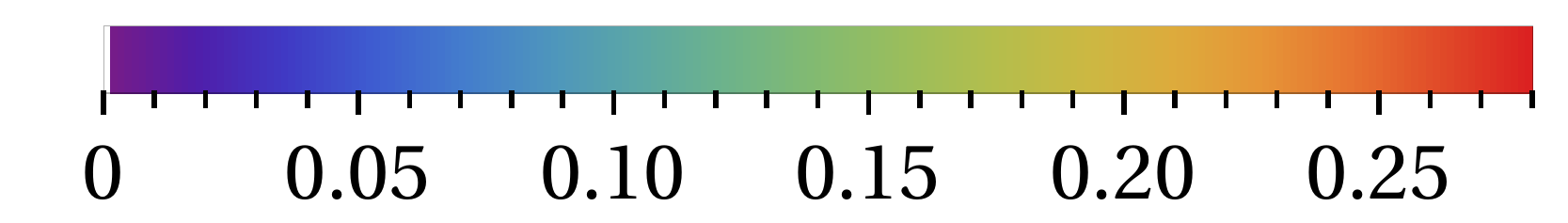}  
    \caption{Floquet diagrams for parametric resonance of the daugher field in the centred potential case, for $p=2$, $3$, $4$, $4.5$, $5$ and $6$, obtained from the solution of Eq.~(\ref{eq:linear2}) and (\ref{eq:inflaton-homeq}). The Floquet index $\mathfrak{Re}[\nu_k]$ is depicted in terms of $\tilde q$ and $\tilde \kappa$. White regions correspond to parameters for which $\mathfrak{Re}[\nu_k]=0$.}  \label{fig:AllFloquetDiagrams}
\end{figure*}

In Fig.~\ref{fig:AllFloquetDiagrams} we show the Floquet charts of the daughter field for $p=2$, 3, 4, 4.5, 5, and 6, in terms of $\tilde \kappa$ and $\tilde q$. We observe a clear structure of bands, meaning that the resonance is much stronger for some values of $\tilde q$ than others. The parameter space can be divided in two regimes depending on the value of $\tilde q$. If $\tilde q < 1$, the bands are so narrow that the resonance is very weak: this is referred to as \textit{narrow resonance}. On the contrary, if $\tilde q \geq 1$, the bands are much broader and the typical Floquet indices are much larger, so the resonance is stronger: this corresponds to \textit{broad resonance}. In the later case, for $p=4$, and within the following $\tilde q$-intervals,
\be \tilde  q \in  \left[ n (2n-1) , n(2n + 1) \right] \, , \hspace{0.3cm} n=1,2\dots \ , \hspace{0.4cm} [p=4] \ , \label{eq:BandsParRes4}\ee
(i.e.~$\tilde q = q_{\star} = [1,3], [6,10]\dots$), the range of excited resonance momenta can be written as $0 \leq \tilde \kappa \lesssim \tilde\kappa_+ \sim \tilde q^{1/4}$. The maximum Floquet index throughout the complete parameter space is $\nu_{k,\rm max}\simeq 0.26$, which is attained at the center of these bands at $\tilde \kappa=0$. A similar structure of bands is observed for other values of $p(\neq 4)$, but their position is not described by Eq.~(\ref{eq:BandsParRes4}).

The value of $\tilde q$ depends on the scale factor, so the system moves along a trajectory in the $\{\tilde q,\tilde \kappa\}$ parameter space as the Universe expands. We can distinguish three situations depending on the value of $p$: 1) If $p<4$, $\tilde q$ decreases with time, so even if the resonance is initially broad, it will eventually become narrow at late times; 2) if $p>4$, $\tilde q$ grows with time, so either the resonance goes from initially narrow to broad at later times (if $q_{\star} \lesssim 1$), or it is always broad (if $q_{\star} \gtrsim 1$); 3) if $p=4$, $\tilde q$ is constant, so the type of resonance will not change, and it will be always either narrow (if $q_{\star} \lesssim 1$) or broad (if $q_{\star} \gtrsim 1$). 

Initially the energy density of the daughter field is small, $\rho_X \ll \rho_{\bar{\phi}}$.  However, $\rho_X$ grows exponentially as the inflaton oscillates, and if the system is in the broad resonance regime long enough, we eventually arrive at a situation with $\rho_X \simeq \rho_{\bar{\phi}}$. The initially homogeneous inflaton then fragments due to backreaction, becoming fully inhomogenous. Similarly to the analysis of self-resonance, we define the corresponding backreaction time $u_{\rm br}$ as the time when the condition $\rho_X (u_{\rm br}) = \rho_{\bar{\phi}} (u_{\rm br})$ holds for the first time. We now estimate this time scale under the assumption that the resonance is always broad ($\tilde q > 1$) during the time interval $0 < u < u_{\rm br}$. 

The energy density of the $X$ field can be written as
\bea \rho_{X} &=& \frac{1}{2 \pi^2 a^3} \int dk k^2 \Omega_{k,X} n_{k,X} \ , \\
&=& \frac{\omega_{\star}^4  }{2 \pi^2}  a^{\frac{12 (2 -p)}{p+2}} \int d {\tilde \kappa} \tilde \kappa^2 \tilde\omega_{k,\chi}  n_{k,X} \ , \label{eq:DaughterIntegral2}\eea
where $\Omega_{k,X} \equiv \sqrt{(k/a)^2 + g^2 \bar{\phi}^2} = a^{\frac{3(2-p)}{p+2}} \tilde\omega_{k,\chi}\,\omega_{\star}$ is the effective frequency of the mode, and $n_{k,X}$ is the occupation number. The amplitude of the daughter field modes grows exponentially for certain ranges of $\tilde \kappa$, which depend in a non-trivial way on $\tilde q$, as seen in Fig.~\ref{fig:AllFloquetDiagrams}. Furthermore, $\tilde q$ evolves with time for $p \neq 4$, so the specific modes being excited change with time. This gives rise to a stochastic behaviour \cite{Kofman:1997yn}. In order to simplify the computation, we assume that all modes below a certain cutoff $\tilde \kappa < \tilde \kappa_+$ are excited exponentially with a constant Floquet index $\bar{\nu}$, so that $n_{k,X} = e^{2 \bar{\nu} u} \theta(1 - \tilde \kappa / \tilde \kappa_+)$. As we shall see, this approximation gives a good order-of-magnitude estimate of the time scale at which backreaction happens.

We can get an estimate for $\tilde \kappa_+$ from the condition of adiabaticity violation,
\be
\frac{\tilde\omega'_{k,\chi}}{\tilde\omega_{k,\chi}^2} \gtrsim 1 \hspace{0.3cm} \rightarrow \hspace{0.3cm} \tilde \kappa^2 \lesssim (\tilde q \bar{\varphi} \bar{\varphi}')^{2/3} - \tilde q \bar{\varphi}^2 \ , \label{eq:adiab-cond} \ee
where we take the approximation $(a'/a)^2, a''/a \ll \tilde \kappa^2$. For simplicity, we expand the field amplitude around the minimum of the potential (at the time $u \simeq u_*$), such that $\bar{\varphi} \sim \beta_{\bar{\varphi}} \Delta u \equiv \beta_{\bar{\varphi}}  (u - u_*) $ and $\bar{\varphi}' \sim \beta_{\bar{\varphi}}$. By substituting this into (\ref{eq:adiab-cond}), we can show that a range of momenta $\tilde\kappa$ are excited during the time interval $\Delta u \simeq \tilde q^{-1/4} \beta_{\bar{\varphi}}^{-1/2}$. The range of excited momenta is the widest at the time $\Delta u \simeq 3^{-3/4} \tilde q^{-1/4} \beta_{\bar{\varphi}}^{-1/2}$, 
with the maximum momentum excited being $\tilde \kappa \lesssim \tilde \kappa_+ \equiv 2^{1/2} 3^{-3/4} \beta_{\bar{\varphi}}^{1/2} \tilde q^{1/4} = 2^{1/2} 3^{-3/4} \beta_{\bar{\varphi}}^{1/2} q_{\star}^{1/4}  a^{\frac{3(p-4)}{2(p+2)}}$. We will use this expression as the upper limit of the integral in (\ref{eq:DaughterIntegral2}). We can also show that most of the excited momenta are non-relativistic,
\bea \frac{m^2_{\chi}}{\tilde \kappa^2_{+}} \sim \frac{\tilde q \sqrt{\langle \bar{\varphi}^2 \rangle}}{\tilde \kappa_{+}^2} \sim \tilde q^{1/2} \gtrsim 1 \ , \eea
where we have used $\langle \bar{\varphi}^2 \rangle \approx 1/2$, see Eq.~(\ref{eq:Varphi2}) from Appendix \ref{App:PeriodAverages}. The energy density $\rho_X$ is finally given by
\bea \rho_{X} &=& \frac{\omega_{\star}^4}{2 \pi^2}  a^{\frac{12 (2 -p)}{p+2}}\int d {\kappa} \kappa^2 \omega_{k, \chi} e^{2 \bar{\nu} u} \theta \left(1 - \frac{\tilde \kappa }{ \tilde \kappa_{+}} \right)  \nonumber \\
&\simeq& \frac{\omega_{\star}^4}{2 \pi^2}  a^{\frac{12 (2 -p)}{p+2}}  \tilde q^{1/2} \langle \bar{\varphi}^2 \rangle^{1/2}   \int_0^{\tilde \kappa_{+}} d {\tilde \kappa} \tilde \kappa^2 e^{2  \bar{\nu} u}  \nonumber \\
&\simeq & \frac{\omega_{\star}^4}{2^{\frac{3}{2}} \cdot 3  \pi^2}   \tilde q^{1/2} a^{\frac{12(2-p)}{p+2}}  \tilde \kappa_{+}^3 e^{2  \bar{\nu} u} \nonumber \\
&=& \frac{\omega_{\star}^4 \, \beta_{\bar{\varphi}}^{3/2} }{27 \cdot 3^{1/4} \pi^2} q_{\star}^{5/4} a^{\frac{-3(4+3p)}{2(p+2)} } e^{2  \bar{\nu} u }\ .
\eea
From the condition $\rho_X= \rho_{\bar{\phi}}$ we get
\bea 
\label{eq:zBR}
u_{\rm br} &\simeq& \frac{1}{2  \bar{\nu}}  \left[ \frac{3(4-p)}{2(p+2)} \ln a(u_{\rm br}) + \ln \left( \frac{\phi_{\star}^{4-p}}{\lambda \mu^{4-p}} \right)\right. \nonumber \\
&& \left.  - \frac{5}{4} \log q_{\star}   + \ln \left( \frac{ 27 \cdot 3^{\frac{1}{4}} \pi^2}{p \beta_{\bar{\varphi}}^{3/2}} \right) \right] \ .\eea

\begin{figure}
    \begin{center}
    \includegraphics[width=0.48\textwidth]{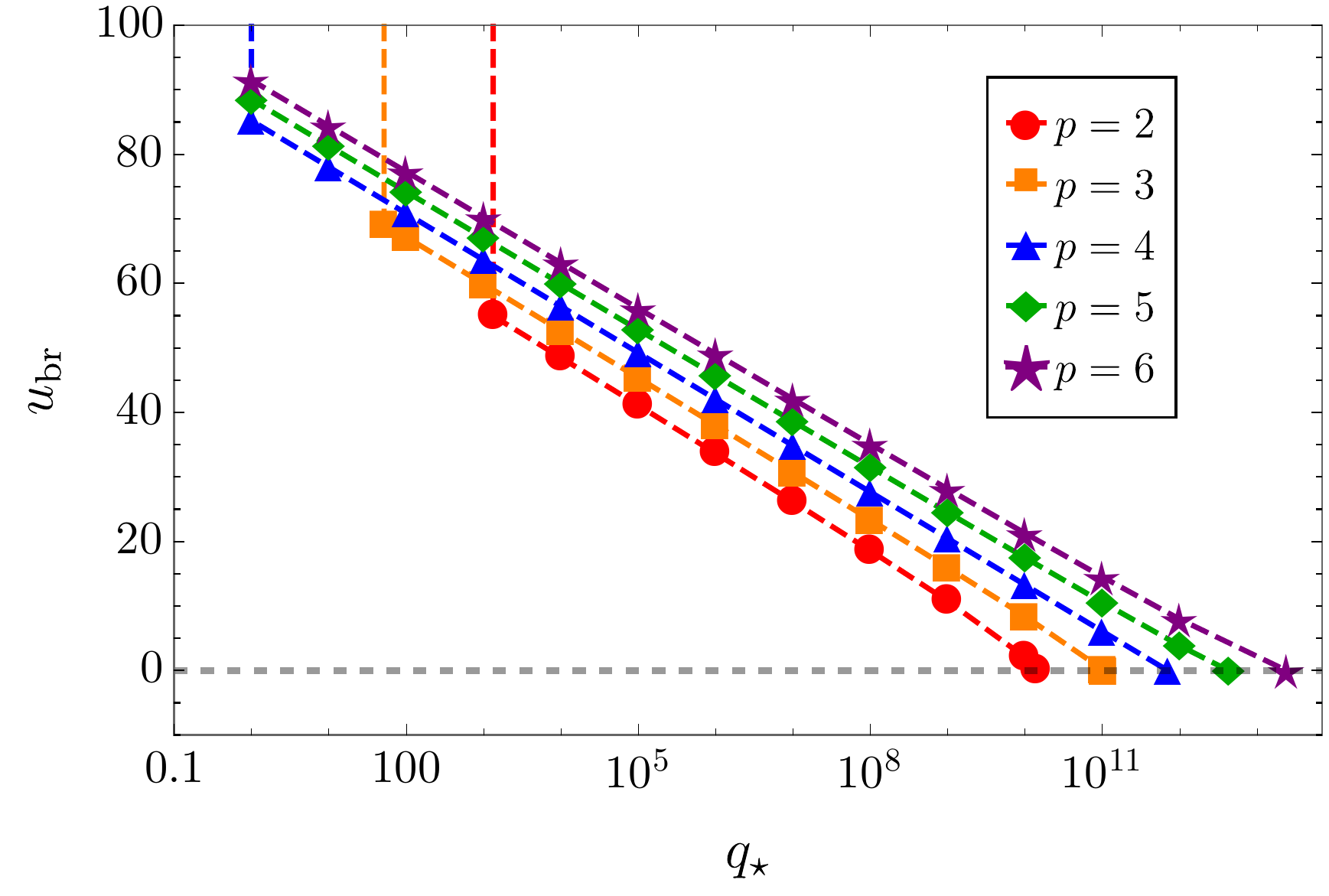}
    \includegraphics[width=0.46\textwidth]{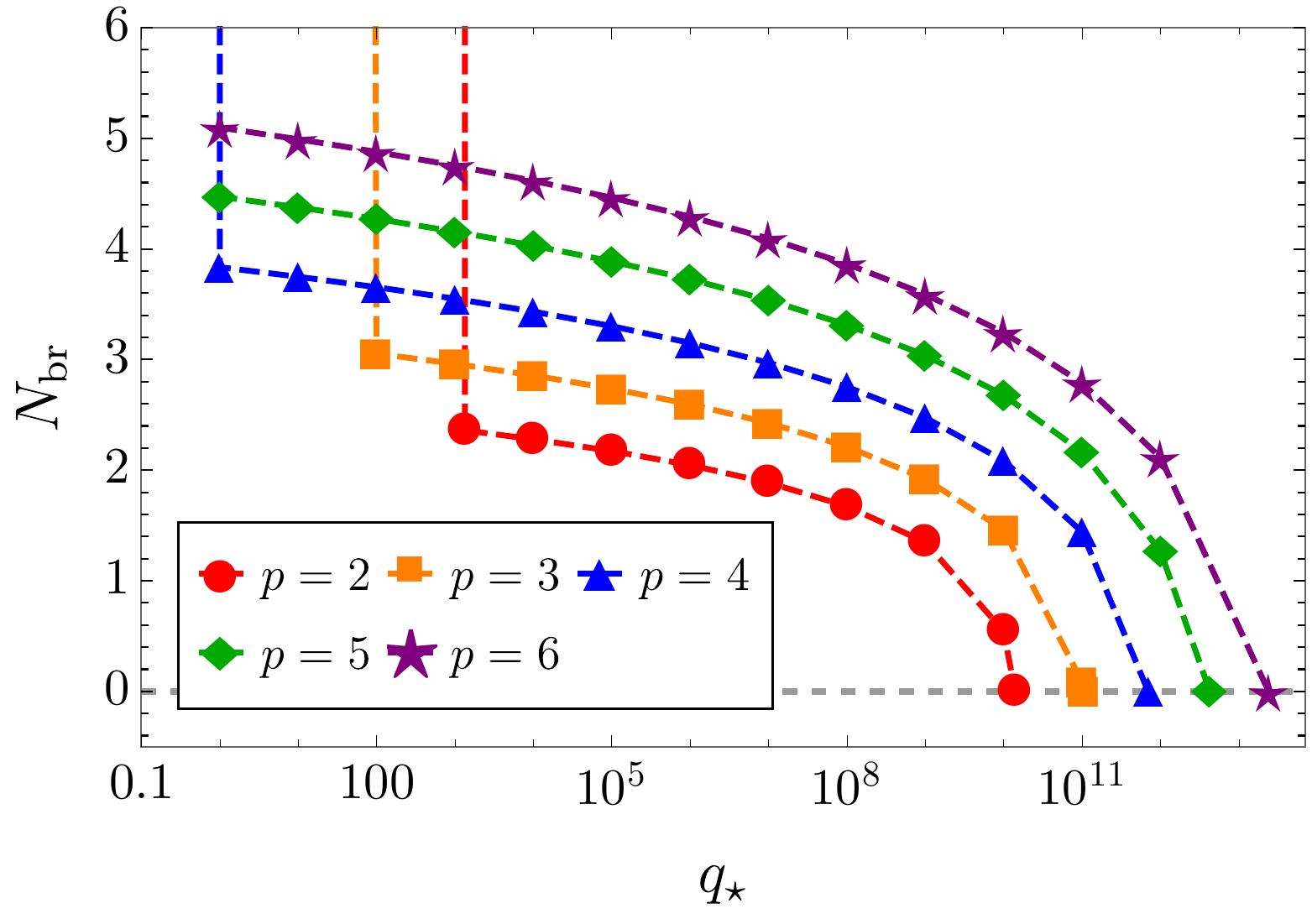}
    \end{center} \vspace*{-0.4cm}
    \caption{Estimates of the backreaction time induced by parametric resonance in the centred potential case. Results are shown in terms of natural time $u_{\rm br}$ (top) and post-inflationary number of e-folds $N_{\rm br}$ (bottom), as a function of $q_{\star}$ and for different values of $p$. Vertical dashed lines for $p=2-4$ show the values of $q_{\star}$ for which the condition (\ref{eq:qstarM}) is no longer valid.}
    \label{fig:Analytic-ParRes}
\end{figure}

For monomial potentials, we can use Eqs.~(\ref{eq:mu}), (\ref{eq:PhiHStar}) and (\ref{eq:sf-corr}) to solve this expression for $u_{\rm br}$. For $p=4$ the expression can be solved explicitly. If we take $\bar{\nu} = 0.2$ and $N_k=60$, we obtain the following logarithmic dependence,
\be u_{\rm br} \simeq 86 - 7.2 \log_{10} (q_{\star}) \ ,  \hspace{0.3cm} [p=4] \ , \label{eq:zbr-p4} \ee
which is similar to the result obtained previously in \cite{Figueroa:2016wxr}. The equation~(\ref{eq:zBR}) is however implicit for $p \neq 4$, and can only be solved numerically. We show in Fig.~\ref{fig:Analytic-ParRes} the value of $u_{\rm br}$ as a function of $q_{\star}$ for $p=2$, 3, 4, 5 and 6, as well as the corresponding number of e-folds $N_{\rm br} \equiv \log [a(u_{\rm br})]$ after inflation. The prediction for $u_{\rm br}$ is very similar for the considered values of $p$, so Eq.~(\ref{eq:zbr-p4}) can in fact roughly approximate the solution in all cases, even for $p \neq 4$. However, when expressed in number of e-folds, we observe that $N_{\rm br}$ grows monotonically with $p$ for a fixed value of $q_{\star}$. This happens because the number of e-folds goes as $N \sim \frac{p+2}{6} {\rm log} (u)$, so $N$ grows with $p$. Note also that the above expression predicts a negative backreaction time  for $q_{\star} \gtrsim 10^{12}$: this signals that our linearized approach is not valid anymore, as backreaction would take place immediately after the daughter field is first excited.

Note that in the above computation we assumed that the system remains in broad resonance ($\tilde q > 1$) from the end of inflation till backreaction. However, there are two situations in which this is not the case. First, for $p<4$, $\tilde q$ decreases as the Universe expands, so the system can enter into narrow resonance before backreaction happens, i.e.~$\tilde q (u_{\rm br}) \leq 1$. We have that $\tilde q =q_{\star} a^{\frac{6 (p-4)}{p+2}} \sim q_{\star} u^{p-4} $, so this condition can be expressed in terms of $q_{\star}$ as follows,
\be \tilde q (u_{\rm br}) \equiv q_{\star} u_{\rm br}^{p-4} (q_{\star}) \geq 1 \longrightarrow  q_{\star} \geq q_{\star}^{\rm (min)} \simeq 10^{1.9 (4-p)} \ , \label{eq:qstarM} \ee
where the expression for $q_{\star}^{\rm (min)}$ given in the right hand side of the inequality is simply a fit to the numerical solution of the corresponding implicit equation. If $ q_{\star} \leq q_{\star}^{\rm (min)}$, the homogeneous inflaton mode never decays due to backreaction of $X$, as the resonance becomes narrow before backreaction. On the other hand, if $p>4$ and $q_{\star} < 1$, $\tilde q$ grows with time, and even though the resonance is initially narrow, the system eventually develops broad resonance and hence $X$ backreacts on the inflaton. This delays the backreaction time (\ref{eq:zBR}) by an extra amount of time $\Delta u \sim q_{\star}^{-\frac{1}{p-4}}$.
    
A rapid inspection of the corresponding Floquet diagrams shows that broad parametric resonance of the daughter field, if present, is always much stronger than inflaton self-resonance. In particular, the maximum Floquet index for inflaton self-resonance is $\mu_k \approx0.035$, while for parametric resonance of the daughter field is $\nu_k \approx  0.26$. Due to this, the predicted $N_{\rm br}$ is always larger for self-resonance (Fig.~\ref{fig:PvszbrNbr}) than for broad parametric resonance (Fig.~\ref{fig:Analytic-ParRes}), for all considered values of $p \in [2,6]$. 

\subsection{Analytical analysis of resonances: Displaced potential} \label{sec:Lin:DisplacedPotential}

Let us turn to the case of an inflaton potential with minimum at $v \neq 0$. In this case, the daughter field $X$ acquires now a mass $m_X$ at the minimum of the potential, given by
\be
m_X^2 \equiv \frac{\partial^2 V}{\partial X^2}\biggr\rvert_{\mathrm{min}} = g^2v^2 =q_{\star} \left(\frac{\omega_{\star} v}{\phi_{\star}}\right)^2 \ ,  \label{eq:massX}\ee
where $q_{\star} = g^2 \phi_{\star}^2 / \omega_{\star}^2$ (c.f.~\ref{eq:ResParam}) is the initial resonance parameter. Considering also the \textit{effective} mass of the inflaton,
\be
m_\phi^2 \equiv \frac{\partial^2 V}{\partial \phi^2} = \lambda \mu ^{4-p}\phi^{p-2}  , \label{eq:massPhi}\ee
it is convenient to define the (time-dependent) ratio $\tilde R$
\bea
\tilde R^2 &\equiv& \frac{m_X^2}{m_\phi^2} =  \frac{g^2v^2}{\lambda \mu ^{4-p}\phi^{p-2}} = R_{\star}^2 a^\frac{6(p-2)}{(p+2)} \ , \label{eq:massratio0} \\  
R_{\star}^2 &\equiv & \left. \frac{m_X^2}{m_\phi^2} \right|_{t=t_{\star}} = \frac{q_{{\star}}v^2}{\phi_{\star}^2} \ ,
\label{eq:massratio} \eea
where $R_{\star}$ refers to the ratio at the initial time $t_{\star}$. For $p=2$, $\tilde{R}$ is constant and can be interpreted as the \textit{mass ratio} between the two fields, whereas for $p>2$ it is a ratio that grows with the expansion of the universe.

Similarly to the centred potential case, it is convenient to work with natural spacetime and field variables, c.f.~ Eqs.~(\ref{eq:newvars1}) and (\ref{eq:newvars2}). The field EOM, including interactions and spatial gradients, read 
\bea
&& \varphi'' - a^{\frac{-(16 - 4p)}{2+p}} \nabla^2_{\vec{y}} \varphi +  |\varphi|^{p-1} + \tilde q \chi^2(\varphi+\tilde v) = 0 \ , \label{eq:eomvev1}\hspace{0.7cm}  \\
&& \chi''- a^{\frac{-(16 - 4p)}{2+p}} \nabla^2_{\vec{y}} \chi + \tilde q(\varphi+\tilde v)^2 \chi = 0 \ , \label{eq:eomvev2}  
\eea
where $\tilde q = q_{{\star}} a^{\frac{6 (p-4)}{p+2}}$ (c.f.~\ref{eq:ResParam}) and
\bea
\tilde v(a) &\equiv& v_{\star} a^{\frac{6}{2+p}} \ , \hspace{0.55cm} v_{\star} \equiv \frac{v}{\phi_{\star}} \label{eq:effecvres} \ . 
\eea

During the early stage of preheating, we can do again a linearized analysis of the inflaton and daughter field fluctuations, by expanding the fields as $\varphi (\vec{y}, u)  \equiv  \bar{\varphi} (u) + \delta \varphi (\vec{y},u)$ and $\chi(\vec{y},u)\equiv\delta \chi (\vec{y},u)$. The evolution of the homogeneous inflaton mode $\bar{\varphi}(u)$ is then described by the EOM $\bar{\varphi}'' + |\bar{\varphi} |^{p-2}\bar{\varphi} \simeq 0$, and its solution is approximately described by Eq.~(\ref{eq:Infl-sol}). The mode equations of the fluctuations are
\bea\delta \varphi^{''}_k + \omega_{k,\varphi}^2 \delta \varphi_k = 0&& \hspace*{-0.55cm}\ , \hspace{0.3cm} \omega_{k,\varphi} \equiv \sqrt{ {\tilde\kappa}^2 + (p-1) |\bar{\varphi}|^{p-2} } \ , \label{eq:linearshift1} \hspace{0.6cm}  \\
\delta \chi^{''}_k +  \omega_{k,\chi}^2  \delta \chi_k = 0&& \hspace*{-0.55cm}  \ , \hspace{0.3cm}  \omega_{k,\chi} \equiv \sqrt{ {\tilde\kappa}^2 + \tilde q (\bar{\varphi}+\tilde v)^2} \ , \label{eq:linearshift2} \eea
where  $\tilde\kappa \equiv \kappa a^{\frac{-(8-2p)}{2+p}}$ is the resonance momentum in natural units (c.f.~\ref{eq:resonant-momentum}). The equation of the inflaton fluctuations is identical to the one of the \textit{centred case} (\ref{eq:linear1}), so the process of \textit{self-resonance} remains unchanged. As discussed, it admits exponential solutions of the type $\delta \phi_k \sim e^{\mu_k u}$ with $\mathfrak{Re}[\mu_k]>0$ for certain combinations of $\{\tilde \kappa, p\}$, which are described by the Floquet chart of Fig.~\ref{fig:FloquetDiagrams}.

The daughter field experiences an exponential growth $\delta \chi_k \sim e^{\nu_k u}$ due to \textit{parametric resonance} for certain combinations of parameters $\{\tilde \kappa,  \tilde v, \tilde q; p \}$. On the one hand, if $\tilde v \ll 1$, we recover the situation of the centred potential, explained in Section \ref{sec:3B}. On the other hand, if $\tilde v \gg 1$, the adiabaticity condition is never violated, i.e. $\tilde\omega_{k,\chi}'/\tilde\omega_{k,\chi}^2<1$ holds at all times, so no excitation of the daughter field takes place. We are therefore interested in studying the post-inflationary dynamics when $\tilde v$ has an intermediate value between both limits.

\begin{figure*}
	\includegraphics[width=0.9\textwidth]{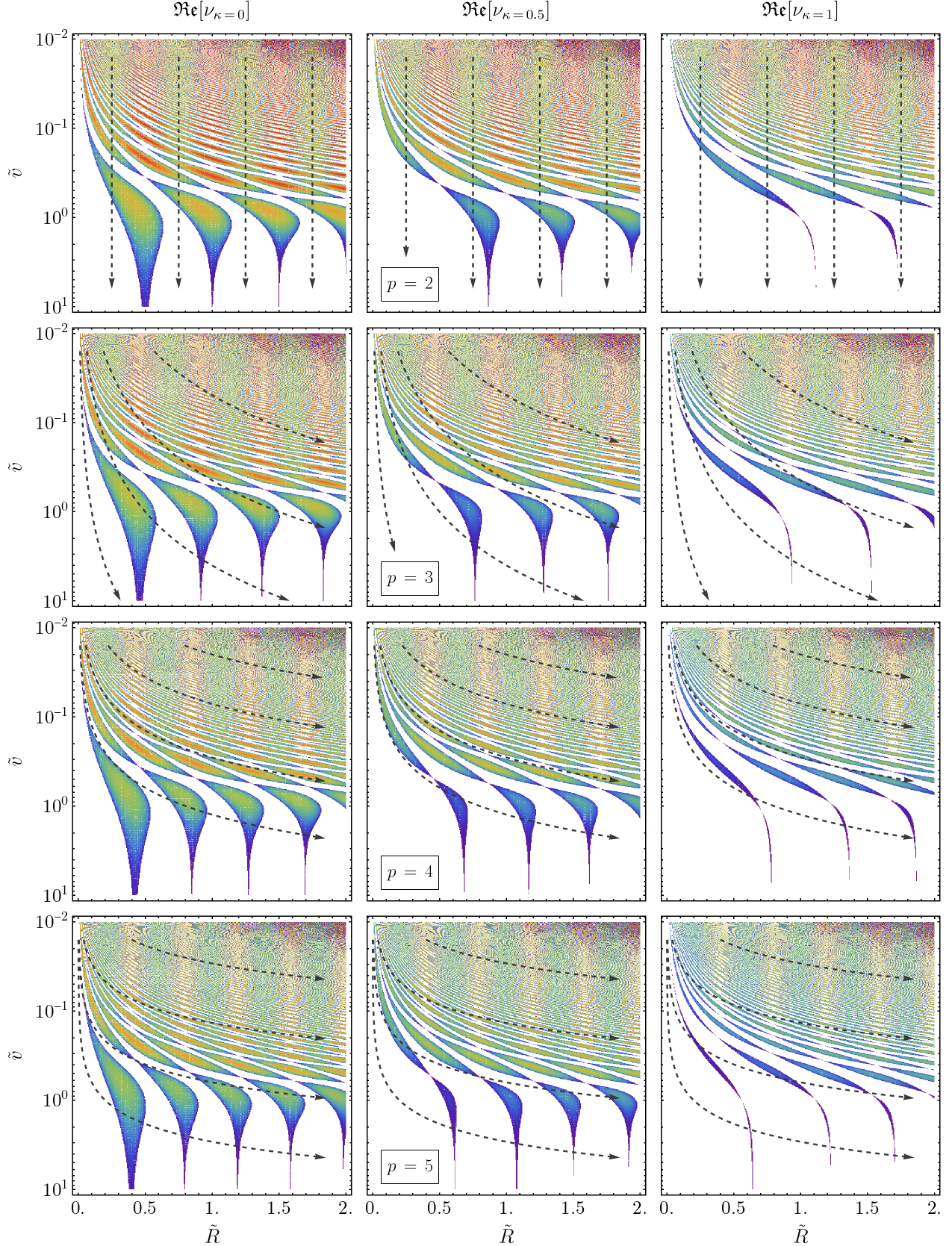}  \vspace{0.1cm} \\
	\centering
	\includegraphics[width=0.35\textwidth]{barlegend_pr_horizontal.pdf}
    \caption{Floquet index $\mathfrak{Re}[\nu_k]$ for parametric resonance of the daughter field in the displaced potential case. From top to bottom, we consider the power-law coefficients $p=2$, $3$, $4$ and $5$, and in each case, we fix the resonance momenta to $\tilde \kappa=0$ (left) $0.5$ (middle) and $1$ (right). Charts are plotted as a function of $\tilde R$ and $\tilde v$. White areas correspond to regions where $\mathfrak{Re} [\nu_k] = 0$. The dashed lines indicate how the system moves in parameter space as the universe expands.}  \label{fig:shifted-floquet} 
\end{figure*}

Let us generalize the previous Floquet analysis to the case of non-vanishing minimum $v$.  For this purpose, we write the equation of $\delta\chi_k$ in a more convenient form,
\be \delta \chi^{''}_k +  \left(\tilde \kappa^2+ \tilde R^2 \left(1+ \frac{\bar{\varphi}}{\tilde v} \right)^2   \right)\delta \chi_k = 0\ , \label{eq:linearshiftmm2}\ee
which allows to study the properties of the solution in terms of $\tilde R$ and $\tilde v$. Fig.~\ref{fig:shifted-floquet} summarizes our results. We have considered the power-law coefficients $p=2$, 3, 4, and 5, and for each case, we show the Floquet charts for specific choices of resonance momenta: $\tilde \kappa=0$, 0.5, and 1. Each panel shows the real part of the Floquet index $ \mathfrak{Re} [\nu_k]$ as a function of $\tilde R$ and $\tilde v$ (note that there is resonance for values of $\tilde R > 2$ even if these are not shown in the figure). Interestingly, we see that parametric resonance can excite daughter particles with mass much larger than the inflaton mass, $m_X \gg m_\phi$, in contrast to perturbative decay processes.

In the regime $\tilde v \lesssim 1$ we can observe that small variations in $\tilde R$ or $\tilde v$ may lead to very different behaviours in the resonance structure. The largest Floquet indices for a particular mode can be found at the ratios $\tilde R = R_{\star} \sim n\beta_{\bar \varphi}/2$ with $n=1, 2, \dots$, where $\beta_{\bar \varphi}$ is given in Eq.~(\ref{eq:Inflaton-period}). The largest Floquet index overall for $p=2$ can be found for $\tilde \kappa=0$, with $\mathfrak{Re}[\nu_\kappa] \approx 0.26$. However, if we consider instead larger values of $p$, the maximum value attained by $\nu_\kappa$ is smaller: for example, for $p=4$ we obtain $\nu_\kappa^{\rm max} \approx 0.235$ for very small $\tilde v$.

Let us now consider the structure of resonances for $\tilde v \gtrsim 1$. As observed in the Floquet charts, in this regime the resonance bands are significantly broader for lower momenta, while they become increasingly narrow for higher values of $\tilde \kappa$. For very large values of $\tilde v$, the bands become thin spikes located around specific values of $\tilde R$. This property of the solutions can be explained by analyzing the mode equation in the limit $\tilde v \gg 1$. In this regime, Eq.~(\ref{eq:linearshiftmm2}) can be written as,
\be \delta\chi_k ''+(\alpha+\beta \bar{\varphi})\delta\chi_k \simeq 0 \ , \hspace{0.3cm}  \alpha \equiv \tilde \kappa^2 + \tilde R^2 \ , \hspace{0.3cm} \beta \equiv \frac{2 \tilde R^2}{\tilde v} \ . \label{eq:DisplEqv} \ee
For $p=2$, the inflaton homogeneous mode evolves exactly as $\bar{\varphi} = \cos (u)$, so Eq.~(\ref{eq:DisplEqv}) corresponds to a Mathieu-like equation. In the limit $\beta \rightarrow 0$, the solution of these equations shows a structure of narrow resonance bands located at $\alpha$-values $\alpha=(n/2)^2$, with $n=1,2,3,...$, which become increasingly narrow as $\beta$ gets closer to zero \cite{maclachlan1964theory}. In our system this means that, in the limit $\tilde v \gg 1$, there exist resonance bands (spikes) roughly centred at the mass ratios
\be \tilde R = R_{\star} \simeq \sqrt{ \left (\frac{n}{2}\right)^2- \tilde \kappa^2 }\ , \hspace{0.3cm} n = 1, 2,\dots \ , \hspace{0.3cm} [p=2] \ . \label{eq:MassRatioSpikes} \ee
for all integer values of $n$ obeying $n > 2 \tilde \kappa$. Therefore, the momentum $\tilde \kappa=0$ only gets excited when the mass ratio is around integer or half-integer values. These bands can be clearly seen for $\tilde v>1$ in the Floquet charts for $p=2$, depicted in the top panels of Fig.~\ref{fig:shifted-floquet}. The mass ratio that gives the strongest excitation (for $\tilde \kappa=0$) is $\tilde R = R_{\star} \simeq0.5$. 
 
For $p>2$, we do not have an exact solution for $\bar{\varphi}$, but it can be approximated by $\bar{\varphi} \approx \cos (\beta_{\bar \varphi} u)$ [see Eqs.~(\ref{eq:Infl-sol}) and (\ref{eq:BetaNumbers})]. In this case, the resonance bands are instead located at
\be \tilde R \simeq \sqrt{ \left (\frac{\beta_{\bar \varphi} n}{2}\right)^2- \tilde \kappa^2 }\ , \hspace{0.3cm} n = 1, 2,\dots \ , \label{eq:MassRatioSpikes2} \ee
again for all integer values of $n$ obeying $n>2 \tilde \kappa$. This structure of bands is also clearly observed for $\tilde v \geq 1$ in the Floquet charts for $p=3$, 4, and 5, see Fig.~\ref{fig:shifted-floquet}.

Note that both $\tilde v$ and $\tilde R$ depend on the scale factor, so the system travels along trajectories throughout parameter space as the Universe expands. These are depicted with dashed arrows in the different stability charts of Fig.~\ref{fig:shifted-floquet}. In the case $p=2$, $\tilde R = R_{\star}$ is constant and $\tilde v = v_{\star} a^{3/2}$ grows with time, so the system travels vertically and downwards. For $p>2$, both $\tilde R\propto a^{3(p-2)/(p+2)}$ and $\tilde v\propto a^{6/(2+p)}$ grow, so the system travels both downwards and rightwards. If the initial value of $\tilde v$ satisfies $\tilde v\ll 1$, the system goes first through a stochastic stage during which short intervals of strong excitation alternate quickly with instances of no excitation, and a wide range of modes gets amplified. Once $\tilde v\gtrsim1$, the system enters a stage where modes are amplified only on specific resonance momentum bands centred around the critical values given by Eqs.~(\ref{eq:MassRatioSpikes}) (for $p=2$) or (\ref{eq:MassRatioSpikes2}) (for $p>4$). In any case, as the expansion of the universe goes on, the Floquet indices decrease for all momenta, and eventually the resonance is completely switched off.

\subsection{Energy distribution and equation of state} \label{subsec:EnEos}

We present expressions for the different energy contributions of the fields and explain how their evolution impacts the equation of state after inflation. In the following we do not differentiate between $v = 0$ and $v \neq 0$, as the expressions are common to both cases.

We start by writing the pressure and energy densities as 
\bea  
p &=& \frac{1}{2} \dot{\phi}^2 +  \frac{1}{2} \dot{X}^2 - \frac{1}{6} |\nabla \phi |^2 - \frac{1}{6} |\nabla X |^2 - V (\phi, X) \ , \label{eq:pTot}\\
\rho &=& \frac{1}{2} \dot{\phi}^2 +  \frac{1}{2} \dot{X}^2  + \frac{1}{2} |\nabla \phi |^2 + \frac{1}{2} |\nabla X |^2 + V (\phi, X)  \ , \hspace{0.6cm} \label{eq:rhoTot}
\eea
where $V(\phi,X)$ is the potential energy given in Eq.~(\ref{eq:shifted-inflaton-potential1}). When written in terms of natural variables (\ref{eq:newvars1}) and (\ref{eq:newvars2}), these can be decomposed as
\bea 
\rho &=& \frac{\omega_{\star}^2 \phi_{\star}^2}{a^{\frac{6p}{2+p}} } ( E_{\rm k}^{\varphi} + E_{\rm k}^{\chi} + E_{\rm g}^{\varphi}  + E_{\rm g}^{\chi}+ E_{\rm p} + E_{\rm i} ) \ , \label{eq:rho} \\
p &=& \frac{\omega_{\star}^2 \phi_{\star}^2}{a^{\frac{6p}{2+p}} }  ( E_{\rm k}^{\varphi} + E_{\rm k}^{\chi} - \frac{1}{3} E_{\rm g}^{\varphi}  - \frac{1}{3} E_{\rm g}^{\chi} - E_{\rm p} - E_{\rm i} ) \hspace{0.3cm} \label{eq:pres} \ , \eea
where $E_{\rm k}^f$ and $E_{\rm g}^f$ ($f=\varphi,\chi$) are the (natural) kinetic and gradient energy densities of each field, $E_p$ is the (natural) potential energy (which can be decomposed as the sum of the inflaton and daughter field contributions, $E_{\rm{p}}\equiv E_{\rm p}^\varphi+E_{\rm p}^\chi$), and $E_{\rm i}$ is the interaction energy (which can also be decomposed in two terms as $E_{\rm{i}}\equiv E_{\rm{i}}^{(1)}+E_{\rm{i}}^{(2)}$). Each of these terms can be written as
\bea
&E_{\rm k}^f &\equiv  \frac{1}{2} \left( f' - \frac{6}{p+2} \frac{a'}{a} f \right)^2 , \hspace{0.3cm} (f=\varphi,\chi) \ ,  \label{eq:energyshift1}  \\ 
&E_{\rm g}^f &\equiv  \frac{1}{2}  a^{\frac{4p-16}{p+2}} |\nabla_{\vec{y}} f |^2 , \hspace{1.2cm} (f=\varphi,\chi) \label{eq:energyshift2}\ ,  \\
&E_{\rm{p}}^\varphi &\equiv  \frac{1}{p} |\varphi|^p \ , \label{eq:energyshift5} \\
&E_{\rm{p}}^\chi &\equiv  \frac{1}{2}a^{\frac{6(p-2)}{p+2}} q_{\star}v_{\star}^2 \chi^2 \ . \label{eq:energyshift6}\\
&E_{\rm{i}}^{(1)} & \equiv  a^{\frac{6(p-3)}{p+2}} q_{\star} v_{\star} \varphi \chi^2 \ , \label{eq:energyshift3} \\
&E_{\rm{i}}^{(2)} & \equiv  \frac{1}{2} a^{\frac{6p-24}{p+2}} q_{\star} \varphi^2 \chi^2 \ ,\label{eq:energyshift4} 
 \eea
Note that for the centred potential case we have $v_{\star} = 0$, so the trilinear interaction and daughter field potential are not present, and $E_{\rm{p}}= E_{\rm p}^\varphi$ and $E_{\rm{i}}=E_{\rm{i}}^{(2)}$. 

We can also define \textit{energy ratios} for each of these terms as $\varepsilon_i \equiv \langle E_i \rangle / \langle \sum_{j} E_{j} \rangle$, where $j$ sums over all energy components and $\langle \dots \rangle$ denotes a volume average. These indicate the different relative contributions to the total energy of the system. By construction, all ratios sum one,

\be \sum_j \varepsilon_j = \varepsilon_{\rm k}^{\varphi} + \varepsilon_{\rm k}^{\chi} + \varepsilon_{\rm g}^{\varphi} + \varepsilon_{\rm g}^{\chi} + \varepsilon_{\rm p} + \varepsilon_{\rm i} = 1 \label{eq:en-const}\ . \ee

The \textit{equation of state} is defined as the ratio between the pressure (\ref{eq:pres}) and the energy density (\ref{eq:rho}), and can be written as
\be w \equiv \frac{p}{\rho} = \varepsilon_{\rm k}^{\varphi} + \varepsilon_{\rm k}^{\chi}  - \frac{1}{3} ( \varepsilon_{\rm g}^{\varphi} + \varepsilon_{\rm g}^{\chi} ) -  (\varepsilon_{\rm p} + \varepsilon_{\rm i} )  \label{eq:EoS-fEn} \ .    \ee
We will be mainly interested in the evolution of the \textit{effective} equation of state $\bar{w}$, which is obtained by averaging the instantaneous equation of state (\ref{eq:EoS-fEn}) over oscillations. The energy distribution and the equation of state will evolve in different ways for different choices of $p$ and $q_{\star}$. 

It has been shown that this kind of field systems \textit{virialize} rapidly after the end of inflation \cite{Boyanovsky:2003tc,Lozanov:2016hid,Figueroa:2016wxr,Lozanov:2017hjm}, with the following relations holding when averaged over both volume and oscillations,
\be \label{eq:EquipIdent1}
\langle \dot{f}^2 \rangle = \langle |\nabla f|^2 \rangle + \left\langle f \frac{\partial V}{\partial f} \right\rangle \ , \hspace{0.3cm} (f=\varphi,\chi) \ . \ee
For the two-field scenario under consideration with potential (\ref{eq:shifted-inflaton-potential1}), these relations can be expressed in terms of energy contributions as
\bea
\langle E_{\rm k}^{\varphi} \rangle & \simeq  & \langle E_{\rm g}^{\varphi}  \rangle + \frac{p}{2} \langle E_{\rm p}^{\varphi} \rangle + \frac{1}{2} \langle E_{\rm i}^{(1)} \rangle + \langle E_{\rm i}^{(2)}  \rangle \  , \label{eq:Virial1}\\
\langle E_{\rm k}^{\chi}  \rangle & \simeq & \langle E_{\rm g}^{\chi}  \rangle + \langle E_{\rm p}^{\chi} \rangle + \langle E_{\rm i}  \rangle \ , \label{eq:Virial2} \eea
Besides, the sum $E_t \equiv \sum_j E_j$ does not change significantly during one oscillation, so we can write analogous relations in terms of energy ratios by simply doing the substitution $\langle E_i \rangle \rightarrow \bar{\varepsilon}_i$ in Eqs.~(\ref{eq:Virial1})-(\ref{eq:Virial2}), where the bar denotes an oscillation average over time.

During the initial stage of preheating, the energy budget is dominated by the oscillatory homogeneous inflaton mode, and the only non-negligible energy ratios are $ \varepsilon_{\rm k}^{\varphi}$ and $\varepsilon_{\rm p}^{ \varphi} $. Using Eqs.~(\ref{eq:en-const}) and (\ref{eq:Virial1}) we get
\be  \bar{\varepsilon}_{\rm k}^{ \varphi}   = \frac{p}{p+2} \ , \hspace{0.4cm}  \bar{\varepsilon}_{\rm p}^{ \varphi}  = \frac{2}{p+2} \ . \label{eq:HomEner} \ee
If we substitute these expressions into (\ref{eq:EoS-fEn}), we recover $\bar{w} \equiv (p-2)/(p+2)$ as expected, c.f.~(\ref{eq:EoSoscillations}). On the other hand, whenever the potential and interaction energies of the system are very small compared to gradient and kinetic energies (say effectively we have $\varepsilon_{\rm p}^{\varphi}, \varepsilon_{\rm i} \ll 1$), we get from (\ref{eq:en-const}) and (\ref{eq:Virial1}) that the kinetic energy of each field equals its gradient energy, $\bar{\varepsilon}_{\rm k}^{f} \simeq \bar{\varepsilon}_{\rm g}^{f} $ $(f=\varphi,\chi)$. By substituting this into (\ref{eq:EoS-fEn}), we get that this configuration gives rise to a radiation-dominated universe $\bar{w}=1/3$. Note that this result is independent on how much energy is transferred between the two fields: only the ratio between the gradient and the potential energies is relevant, which must obey $\varepsilon_{\rm p}  /  \varepsilon_{\rm g}^f \ll 1$ and $\varepsilon_{\rm i}  /  \varepsilon_{\rm g}^f \ll 1$. \pagebreak

\section{Lattice results} \label{Sec:LatResults}

We now present our numerical results on the post-inflationary dynamics. The aim of this section is to properly study the later non-linear regime of the field evolution with lattice simulations, beyond the limitations of the linearized analysis carried out before. We are interested in how the energy ratios $\varepsilon_i$ evolve after inflation, and how they affect the post-inflationary equation of state, sourced by the different $\varepsilon_i$'s, c.f.~Eq.~(\ref{eq:EoS-fEn}). We are particularly interested in their values at \textit{very} late times, to see if a radiation-dominated stage is eventually achieved. These results are accompanied with a spectral analysis of the centred potential case, which is presented in Appendix \ref{app:Spectra}.

We simulate the post-inflationary dynamics for two-field system with total potential
\begin{eqnarray}\label{eq:TotPot}
V(\phi,X) &=& V_{\rm t} (\phi) + V_{\rm int}(\phi,X)\\
&=& \frac{1}{p} \Lambda^4{\rm tanh}^{p} \left( \frac{|\phi|}{M} \right) + \frac{1}{2}g^2(\phi+v)^2X^2\ ,\nonumber
\end{eqnarray}
where, as usual, $\phi \equiv \Phi - v$. For small field values around the origin $|\phi| \ll M$, Eq.~(\ref{eq:TotPot}) reduces to the potential Eq.~(\ref{eq:shifted-inflaton-potential1}) used in our analytical calculations. We have initialized the simulations at the end of inflation, when the field value of the inflaton is $\phi_*$,  and only consider cases in which inflation ends in the positive-curvature region of the potential, i.e.~$\phi_{\rm i}>\phi_*$.\footnote{This condition holds for $M \gtrsim 1.633 m_{\rm pl}$ for all values of $p$ [see Eqs.~(\ref{eq:inflection-potential}) and (\ref{eq:end-of-inflation})]. For $M \lesssim 1.633 m_p$, the inflaton will enter the tachyonic region of the potential during at least the first oscillations, which will trigger a growth of the inflaton fluctuations in a process of self-resonance, but different to the one studied in Section \ref{sec:3A}. In this case, the inflaton fragments into long-lived \textit{oscillons}, as long as the inflaton is sufficiently weakly coupled to daughter fields. For $p>2$, the inflaton fragments into \textit{transients}, which are similar to oscillons but with significantly shorter lifetimes \cite{Lozanov:2016hid,Lozanov:2017hjm}.} 

In the following we explore the dynamics for different values of the power-law coefficient $p$, the resonance parameter $q_*$ for the centred potential (see Eq.~\ref{eq:ResParam}), and the mass ratio $R_*$ for the displaced potential (see Eq.~\ref{eq:massratio}). The inflaton potential in Eq.~(\ref{eq:TotPot}) is well approximated by the monomial function (\ref{eq:powlaw-pot}) during the oscillatory regime, and hence the initial preheating stage on the lattice is well described by the linearized analysis presented in Section \ref{Sec:LinearAn}. For illustrative purposes, we fix the value of $M$ in the simulations we present in this section. We choose $M = 10 m_{\rm pl}$, corresponding approximately to the largest value compatible with the upper bound of the tensor-to-scalar ratio, see Fig.~\ref{fig:nsr-M}. The qualitative results we  present here for $M=10m_{\rm pl}$ hold for other choices of $M$ as long as the inflaton potential can be properly approximated by a monomial function during preheating. In Section~\ref{Sec:CMB} we take simulations into account for the whole range $M/m_{\rm pl} = 2 - 10$, to constrain the inflationary observables $n_s$ and $r$, as a function of $M$. 

Simulations have been carried out with {\it Velocity-Verlet} integration~\cite{Figueroa:2020rrl} in \CLns~\cite{Figueroa:2021yhd}, a recent package for lattice simulations of interacting fields in an expanding universe.\footnote{Different integration methods, such as the staggered leapfrog algorithm used in Ref.~\cite{Antusch:2020iyq}, give equivalent results as long as numerical errors are under control and, in particular, energy is sufficiently well conserved.} We have run simulations in $2+1$ dimensions (see footnote \footref{fn2Dsims} for an explanation of the meaning of ``$(2+1)$-dimensional simulations''), but do not expect relevant differences compared to $(3+1)$-dimensional simulations, based on our quantitative comparison already presented in our previous work \cite{Antusch:2020iyq}. There we confirmed that the post-inflationary dynamics of the system are qualitatively and quantitatively similar in both cases, by direct comparison in the parameter regions were both kinds of simulations are possible. The main reason for using simulations in $2+1$ dimensions is that it reduces the simulation time by a factor $\sim 10^2 - 10^3$ in comparison to three-dimensional ones. This allows us to properly explore the late time regime, which requires extremely long simulations, sometimes covering more than $\sim 10$ e-folds of expansion after inflation. Furthermore, as described in Section \ref{Sec:LinearAn}, the inflaton and daughter fields are excited at different momentum scales, and their field spectra propagate towards ultraviolet (UV) scales during and after backreaction. Therefore, in certain cases we had to use relatively large lattices, up to $N = 1024$ points/dimension for values of $p$ slightly larger than $p=2$, e.g.~$p=2.3$.

Very long simulations may pose an important problem due to the accumulation of errors after many time-steps of evolution. Our algorithm uses the acceleration Friedmann equation to evolve the scale factor, while it uses the Hubble rate equation to check the accuracy of the solution, as it is a direct constraint of the field dynamics (this is typically referred to as `energy conservation' in our context of field evolution in an expanding background). Results for the energy distribution and equation of state cannot be trusted when the relative difference between the (volume-averaged) left- and right-hand sides of the Hubble Friedmann equation becomes too large, $\Delta_e \equiv ({\rm LHS} - {\rm RHS})/({\rm LHS} + {\rm RHS}) \gtrsim \mathcal{O} (10^{-2})$. For many parameter regions this happens before the equation of state and/or the energy ratios achieve a stationary regime. In particular, the time scale at which this stationary regime is achieved grows as $z \sim q_{*}^\delta$ with $\delta \in [0, 1]$ (see  Ref.~\cite{Figueroa:2016wxr}), so it becomes increasingly difficult to study very large values of $q_{*}$. We have partially alleviated this issue by simulating the system with higher-order Velocity-Verlet algorithms~\cite{Figueroa:2020rrl}, which are implemented in \CL up to $O(dt^{-10})$. For the cases we present in the following, the violation of energy conservation does not exceed $\Delta_e \sim \mathcal{O} (10^{-3})$ at the end of the simulations.

\subsection{Lattice analysis: Centred potential} \label{Sec:LatResults1}

We begin by explaining the results from our lattice simulations with a centred potential ($v=0$).  We consider in detail the following three scenarios, according to the choice of power-law coefficient:  $p=2$, $p \in (2,4)$, and $p\geq4$.  \vspace{0.3cm}

\textbf{i)} $\boldsymbol{p=2}$. In this case, the inflaton does not develop fluctuations via self-resonance, but the daughter field gets excited via parametric resonance if a quadratic-quadratic coupling between both fields is sufficiently large. This may trigger the decay of the inflaton homogeneous mode due to backreaction. The strength of the resonance is set by the effective resonance parameter $\tilde q \equiv q_{*} a^{-3}$ [c.f.~\ref{eq:ResParam}], which decreases as the Universe expands. For parameters in the range $1 < q_{*} < q_{*}^{(max)}$ [where $q_{*}^{(max)} = 10^{3.8} \approx 6.3 \cdot 10^3$, recall Eq.~(\ref{eq:qstarM})], the resonance is initially broad but it becomes narrow before backreaction effects take place. Once narrow resonance is set, the energy of the daughter field decays as $\rho_X \sim a^{-4}$, while the one of the inflaton homogeneous mode goes as $\rho_{\phi} \sim a^{-3}$. As a result, the homogeneous inflaton eventually dominates the energy budget again. The gradient energy of both fields remains always subdominant, so the deviation of the equation of state from the homogeneous prediction $\bar{w} \simeq \bar{w}_{\rm hom} = 0$ is negligible.

\begin{figure*}
    \centering
    \includegraphics[width=0.47\textwidth]{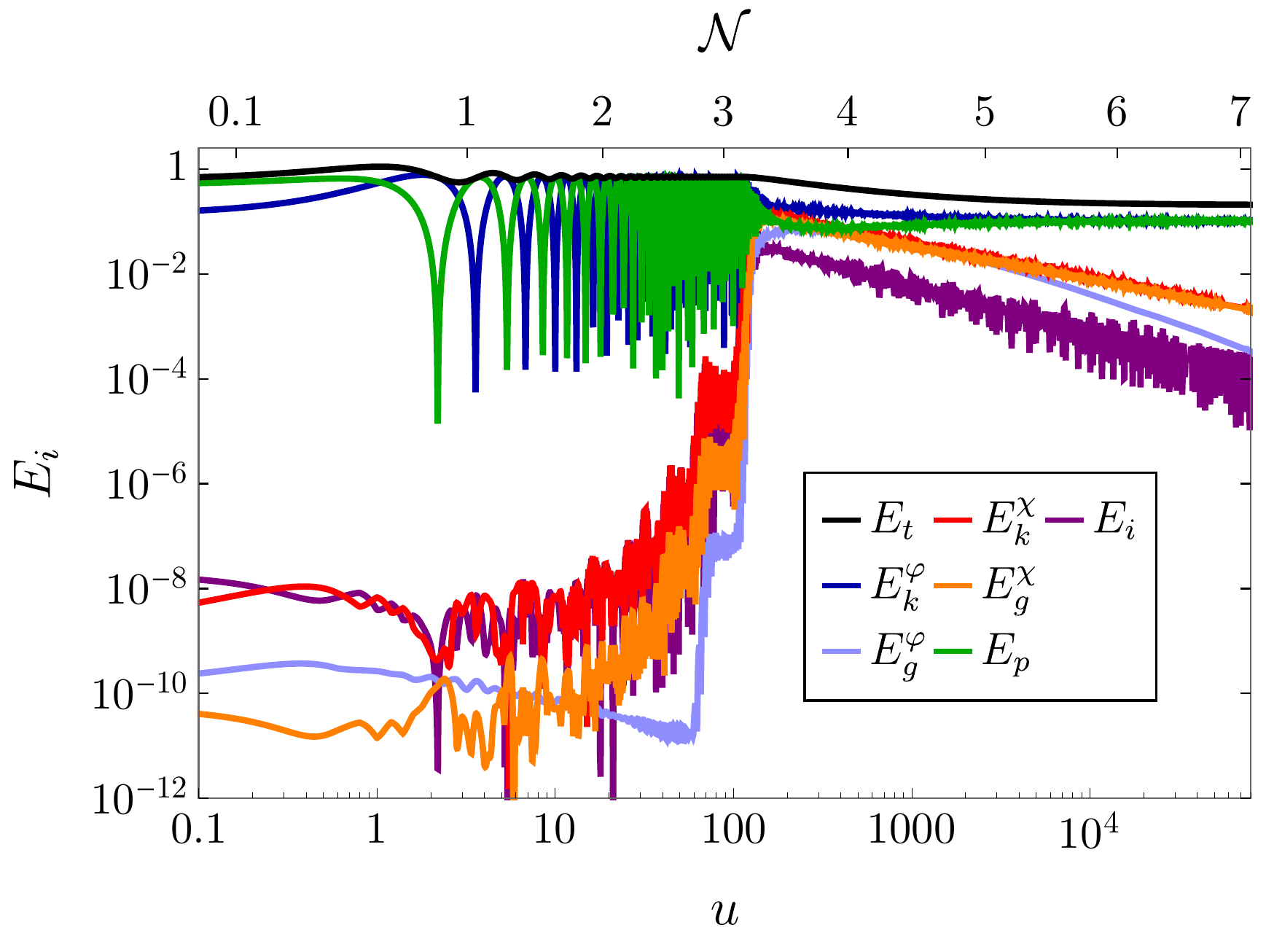} \,\,\,\,\,
    \includegraphics[width=0.47\textwidth]{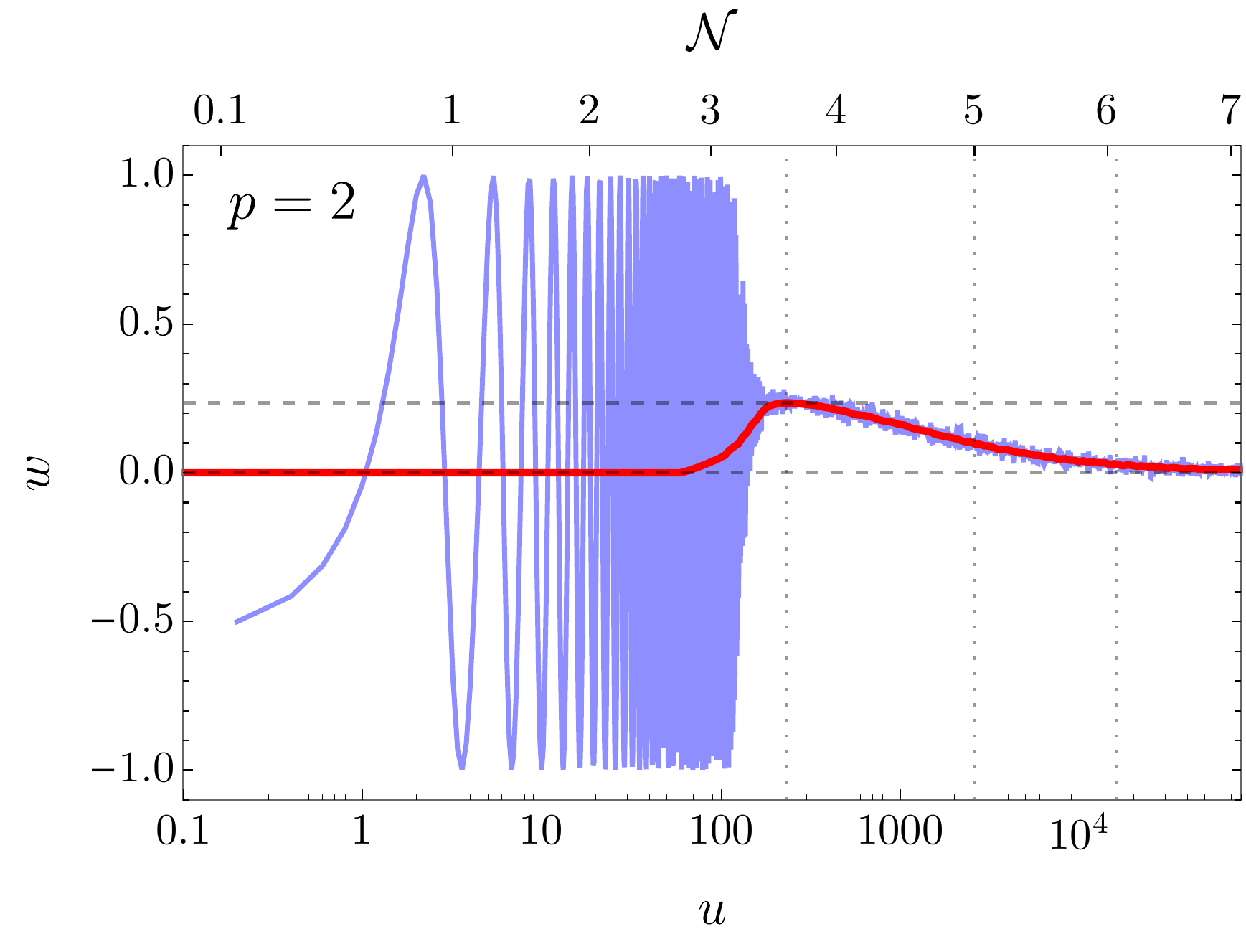}  \vspace{-0.2cm}
    \caption{[$v=0$] Results from lattice simulations with $p=2$, $q_*=2.4 \cdot 10^4$, and $M= 10m_{\rm pl}$. Left: Evolution of the different energy contributions (\ref{eq:energyshift1})-(\ref{eq:energyshift5}) and (\ref{eq:energyshift4}) and their sum $E_t \equiv \sum_i E_i$ as a function of time and post-inflationary number of e-folds. Right: Instantaneous equation of state $w$ (blue line) and its effective oscillation-averaged approximation $\bar{w}$ (red thick line). The horizontal dashed line indicates the maximum equation of state $\bar{w}_{\rm max}$, while the different vertical lines indicate the times at which the equation of state attains $\bar{w}=\bar{w}_{\rm max}$, $0.1$, and $0.03$.}  \label{p2-EosEn}
\end{figure*} 

The case $q_{*} > q_{*}^{(max)}$ is more interesting. In Fig.~\ref{p2-EosEn} we show, for $p=2$ and $q_{ *}=2.4 \cdot 10^4$, the post-inflationary evolution of the different energy ratios and equation of state. Contrary to the previous case, now the inflaton homogeneous mode decays via backreaction effects, at approximately the \textit{backreation time} $u_{\rm br}$ given in Eq.~(\ref{eq:zBR}). For the range of parameters considered in this work, we typically have $u_{\rm br} \sim 30 - 100$, in agreement with the analytical estimation of Fig.~\ref{fig:Analytic-ParRes}. Approximately at this time, the fraction of energy stored in the gradients of both fields becomes sizeable, and due to this, the effective equation of state deviates from the homogeneous solution $\bar{w} \simeq \bar{w}_{\rm hom} \equiv 0$ as time approaches $u_{\rm br}$, reaching a local maximum $\bar{w} = \bar{w}_{\rm max} < 1/3$ at some later time $u_{\rm max} > u_{\rm br}$. However, as $\tilde q = q_{*} a^{-3}$ decreases with time, the resonance eventually becomes narrow and the interaction negligible. From then on, the total inflaton energy decays as $\rho_\phi \sim a^{-3}$ due to its mass, while the energy of the (massless) daughter field dilutes as $\rho_X \sim a^{-4}$. The total gradient energy also becomes very small at late times, so the homogeneous mode of the inflaton eventually dominates the energy budget of the system again. This behaviour gets reflected in the effective equation of state, which slowly decays from the maximum $\bar{w} = \bar{w}_{\rm max}$ towards $\bar{w} \rightarrow 0$.

\begin{figure}
    \centering
       \includegraphics[width=0.47\textwidth]{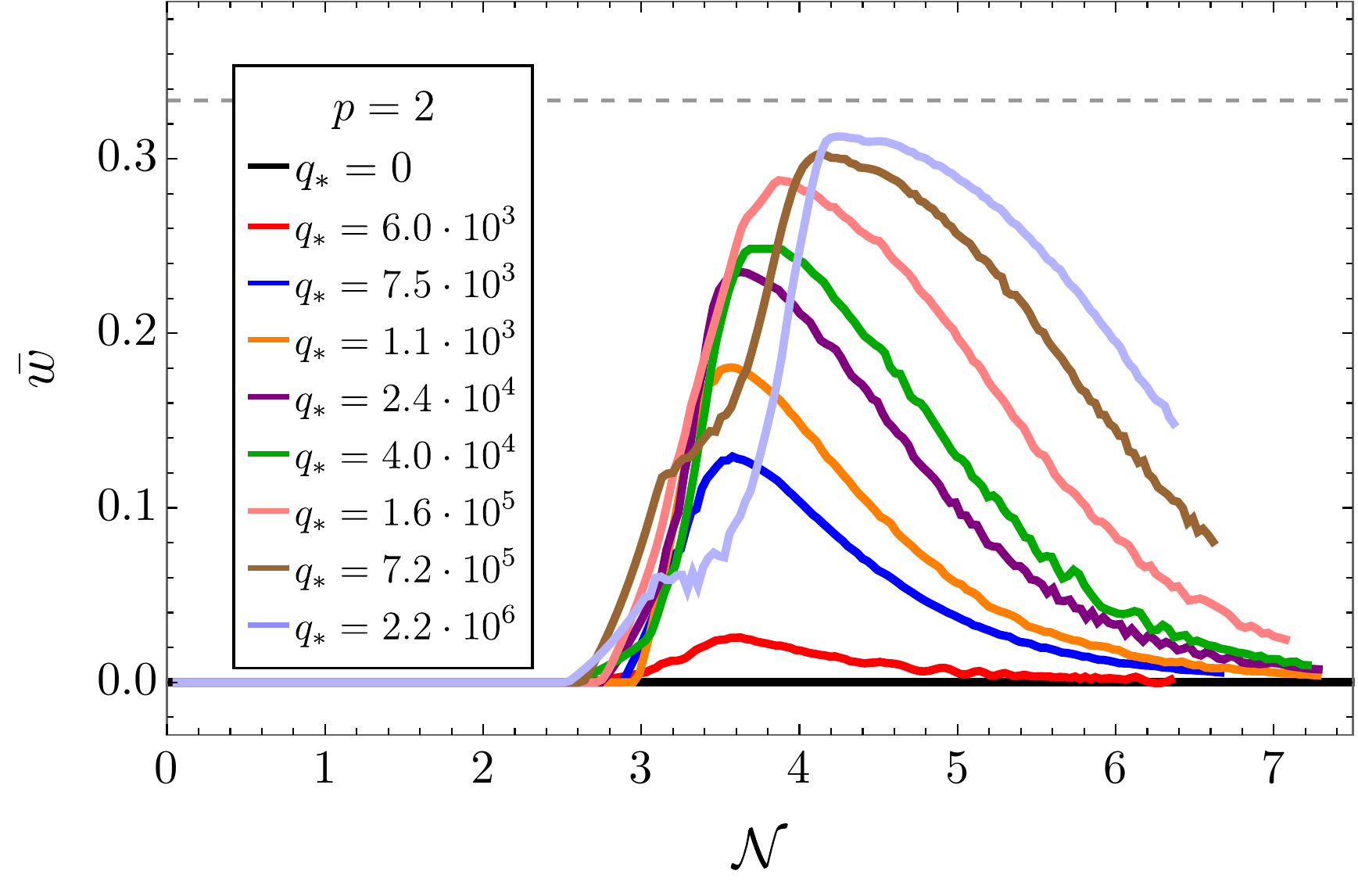} \vspace{-0.1cm}
    \caption{[$v=0$] Evolution of the effective equation of state for $p=2$, $M=10m_{\rm pl}$, and different choices of $q_*$, as a function of the post-inflationary number of e-folds.}  \label{fig:p2qParam}
\end{figure}

\begin{figure}
    \centering
    \includegraphics[width=0.43\textwidth]{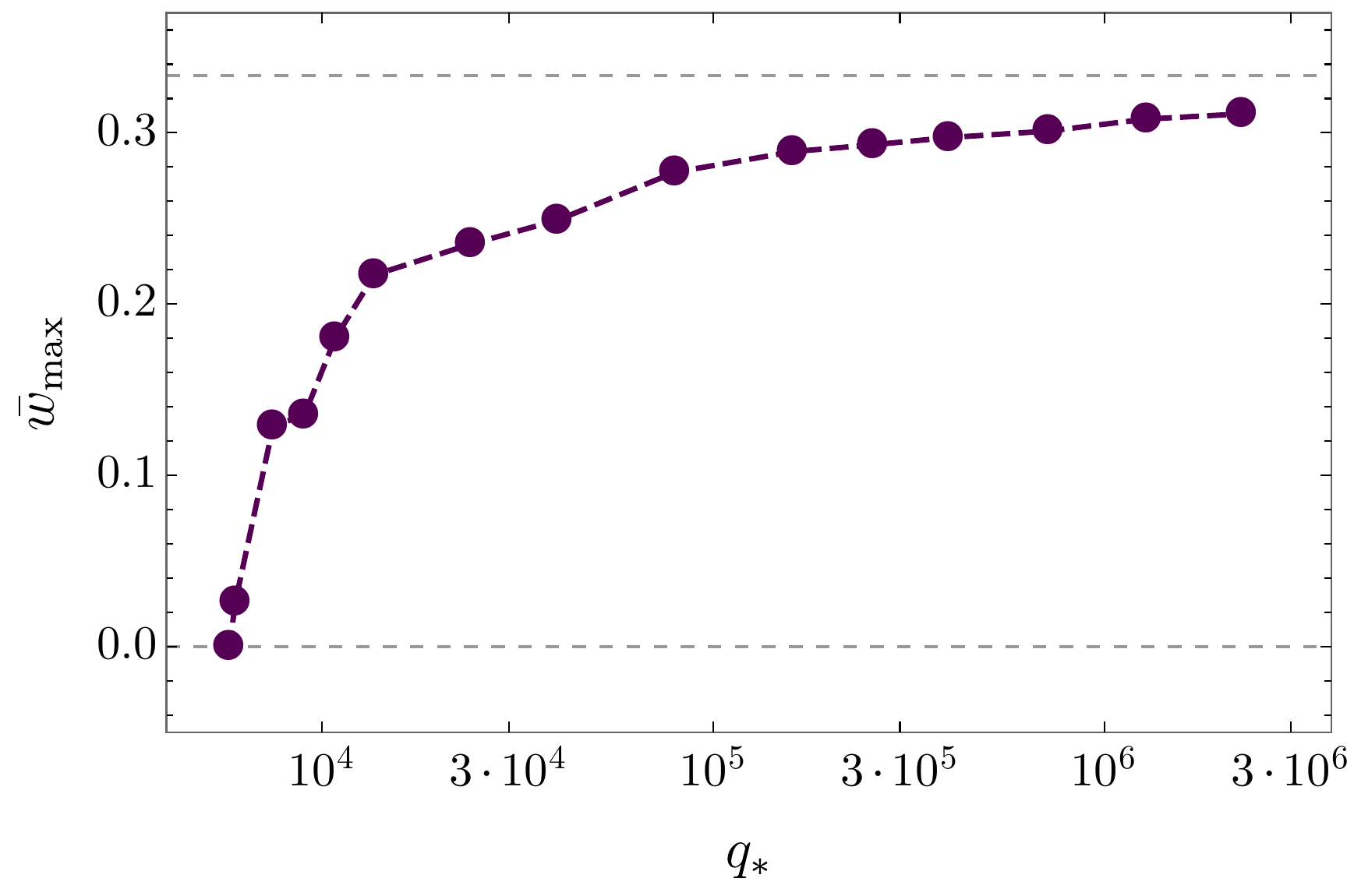}
    \includegraphics[width=0.43\textwidth]{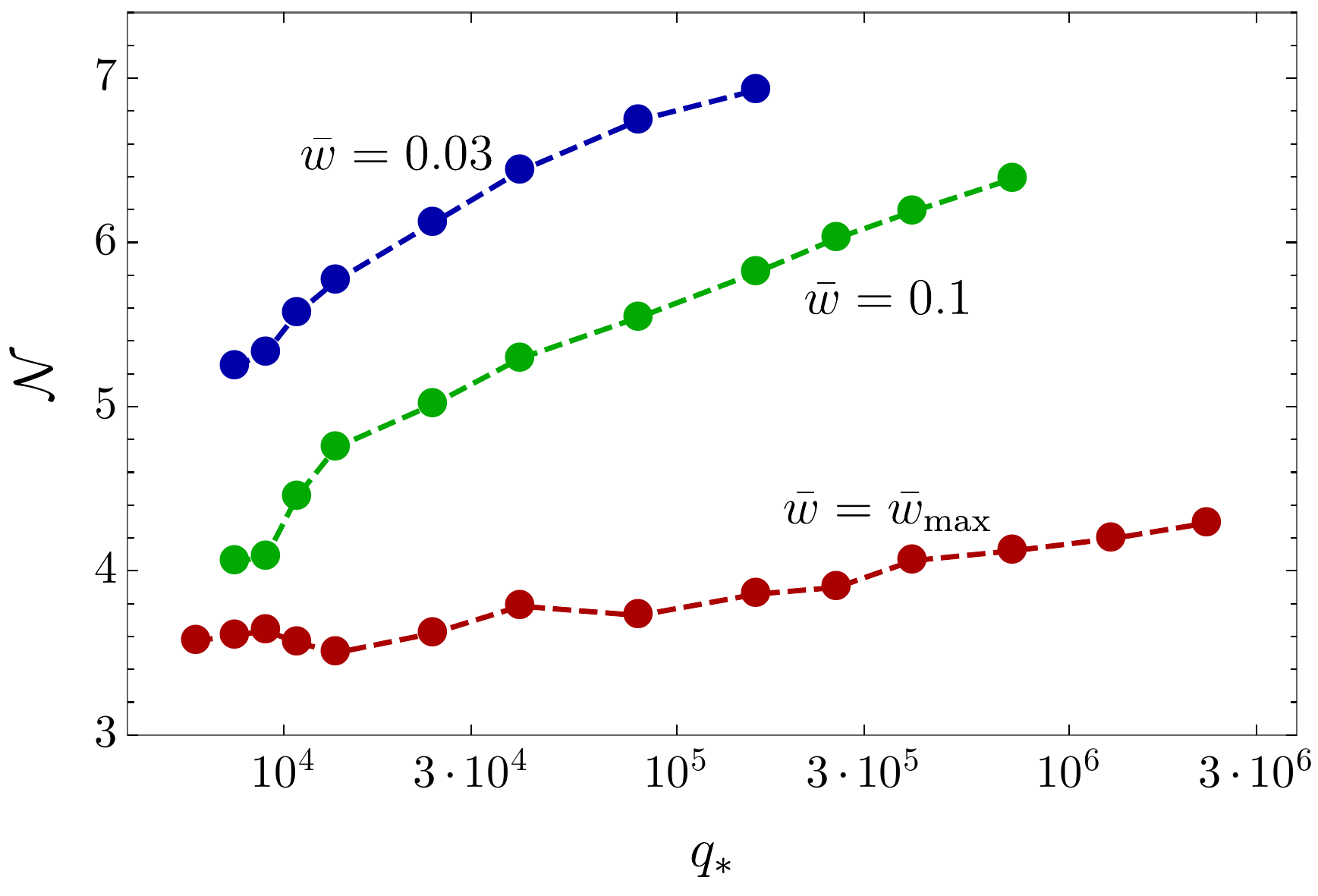}
    \caption{[$v=0$] Parametrization of the effective equation of state for $p=2$ and $M=10m_{\rm pl}$. The top panel shows the maximum value attained after inflation $\bar{w}_{\rm max}$ for different choices of $q_*$. The bottom panel shows the corresponding post-inflationary number of e-folds of expansion at which $\bar{w} = \bar{w}_{\rm max}$ is attained, as well as $\bar{w} = 0.1$, 0.03 during the subsequent relaxation process.} \label{p2-Fits}
\end{figure}
   
The qualitative evolution of the equation of state is very similar for all resonance parameters obeying $q_* \gtrsim q_*^{\rm (min)}$, though the specific details depend on the  choice of $q_*$. This can be seen in Fig.~\ref{fig:p2qParam}, where we show the evolution of $\bar{w}$ obtained from simulations with different values of $q_*$. We can observe that, as $q_*$ increases, the maximum value that the equation of state attains, $\bar{w}_{\rm max}$, becomes larger, and the whole growth-and-decay process also takes longer. We have quantified this in Fig.~\ref{p2-Fits}. In the top panel we show that the larger the value of $q_{*}$ is, the larger $\bar{w}_{\rm max}$ becomes, slowly approaching the radiation-dominated value $\bar{w}_{\rm max} = 1/3$ for very large $q_*$. Similarly, in the bottom panel we show the number of post-inflationary e-folds it takes for the equation of state to reach $\bar{w}=\bar{w}_{\rm max}$, as well as to decay down to $\bar{w} = 0.1$ and $0.03$ during the subsequent relaxation process. For the range of resonance parameters considered, $\bar{w}_{\rm max}$ is attained $\sim3-4$ e-folds after the end of inflation, while the relaxation process may take several more e-folds. \vspace{0.3cm}

\textbf{ii) $\boldsymbol{2<p<4}$}. According to our linearized analysis, in this case there are two relevant post-inflationary resonant phenomena dictating the evolution of the energy distribution and equation of state: self-resonance of the inflaton and parametric resonance of the daughter field.

\begin{figure*}
    \centering
    \includegraphics[width=0.42\textwidth]{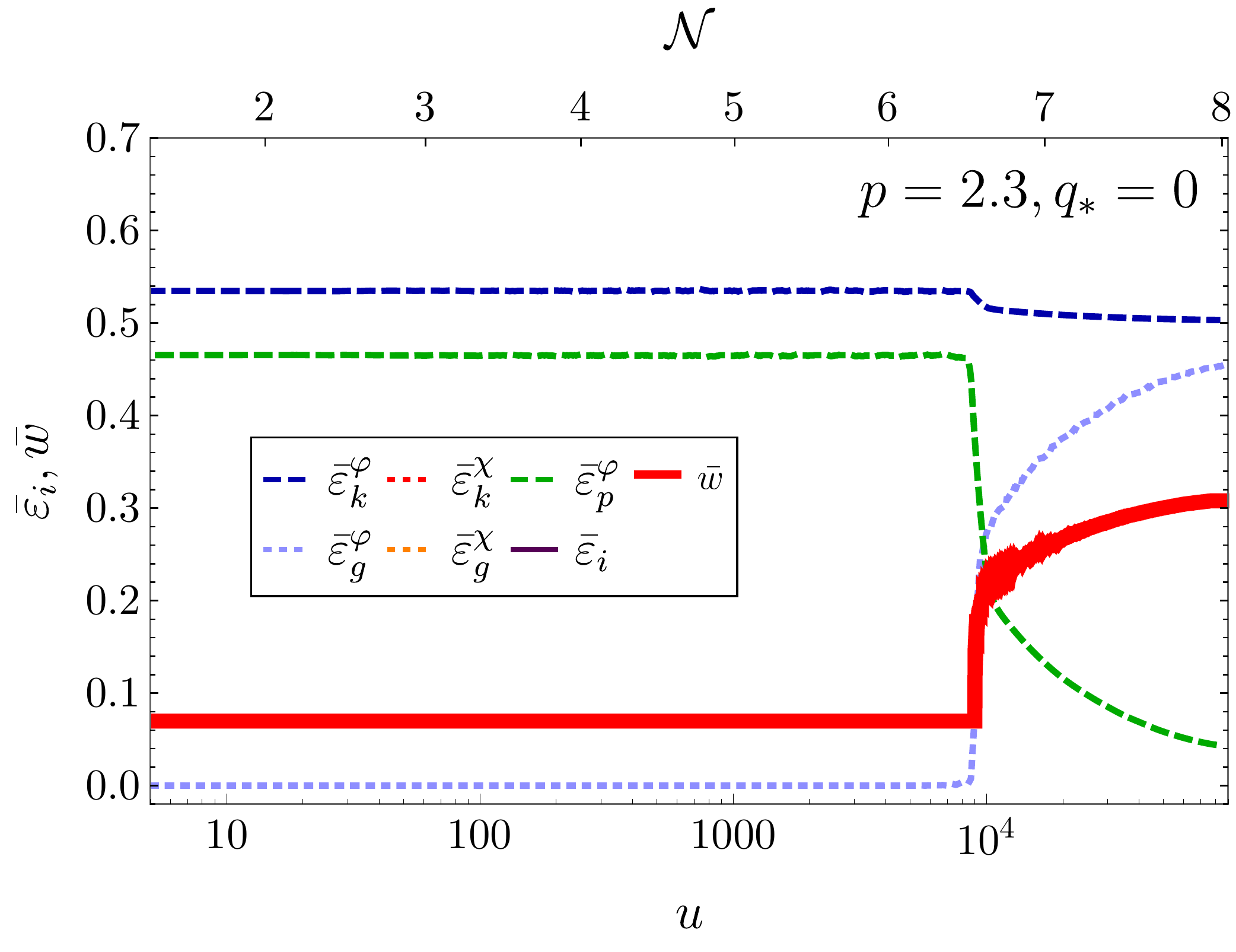}  
        \includegraphics[width=0.42\textwidth]{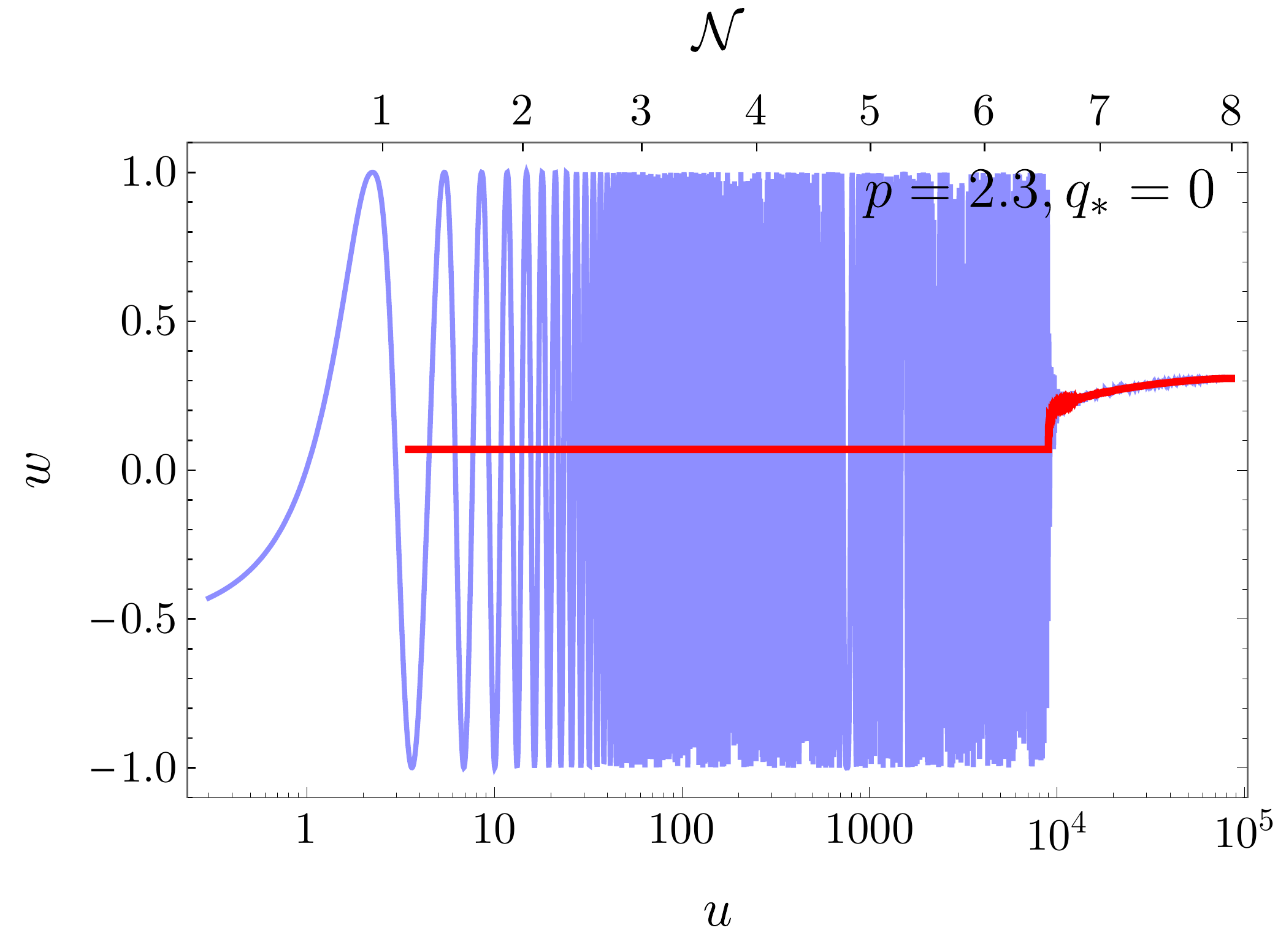}  \\
    \includegraphics[width=0.42\textwidth]{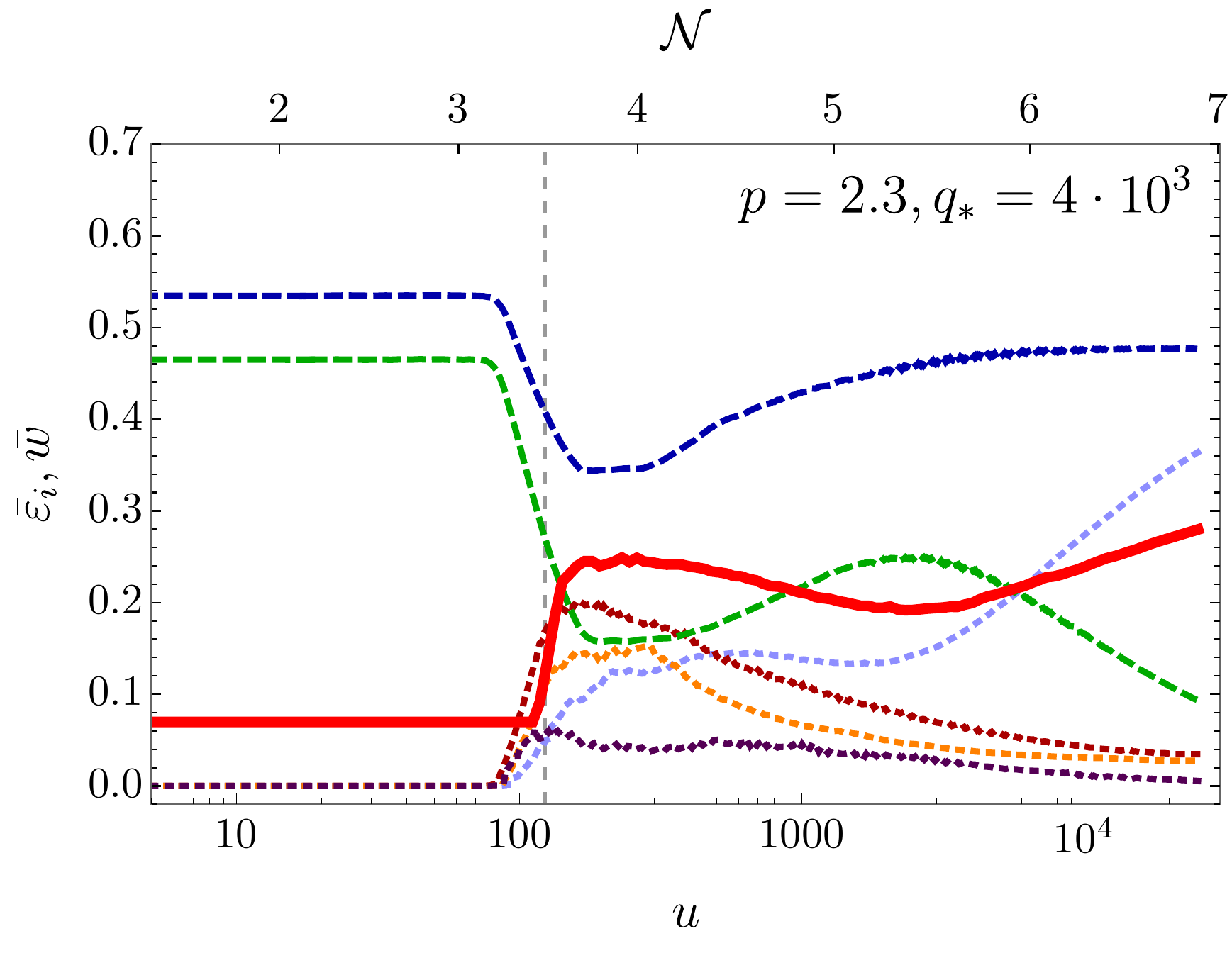} 
        \includegraphics[width=0.42\textwidth]{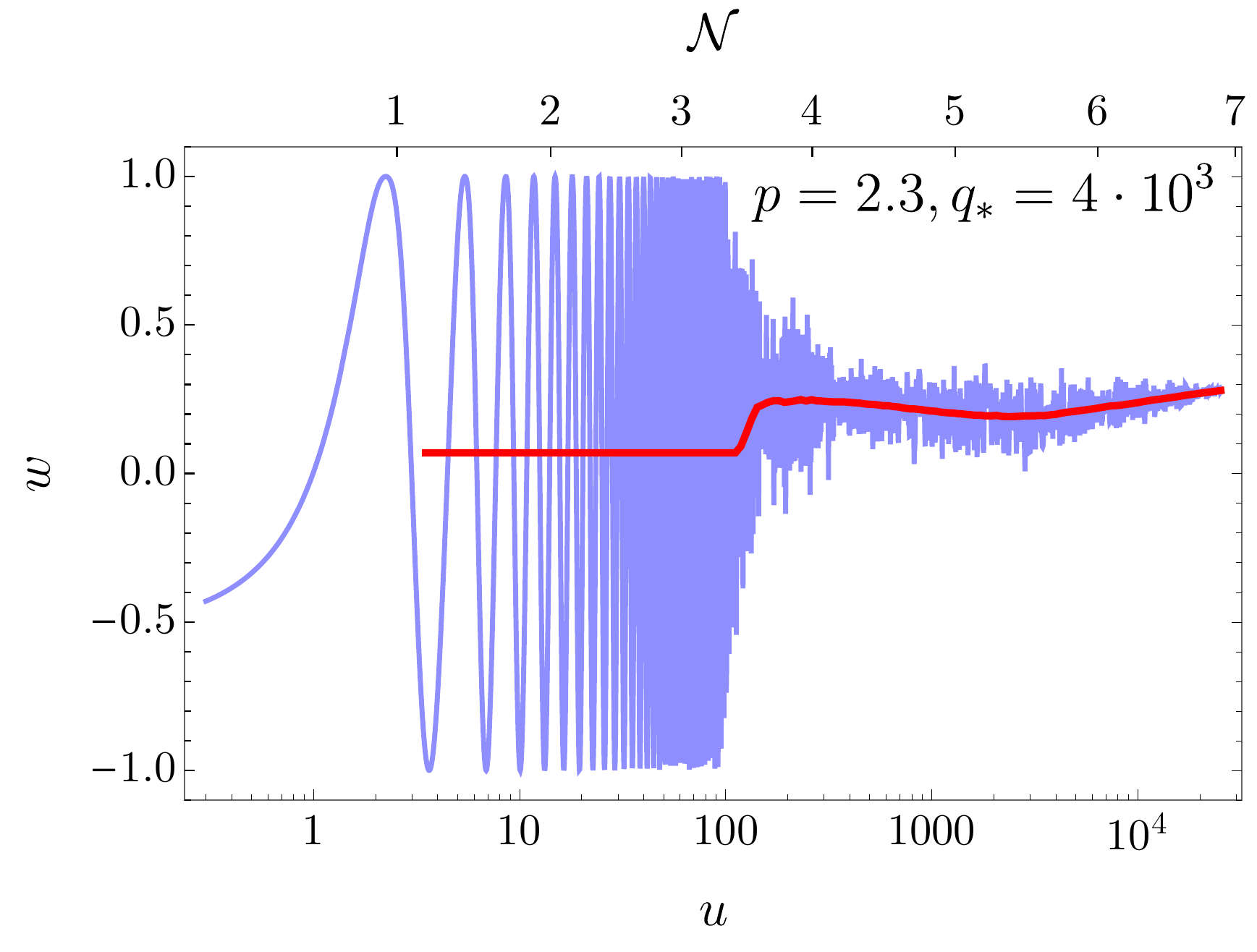} \\  
    \includegraphics[width=0.42\textwidth]{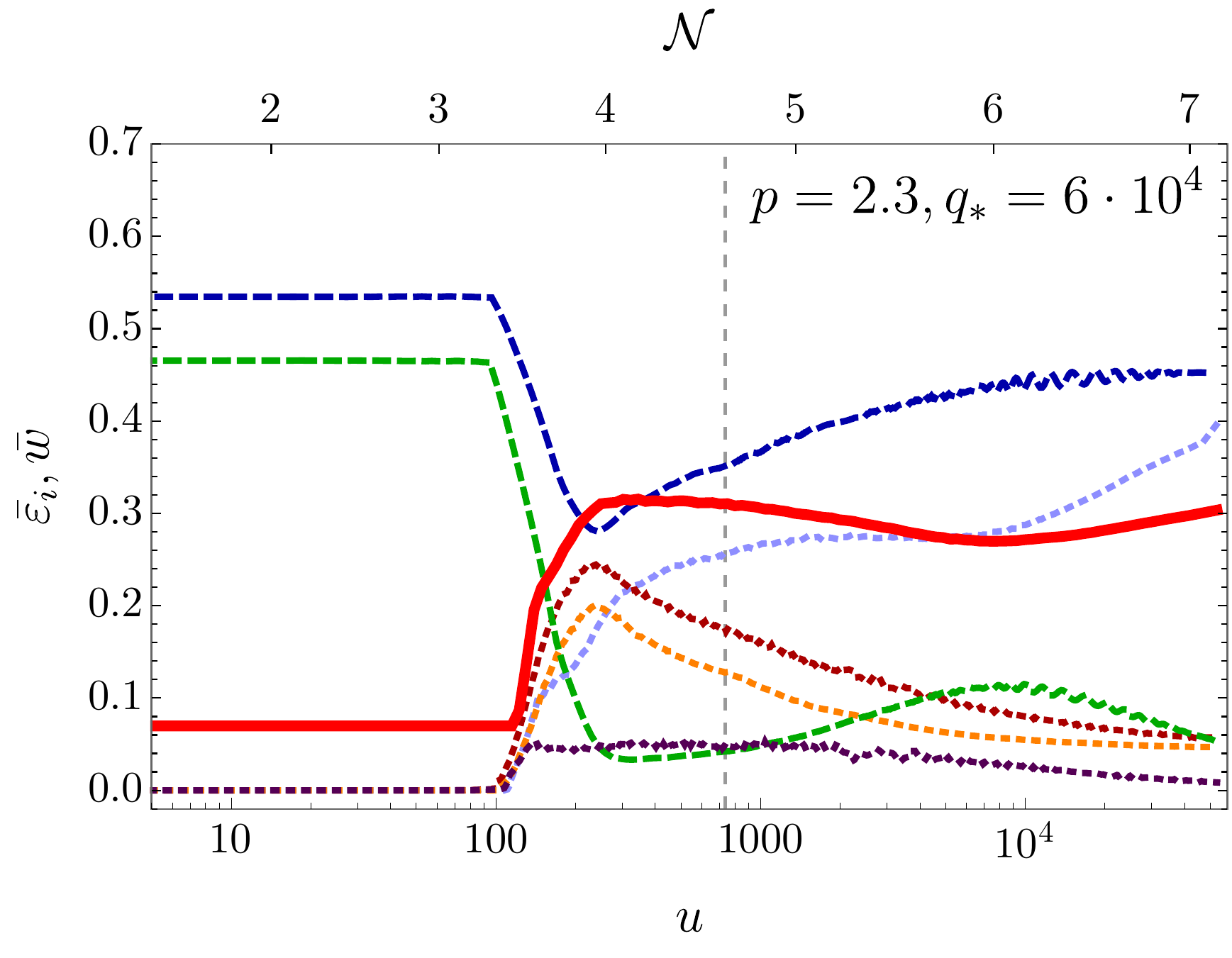}
    \includegraphics[width=0.42\textwidth]{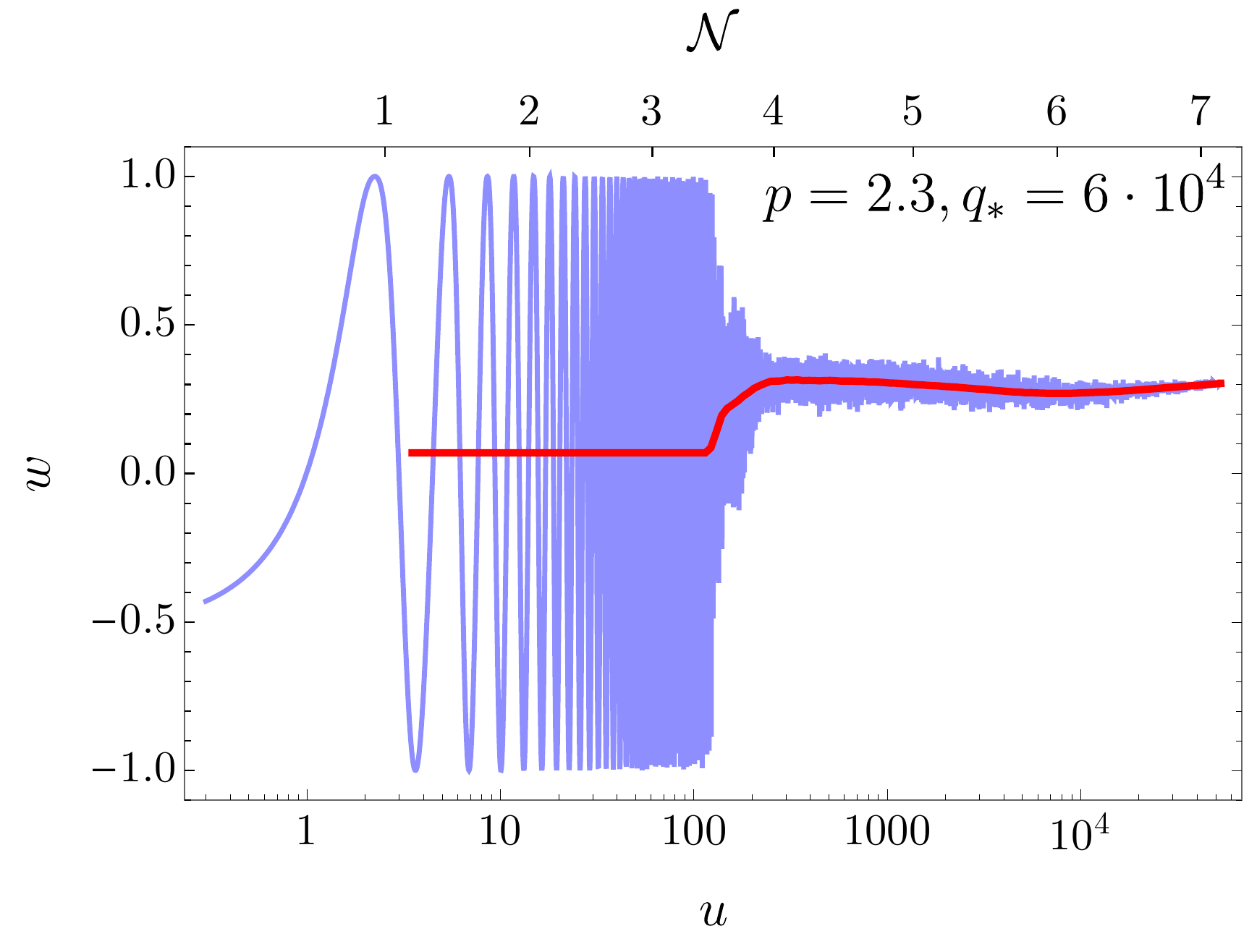}
    
    \caption{[$v=0$] Left panels: Evolution of the energy ratios for $p=2.3$, $M=10 m_p$, and $q_*=0$ (top), $4 \cdot 10^3$ (middle), and $6 \cdot 10^4$ (bottom). Right panels: Evolution of the instantaneous and oscillation-averaged equation of state (depicted in blue and red respectively).}  \label{p23-EosEnergy}
\end{figure*}

Let us consider first the case $p=2.3$, i.e.~a value slightly larger than the $p=2$ case considered just above. This choice allows to illustrate very clearly the different time scales at which parametric resonance and self-resonance are effective. In Fig.~\ref{p23-EosEnergy} we have plotted the evolution of the energy ratios and equation of state for three different resonance parameters: $q_{*}=0$, $4 \cdot 10^3$, and $6 \cdot 10^4$. Initially, the inflaton homogeneous mode dominates the energy budget, with its kinetic energy representing $\sim$53\% of the total, its potential energy representing the other $\sim$47\%, c.f.~Eq.~(\ref{eq:HomEner}), and with negligible gradient energy density. The equation of state at this stage is $\bar{w}_{\rm hom} \simeq 0.07$, as expected from Eq.~(\ref{eq:EoSoscillations}) for $p = 2.3$. This is well observed in the three panels depicted in the figure. However, the energy ratios and equation of state evolve in various manners after backreaction effects kick in, which we explain in the following.

If $q_{*} = 0$, there is no transfer of energy to the daughter field, but the amplitude of the inflaton fluctuations grows exponentially via self-resonance, according to the linearized analysis of Section \ref{sec:3A}. Approximately $N_{\rm br} \sim 6-7$ e-folds after the end of inflation, its gradient energy becomes comparable to the one of the inflaton homogeneous mode, i.e.~$\bar{\varepsilon}_{\rm g}^{\varphi} \approx  \bar{\varepsilon}_{\rm p}^{\varphi}$. This can be seen in the left panel of Fig.~\ref{p23-EosEnergy}, and it is in qualitative agreement with the analytical prediction of Fig.~\ref{fig:PvszbrNbr}. Correspondingly, at this time there is a deviation of the equation of state from $\bar{w} = \bar{w}_{\rm hom} \simeq 0.07$ towards $\bar w \rightarrow 1/3$. Remarkably, the self-interactions keep exciting modes of increasingly higher comoving momenta even after backreaction, which leads to a complete fragmentation of the inflaton homogeneous mode at very late times. This effect was noted first in \cite{Lozanov:2016hid,Lozanov:2017hjm}, and here we confirm the result. Due to this, the ratio $ \bar \varepsilon_{\rm g}^{\varphi}  /  \bar \varepsilon_{\rm p}^{\varphi} $ keeps growing during the later non-linear stage, and goes to $  \bar \varepsilon_{\rm g}^{\varphi}  / \bar \varepsilon_{\rm p}^{\varphi}  \rightarrow \infty$ at very late times. Consistent with the Virial identity (\ref{eq:Virial1}), we have $\bar \varepsilon_{\rm k}^{\varphi}, \bar \varepsilon_{\rm g}^{\varphi} \rightarrow 1/2$ in this regime, and the equation of state (\ref{eq:EoS-fEn}) goes to  $\bar w \rightarrow 1/3$ at very late times.

\begin{figure*} \centering    
    \includegraphics[width=0.4\textwidth]{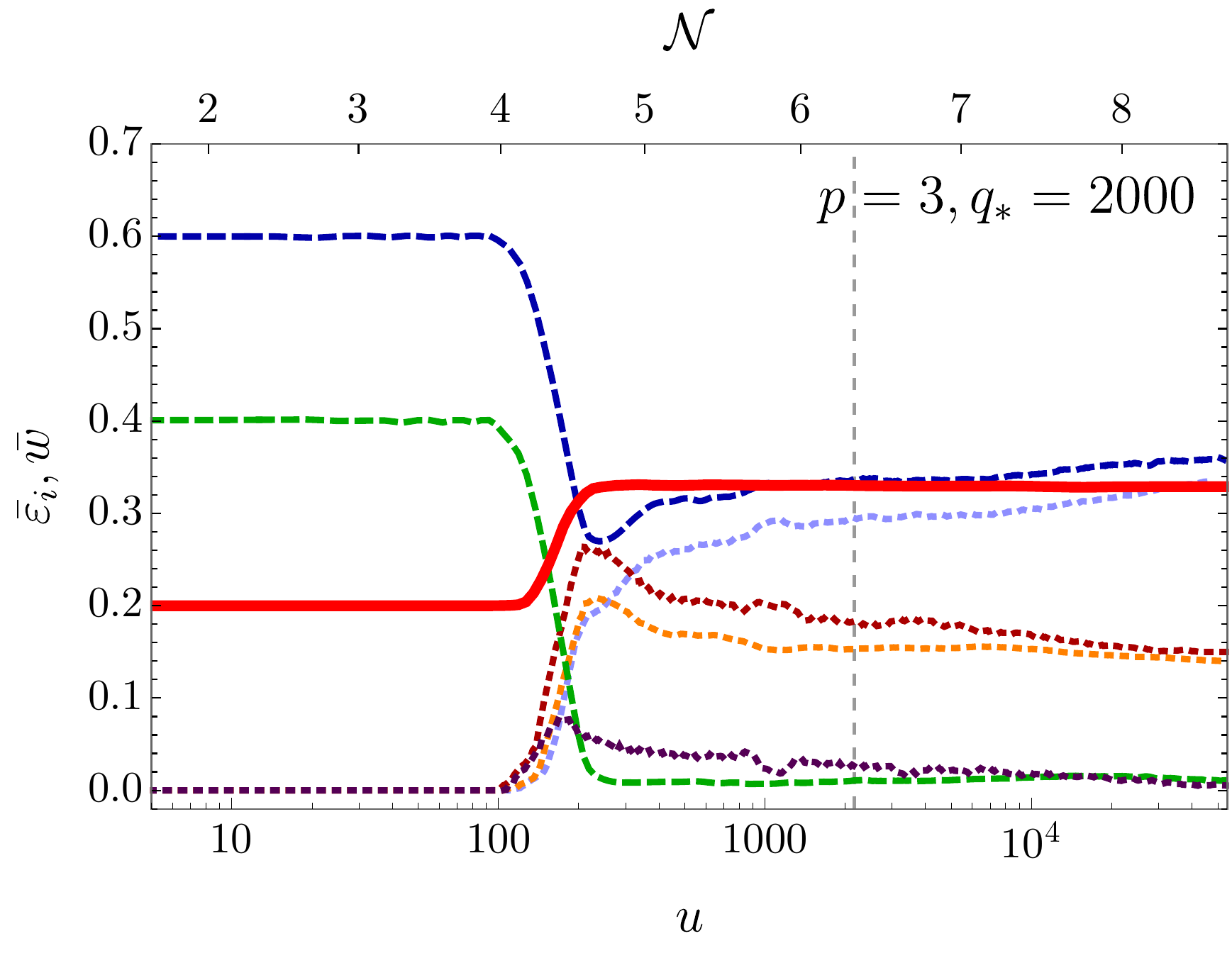} \,\,
    \includegraphics[width=0.4\textwidth]{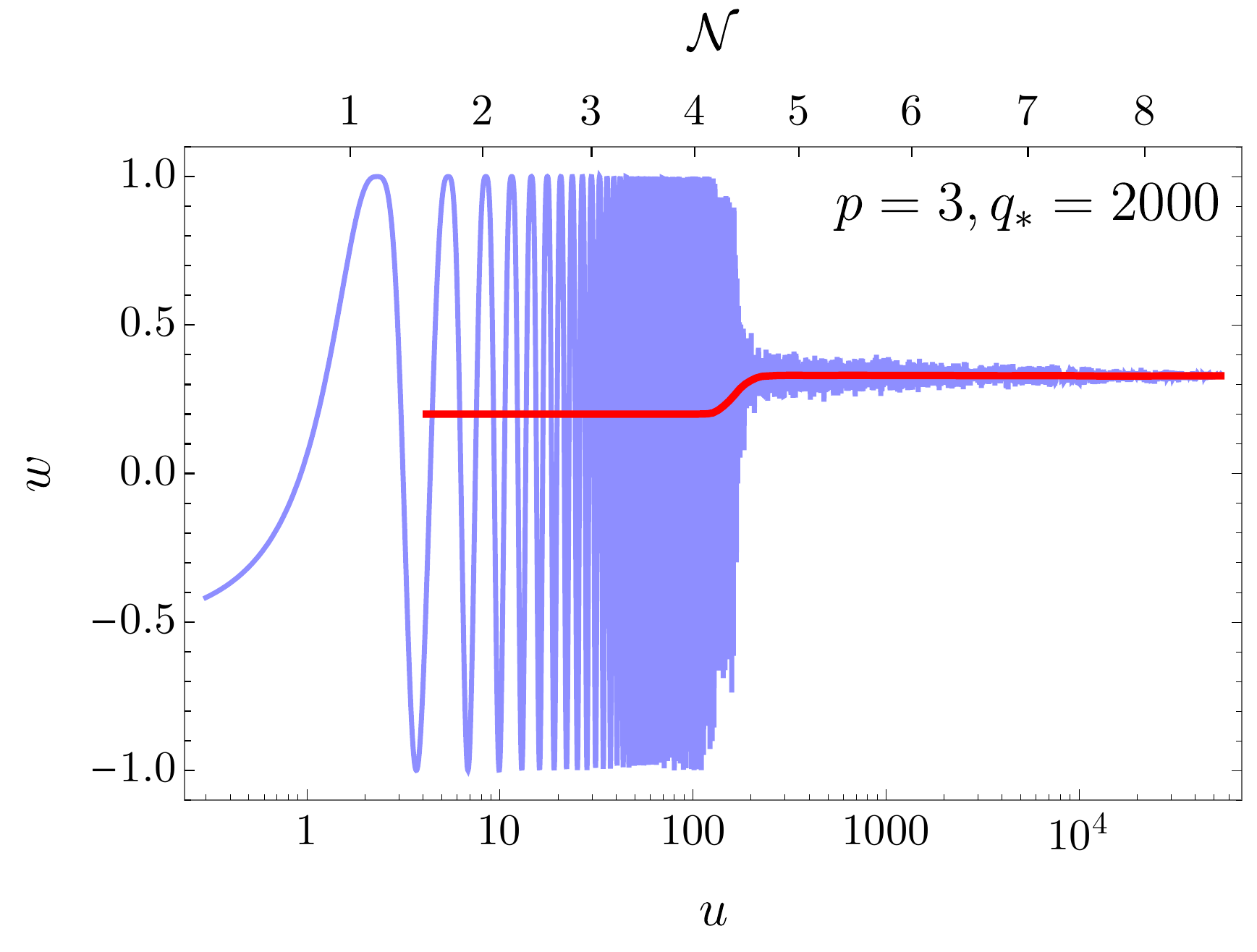}  \\ 
    \includegraphics[width=0.4\textwidth]{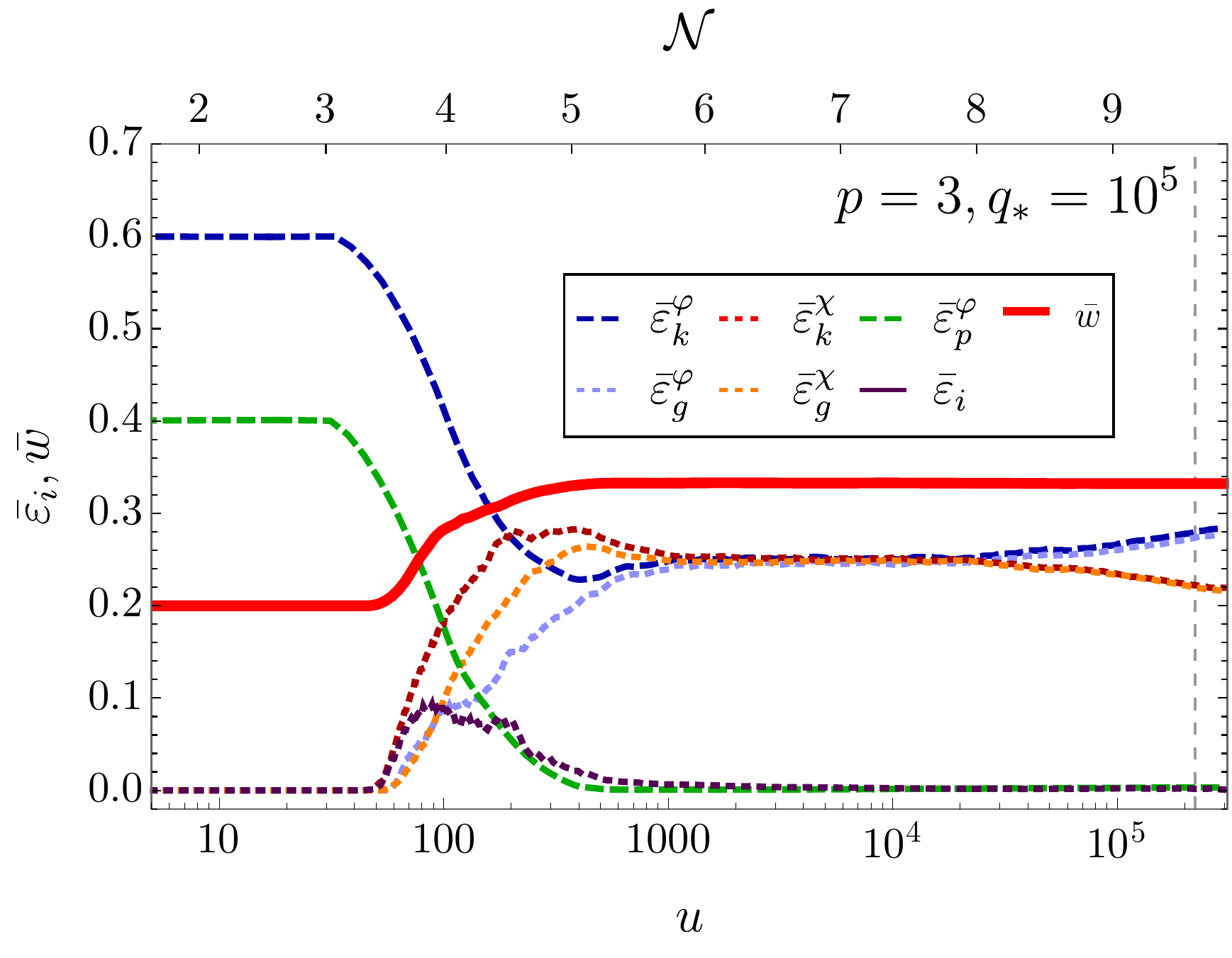} \,\,
    \includegraphics[width=0.4\textwidth]{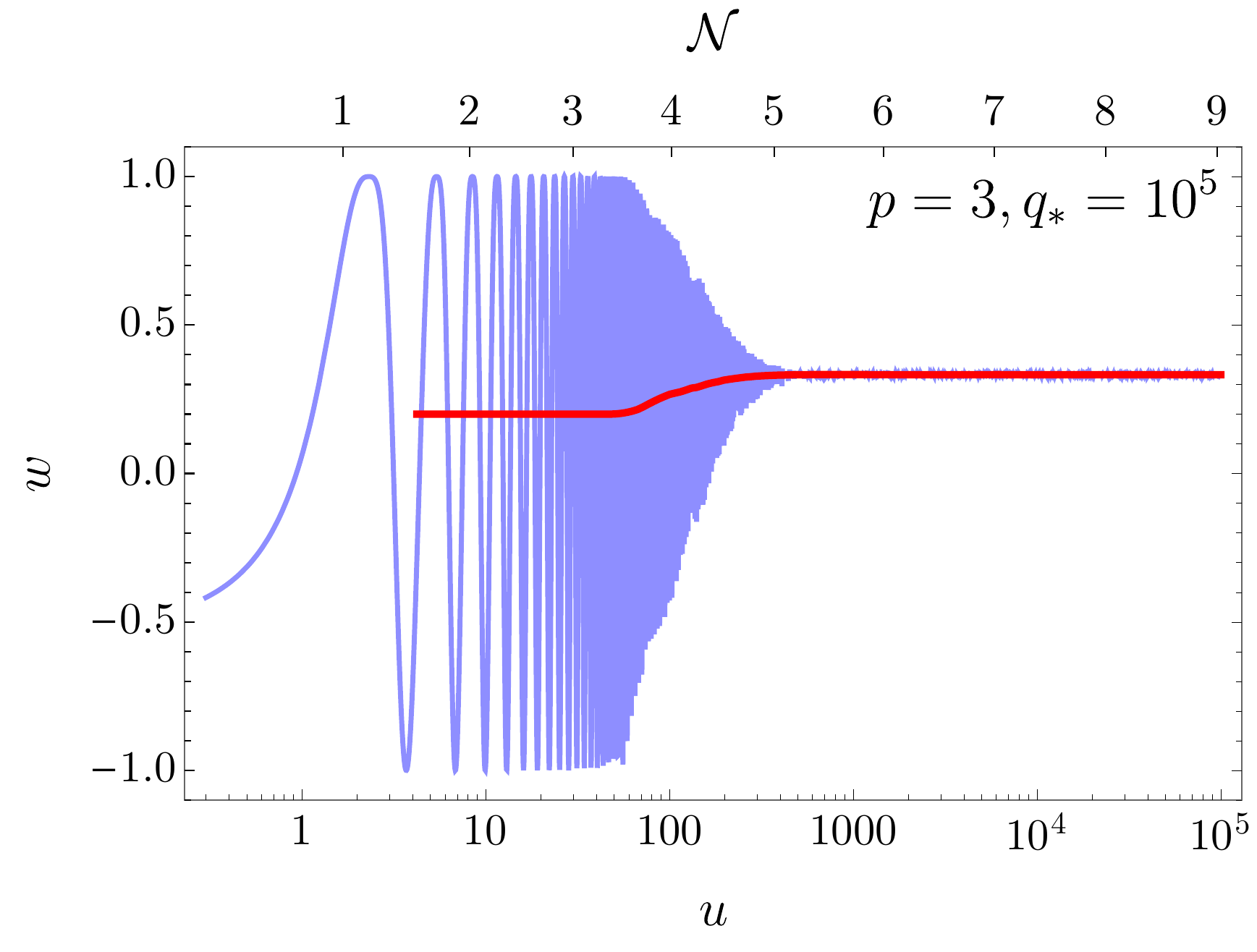}  \\ 
    \caption{[$v=0$] Post-inflationary evolution of the energy ratios and effective equation of state for $p=3$, $M=10m_\mathrm{pl}$, and two different values of $q_*$. The dashed vertical lines in the left panels show when $\tilde q = 1$. } \label{fig:Energiesp3}
    \includegraphics[width=0.4\textwidth]{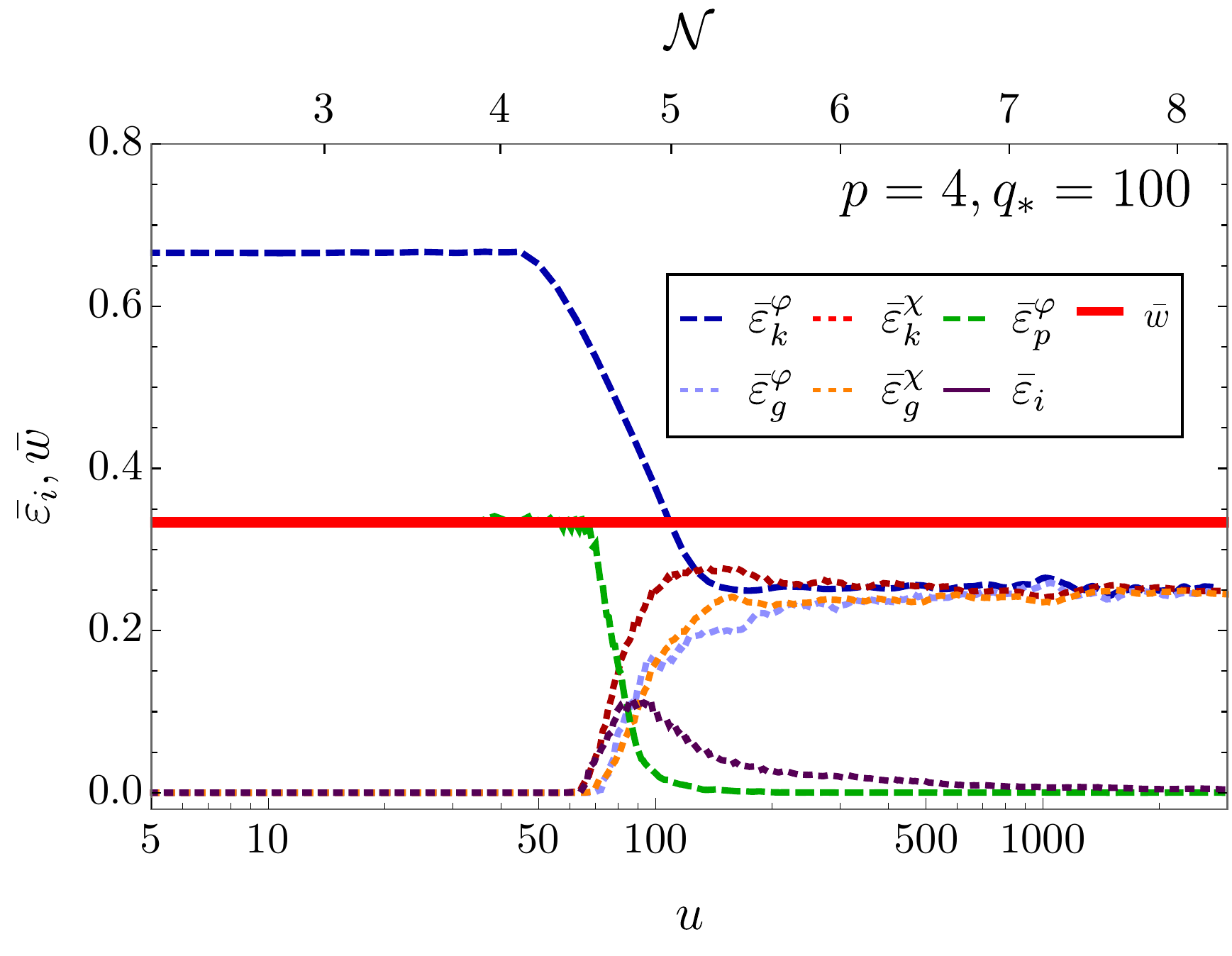} \,\,
    \includegraphics[width=0.4\textwidth]{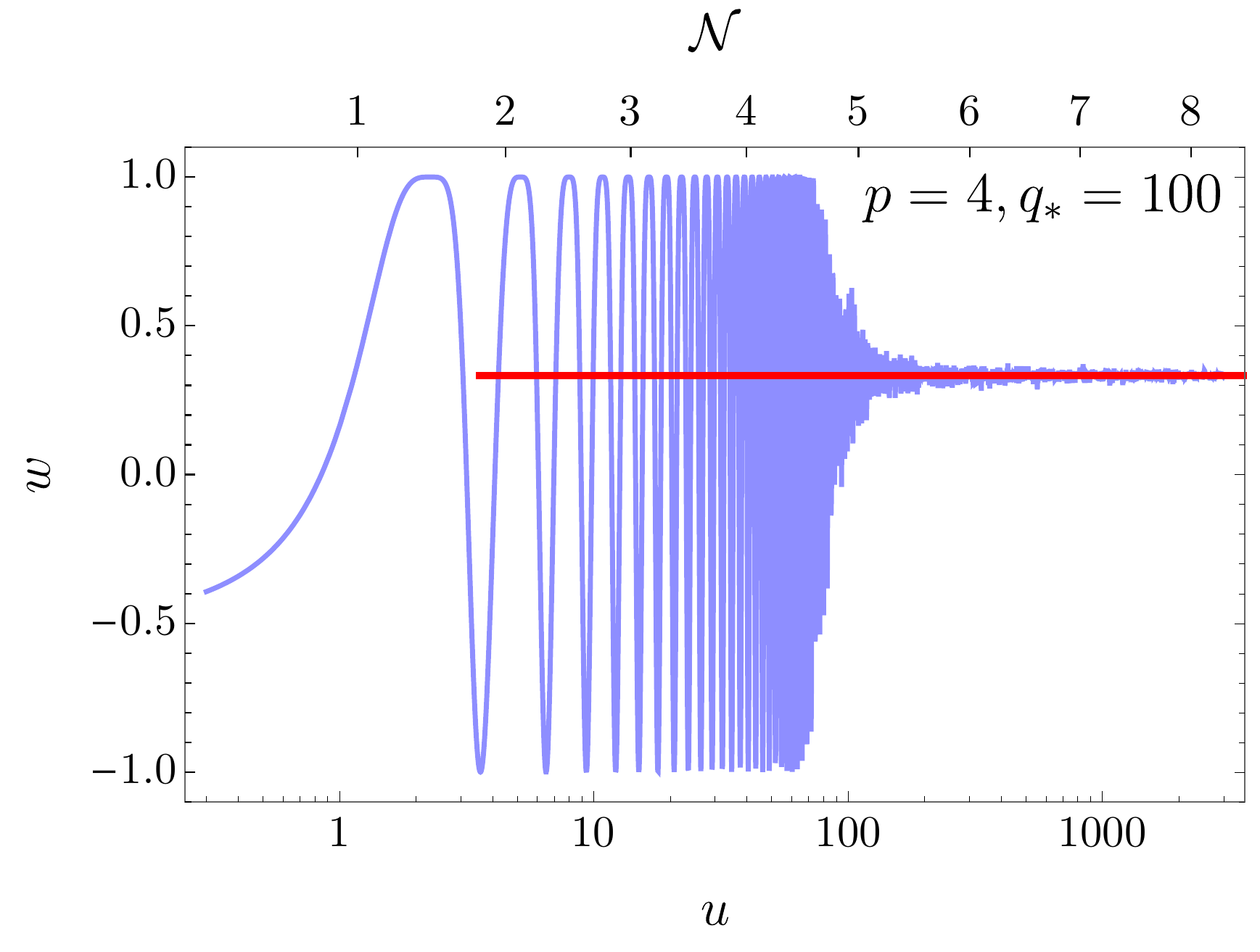} \\
    \includegraphics[width=0.4\textwidth]{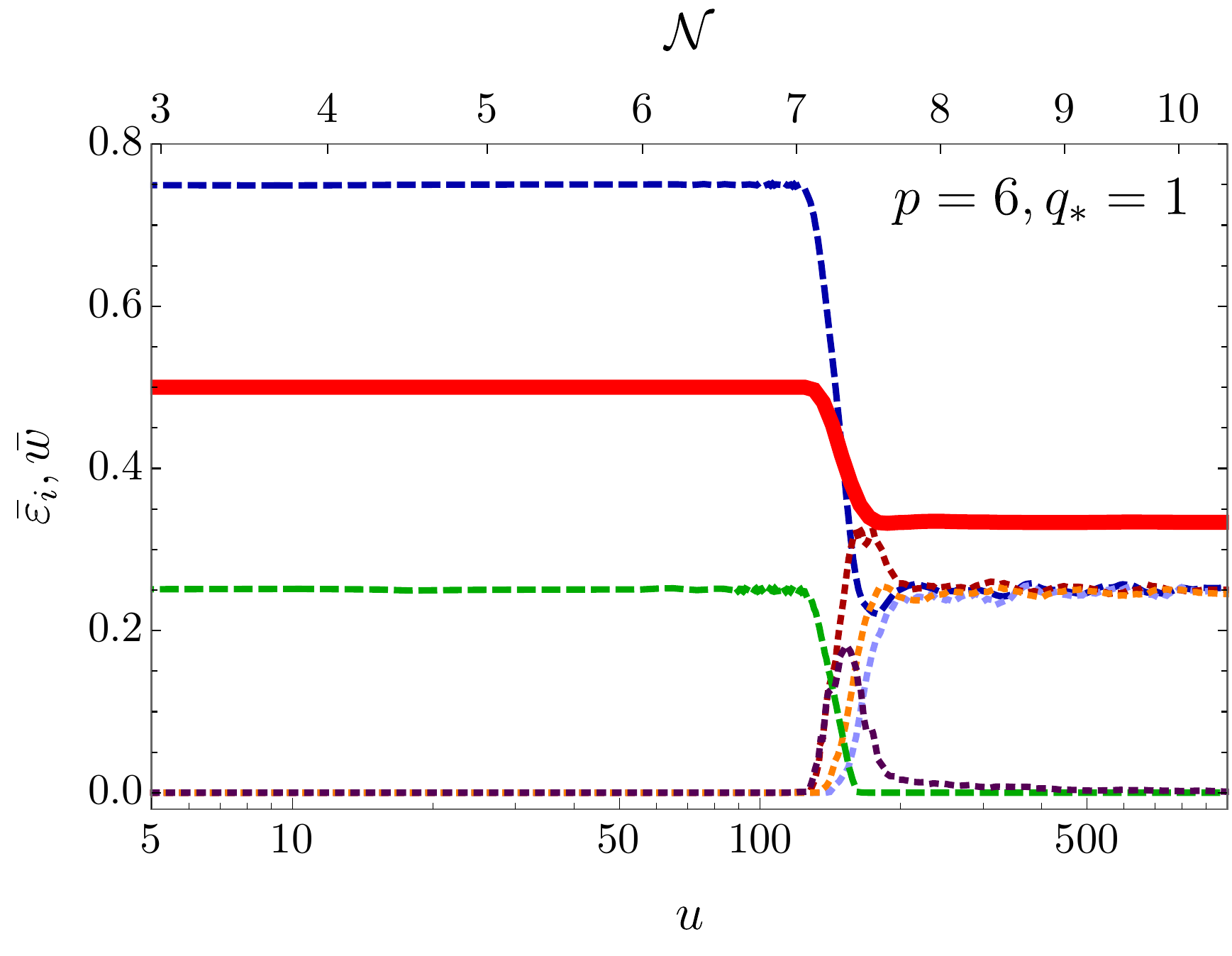} \,\,
    \includegraphics[width=0.4\textwidth]{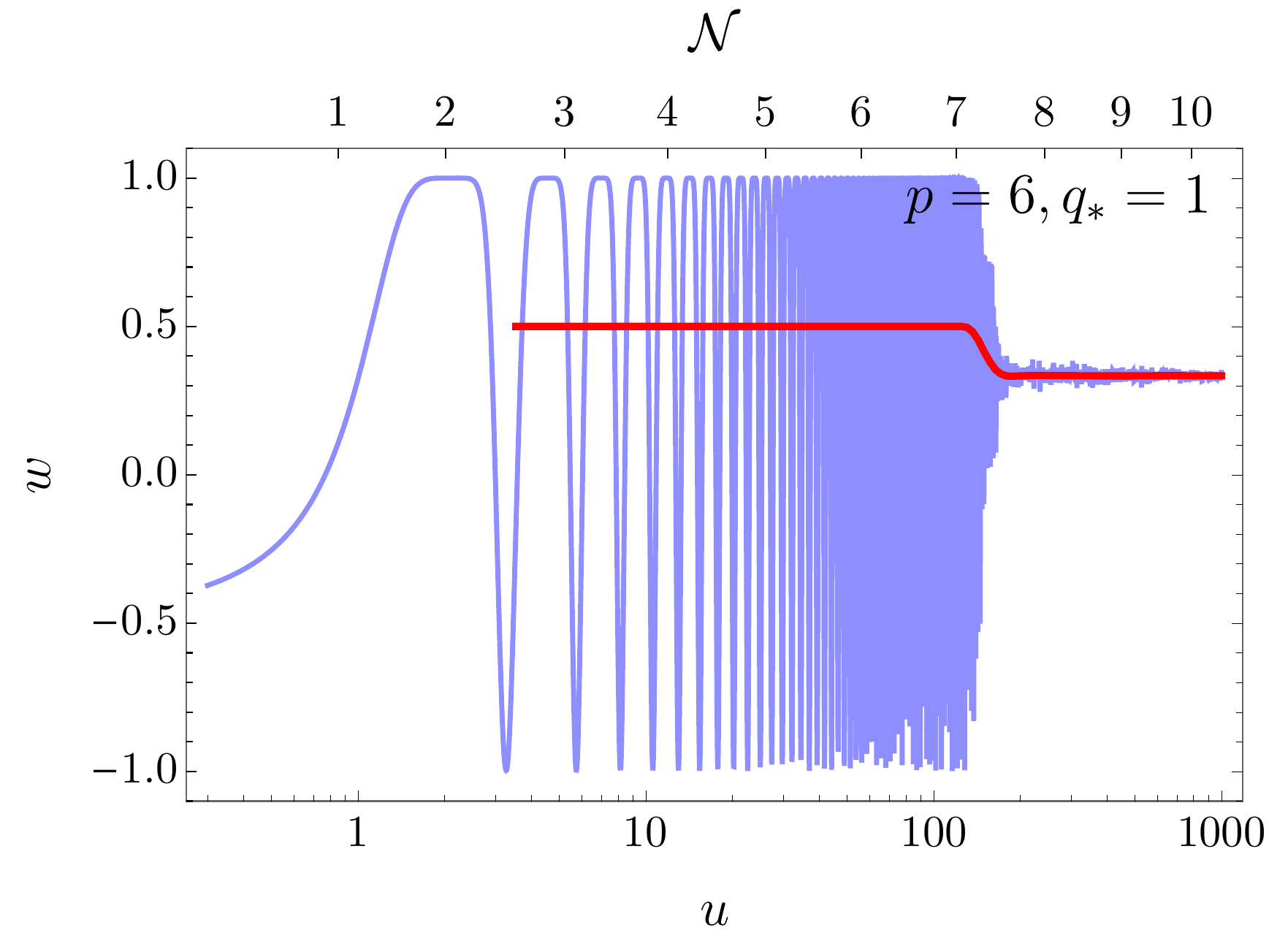} 
    \caption{[$v=0$] Evolution of the energy ratios and equation of state for $p>4$, $M=10m_\mathrm{pl}$, and two different values of $q_*$ as a function of time and number of e-folds. } \label{fig:Energiespg4}
    \end{figure*}

Let us now consider the results for $q_{\rm *} = 4 \cdot 10^3$, which is depicted in the middle panel of Fig.~\ref{p23-EosEnergy}. In this case, there is a significant transfer of energy to the daughter field during the initial linear regime via broad parametric resonance. The inflaton also gets excited via self-resonance, but the effect is always much weaker as discussed in Section \ref{Sec:LinearAn}\footnote{Note that although the gradient energy ratio of the inflaton grows during the linear regime, this is mainly due to backreaction effects from the daughter field, and only subdominantly sourced by the inflaton self-resonance.}. Therefore, backreaction effects induced by the daughter field modes fragment the inflaton condensate sooner than in the previous case: only $N_{\rm br} \sim 4$ e-folds of expansion after the end of inflation, in agreement again with our prediction in  Fig.~\ref{fig:Analytic-ParRes}. In any case, the ratio $( \bar\varepsilon_{\rm g}^{\varphi}  + \bar\varepsilon_{\rm g}^{\chi} ) / \bar\varepsilon_{\rm p}^{\varphi} $ keeps growing both during the linear regime and the early phase of the non-linear stage. Correspondingly, the equation of state deviates from $\bar{w}= \bar{w}_{\rm hom} \simeq 0.07$ towards some maximum value $\bar{w}_{\rm max} < 1/3$. However, the effective resonance parameter decreases with time as $\tilde q = q_{*} a^{\frac{6(p-4)}{p+2}} \sim a^{-2.37}$, so the resonance eventually becomes narrow ($\tilde q < 1$). At that time, something similar to the $p=2$ scenario happens: the daughter (kinetic and gradient) energy ratios stop their growth and start decreasing ($\bar\varepsilon_{\rm k}^{\varphi}, \bar\varepsilon_{\rm g}^{\varphi} \rightarrow 0$) at different dilution rates, and the inflaton gradient energy, which was mainly excited due to interactions with $X$, also stops its growth and gets smaller. In the equation of state, this is manifested as a transitory recovery process from the local maximum $\bar{w} = \bar{w}_{\rm max}$ towards $\bar{w} \rightarrow \bar{w}_{\rm hom}$.  However, unlike the $p=2$ case, now the inflaton field is also excited via self-resonance, and as in the $q_*=0$ case, this process is present \textit{even after the breaking of the initially homogeneous mode}, and it never ceases. Therefore, the inflaton fluctuations slowly pile up, and at a later time scale we obtain $\bar\varepsilon_{\rm g}^{\varphi} / \bar\varepsilon_{\rm p}^{\varphi} \gg 1$. Consequently, the equation of state then starts increasing again, and goes towards $\bar{w} \rightarrow 1/3$ at late times.

The energies also evolve in a similar way for other choices of $p \in [2,4)$ and $q_* > q_*^{\rm (min)}$, but the `oscillatory' pattern of the equation of state is not always clearly seen. For example, in the case of $p=2.3$ and the larger resonance parameter $q_* = 6 \cdot 10^4$ (depicted in the bottom panel of Fig.~\ref{p23-EosEnergy}), the transition from broad to narrow resonance takes place at later times, so the growth of the gradient energy during the linear regime is much larger than for $q_*=4 \cdot 10^3$. The equation of state also becomes very close to $\bar{w} = 1/3$ after the initial raise, so the transitory decrease before inflaton self-resonance becomes relevant is less remarkable. In any case, we have $\bar \varepsilon_{g}^{\varphi} / \bar \varepsilon_p^{\varphi} \gg 1$ at late times due to the self-resonance, so the equation of state goes to $\bar w \rightarrow 1/3$ as well.

Interestingly, if $q_*$ is large enough (and $2\leq p < 4$), we can observe a transitory \textit{equipartition} regime during which the energy is distributed equally between the inflaton and daughter fields. This can be observed for example in Fig.~\ref{fig:Energiesp3}, where the energy distribution and equation of state are depicted for $p=3$, and two different choices of $q_*$. For $q_* = 2 \cdot 10^3$, parametric resonance becomes narrow (i.e. $\tilde q < 1$) around backreaction time, and the daughter field energy density starts to decrease immediately after parametric resonance has terminated. However, for $q_* = 10^5$, the transition to narrow resonance happens at later times, and an equipartition regime emerges with $\bar{\varepsilon}_{\rm k}^{\varphi} \simeq \bar{\varepsilon}_{\rm g}^{\varphi} \simeq \bar{\varepsilon}_{\rm k}^{\chi} \simeq \bar{\varepsilon}_{\rm g}^{\chi} \approx 0.25$. In any case, the exchange of energy between the two fields gets strongly suppressed once $\tilde q < 1$, and the energy of the daughter field becomes negligible at late times, $\bar\varepsilon_{\rm k}^{\chi}$, $\bar\varepsilon_{\rm g}^{\chi} \rightarrow 0$. \vspace{0.3cm}

\textbf{iii)} $\boldsymbol{p\geq4}$. Finally, let us consider the cases in which the inflaton potential is quartic ($p=4$) or steeper than quartic ($p>4$). In both cases the inflaton is excited via self-resonance, which remains always active. As for the daughter field, the effective resonance parameter $\tilde q$ [Eq.~(\ref{eq:ResParam})] either remains constant for $p=4$ or grows for $p > 4$. Therefore, for $p=4$ the resonance always remains broad as long as $q_* > 1$, while for $p>4$ the resonance will be broad at late times even if $q_* < 1$ initially. In these cases, energy is continuously exchanged between the two fields, and eventually an equipartition state is achieved, with the energy equally distributed between the inflaton and the daughter field,  $\bar{\varepsilon}_{k}^{\varphi} \approx \bar{\varepsilon}_{g}^{\varphi} \approx \bar{\varepsilon}_{k}^{\chi} \approx \bar{\varepsilon}_{g}^{\chi} \approx 1/4$. This can be seen in Fig.~\ref{fig:Energiespg4}, where we show the energy distribution and equation of state for the particular examples $p=4$ and $p=6$. As before, the strongly oscillating equation of state breaks at $u_{\rm br}$, and a radiation dominated state is quickly achieved. While for $p=4$ the effective equation of state is always $\bar{w} = 1/3$, for $p=6$ it jumps quickly from $\bar{w} = \bar{w}_{\rm hom}=1/2$ to $\bar{w} = 1/3$. \vspace{0.3cm}

\begin{center}
\textbf{-- Overview of results --}
\end{center}
\vspace*{-1mm}

\begin{figure}
    \centering
    \includegraphics[width=0.45\textwidth]{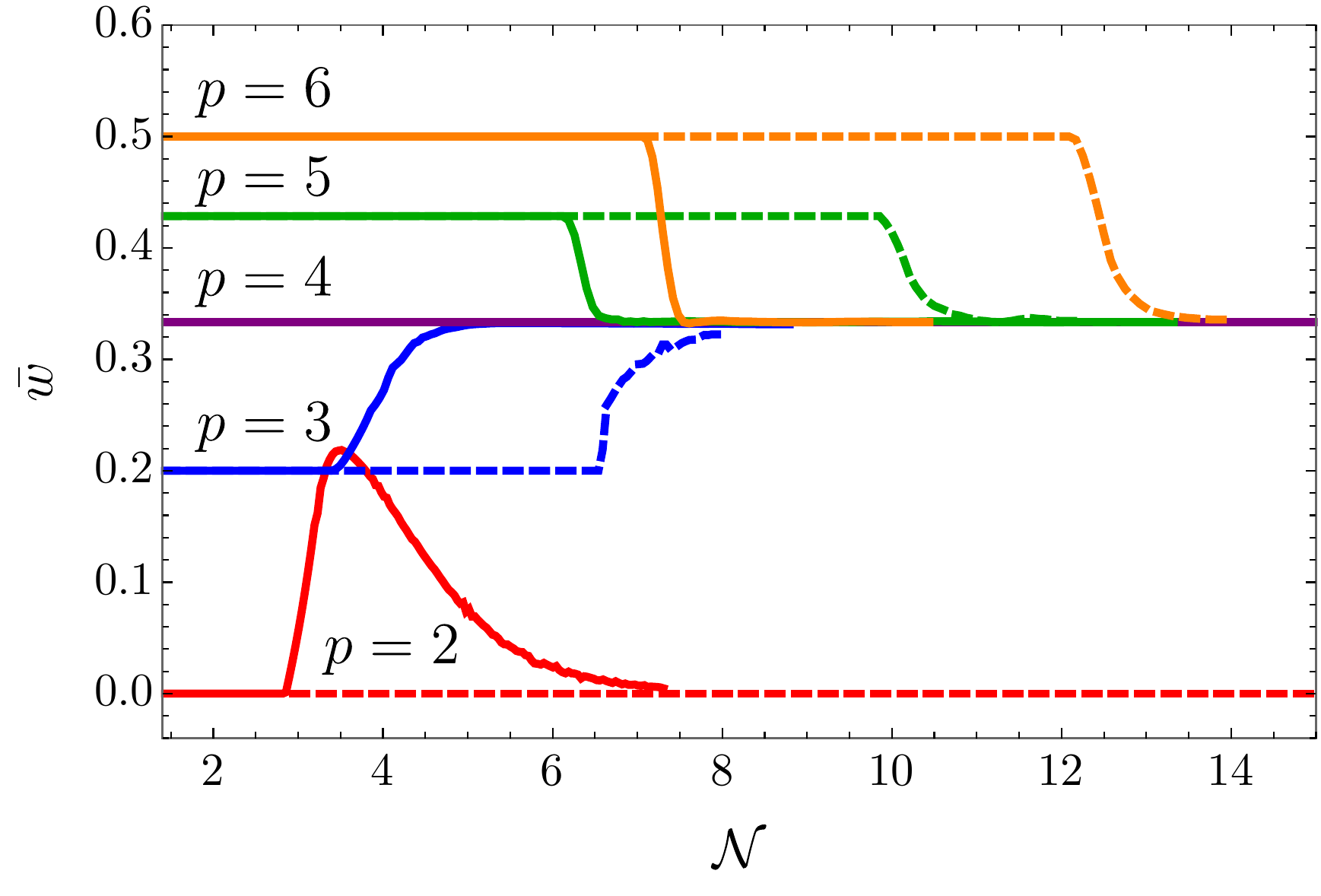}
    \caption{[$v=0$] Evolution of the effective equation of state for $p=2$, 3, 4, 5, 6, extracted form lattice simulations. The dashed lines correspond to cases in which there is no coupling between the inflaton and the daughter field ($q_*=0$), while for the continuous lines such coupling is included. The resonance parameters chosen for this second set of simulations are $q_* = 8 \cdot 10^4$, $2 \cdot 10^4$, $10^2$, $10^{-2}$, and $1$,  for $p=2$, $3$, $4$, $5$ and $6$ respectively.}  \label{Fig:EoSvsP}
\end{figure}

The dependence of the equation of state for different choices of $p$ and $q_*$ is summarized in Fig.~\ref{Fig:EoSvsP}. There we show its evolution for $p=2$, 3, 4, 5, and 6, and for each coefficient we consider two scenarios: one in which there is no coupling between the inflaton and the daughter field, and another in which a sizeable quadratic-quadratic coupling is present. In this second case, we have chosen resonance parameters $q_* > q_*^{\rm (min)}$, so that the initial decay of the inflaton homogeneous mode is triggered by broad parametric resonance of the daughter field. Remarkably, the value that the equation of state attains at very late times is independent on the absence or presence of a coupling, and in the latter case independent as well of its strength: it always goes to $\bar w \rightarrow 0$ for $p=2$, and to $\bar w \rightarrow 1/3$ for $p>2$. However, there are important differences in its evolution before reaching the final state. On the one hand, for $p=2$ and a non-zero coupling, $\bar{w}$ shows a transitory deviation from the homogeneous solution and attains a local maximum $\bar{w}=\bar{w}_{\rm max} < 1/3$. On the other hand, for $p>2$, the transition from $\bar{w} = \bar{w}_{\rm hom} \equiv (p-2)/(p+2)$ to $\bar{w} = 1/3$ takes place several e-folds earlier when a coupling exists. Note also that for the range of values $2 < p \lesssim 3$, we sometimes observe an oscillatory pattern in the averaged equation of state, see e.g.~the middle and bottom panels of Fig.~\ref{p23-EosEnergy}.

At late times, the energy density is distributed between its components in different ways for different choices of $p$ and $q_*$. We show the final values attained by the energy ratios in Table \ref{Tab:ResultsCentred}. For $p=2$, the energy at late times is dominated by the inflaton homogeneous mode, so $\bar{\varepsilon}_{k}^{\varphi} \simeq \bar{\varepsilon}_p^{\varphi} \simeq 1/2$, and no significant amount of energy remains stored in the daughter field or in the field gradients. For $p \in [2,4)$, there is also no significant transfer of energy to the daughter field at late times, but we get $\bar{\varepsilon}_{k}^{\varphi} \approx \bar{\varepsilon}_{g}^{\varphi} \approx 1/2 \gg \bar{\varepsilon}_p^{\varphi}$ due to inflaton self-resonance. Only for $p \geq 4$, the system transfers a sizeable fraction of energy into the daughter field: approximately $\sim$ 50\% of the total ($\bar{\varepsilon}_{k}^{\varphi} \approx \bar{\varepsilon}_{g}^{\varphi} \approx \bar{\varepsilon}_{k}^{\chi} \approx \bar{\varepsilon}_{g}^{\chi} \approx 0.25$).

\begin{table}
    \centering \textbf{Final energy ratios for $V(\phi) \propto |\phi|^p$, $v = 0$}
    \def\arraystretch{1.7}
    \begin{center}
        \begin{tabular}{ |p{2.1cm}|c|c|c|c|c|c|c| } 
            \hline
            \centering  $p$, $q_*$ & {\rm EoS} & $\bar{\varepsilon}_k^{\varphi}$ & $\bar{\varepsilon}_g^{\varphi}$ & $\bar{\varepsilon}_k^{\chi}$ & $\bar{\varepsilon}_k^{\chi}$ & $\bar{\varepsilon}_p^{\varphi}$ & $\bar{\varepsilon}_i$\\   \hline
            \centering $p=2, \forall\, q_*$ & {\rm MD} & 1/2 &  0 & 0 & 0 & 1/2 & 0 \\   \hline
            \centering $2 < p < 4, \forall\, q_*$ & {\rm RD}  & 1/2 & 1/2 & 0 & 0 & 0 & 0 \\   \hline
            \centering $p \geq 4$, $q_* = 0$ & {\rm RD} & 1/2 & 1/2 & 0 & 0 & 0 & 0 \\   \hline
            \centering $p \geq 4$, $q_*>0$ & {\rm RD} & 1/4 & 1/4 & 1/4 & 1/4 & 0 & 0 \\   \hline
        \end{tabular} 
    \end{center}
    \caption{[$v=0$] Final equation of state and energy ratios at asymptotic late times, for different combinations of $p$ and $q_*$, as observed in the simulations. \label{Tab:ResultsCentred}} 
\end{table}

\begin{figure}
    \centering
    \includegraphics[width=0.45\textwidth]{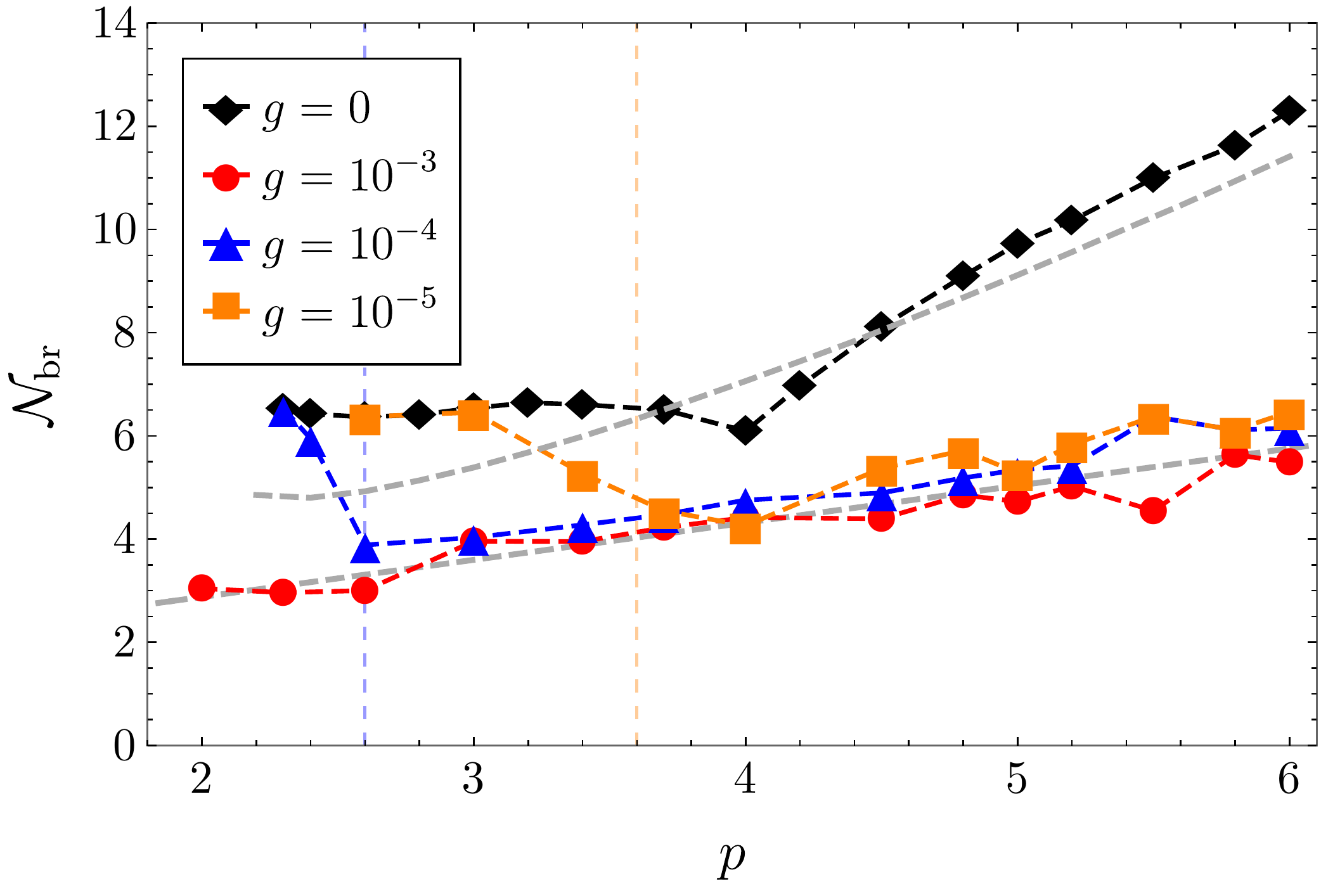} 
    \includegraphics[width=0.45\textwidth]{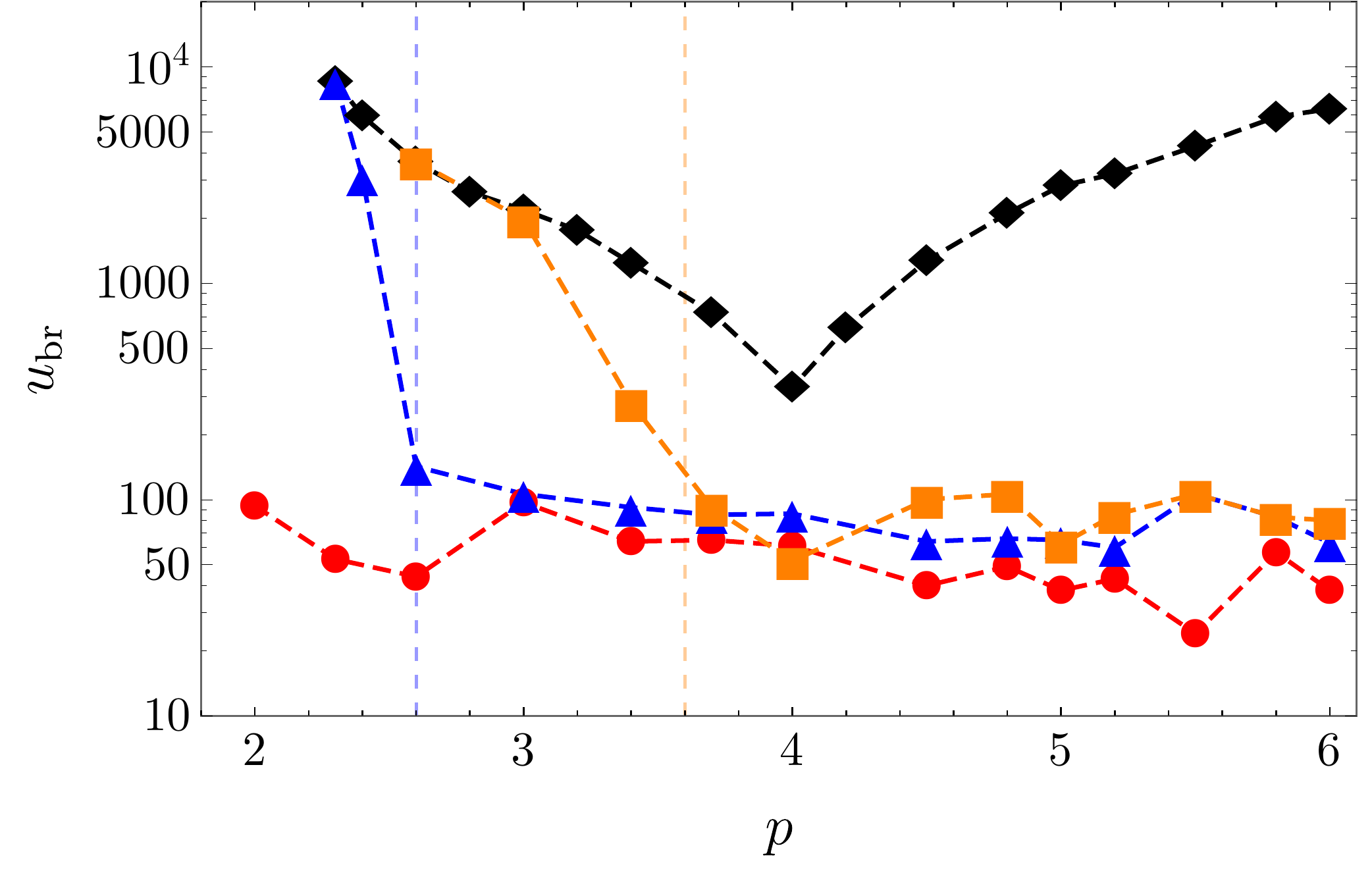}
    \includegraphics[width=0.45\textwidth]{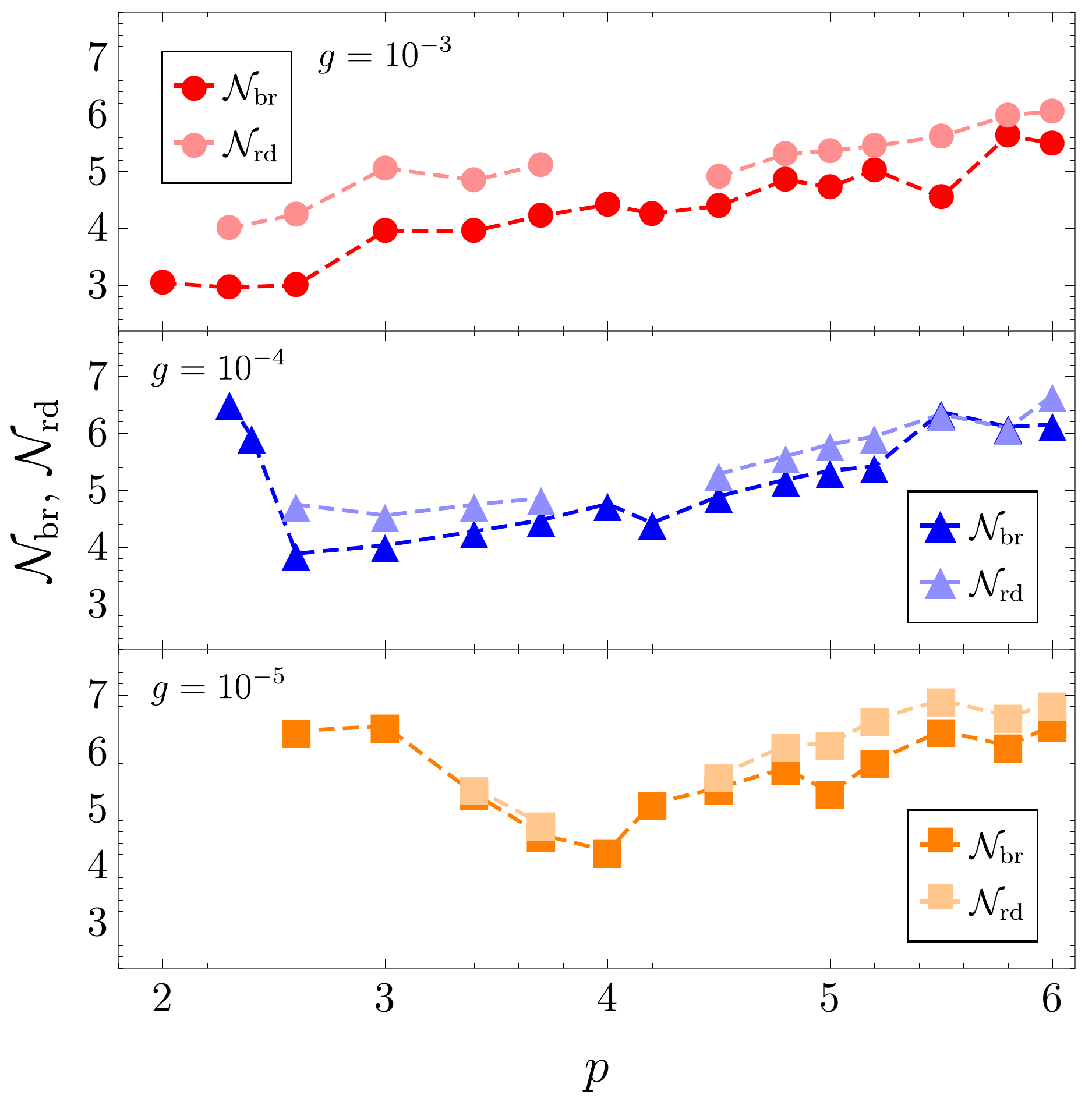}
    \caption{[$v=0$] Backreaction time from lattice simulations in terms of number of e-folds after inflation $\mathcal{N}_{\rm br}$ (top panel), and natural time $u_{\rm br}$ (middle panel). We consider coupling strengths $g^2 = 0$, $10^{-5}$, $10^{-4}$, and $10^{-3}$, for each value of $p$. In the top panel, upper and lower gray dashed lines represent, respectively, the analytical estimation for inflaton self-resonance (see Fig.~\ref{fig:PvszbrNbr}), and the approximation $u_{\rm br} = \frac{p+2}{6} \log (75)$ for parametric resonance [based on Eq.~(\ref{eq:zbr-p4})]. The bottom panel shows the number of e-folds $\mathcal{N}_{\rm rd}$ till the onset of RD for each choice of $g^2 \neq 0$ as a function of $p$. For comparison, it also shows $\mathcal{N}_{\rm br}$.}
    \label{fig:PvsN}
\end{figure}

Finally, we show in Fig.~\ref{fig:PvsN} the backreaction time scale in terms of natural time $u_{\rm br}$ and post-inflationary number of e-folds $N_{\rm br}$. This time describes approximately when the averaged equation of state starts to deviate from the homogeneous solution $\bar{w} = \bar{w}_{\rm hom}$. We have considered four different coupling strengths: $g = 0$ (i.e.~no coupling between the inflaton and the daughter field), $10^{-5}$, $10^{-4}$, and $10^{-3}$. In each case, we have carried out several simulations with different values of $p$, and extracted $N_{\rm br}$ from them. For $g = 0$ we get that $N_{\rm br} \approx 6$ is approximately constant for $2 < p < 4$, but grows with $p$ for $p \geq 4$, up to $N_{\rm br} \approx 12$ for $p=6$. This is in qualitative agreement with the analytical estimation of $N_{\rm br}$ coming from inflaton self-resonance\footnote{Note that the numbers for $N_{\rm br}$ as a function of $p$ are very similar to the ones shown in letter \cite{Antusch:2020iyq}, but are not identical because they come from different sets of lattice simulations with slightly different model parameters. In particular, the simulations of this work have been carried out for $N_k = 60$, while we fixed $N_k = 50$ in the letter.}, see Fig.~\ref{fig:PvszbrNbr}. However, the estimation of $N_{\rm br}$ if a coupling is added is different. For example, in the case $g = 10^{-5}$ we observe that, for $p \gtrsim 3.4$, $N_{\rm br}$ is always several e-folds less than in the $g = 0$ case for the same value of $p$. However, for $p \lesssim 3.4$, $N_{\rm br}$ becomes larger and approximately equal to the $g = 0$ case.  We can understand this by noting that the resonance parameter $q_* = (g^2 / \lambda) (\phi_* / \mu )^{4-p} $ is a decreasing function of $p$ for a fixed value of $g$. For sufficiently low values of $p$ we have $q_* \lesssim q_*^{\rm (min)}$, so the stage of broad parametric resonance is not long enough to trigger the decay of the inflaton homogeneous condensate, and instead the condensate fragments eventually due to inflaton self-resonance (as in the $q_* = 0$ case). The same behaviour can be observed for $g=10^{-4}$, but in this case the transition happens at $p \approx 2.6$. For $g = 10^{-3}$ no transition is observed for the range of considered values of $p$.

The bottom panel of Fig.~\ref{fig:PvsN} shows, for $g = 10^{-5}$, $10^{-4}$, and $10^{-3}$, the post-inflationary number of e-folds since the end of inflation at which the final radiation-dominated stage is approximately achieved (in those cases where such thing happens, i.e.~for $p>2$). More specifically, $N_{\rm rd}$ is defined as the time when the relative difference between $\bar{w}$ and $\bar{w}=1/3$ is 5\%. We can see that, when the inflaton homogeneous mode decays via broad parametric resonance, the difference between $N_{\rm br}$ and $N_{\rm rd}$ is of at most $\sim1$ 1 e-fold, showing that the RD stage is attained quite fast after backreaction time.

\subsection{Lattice analysis: Displaced potential} \label{Sec:LatResults2}

We now present results from lattice simulations of the preheating process in the case of the \textit{displaced} potential, with $v > 0$. The results of the centred potential case are recovered in the limit $v \ll m_{\rm pl}$, while the resonance gets extremely weak for $v\gg m_{\rm pl}$ and the post-inflationary dynamics becomes trivial. Therefore, in our simulations we need to choose some intermediate value. We choose $v=10^{-2}m_\mathrm{pl}$ for the simulations presented in this section, and we comment briefly afterwards about the results expected for other choices of $v$. We set again $M=10m_{\mathrm{pl}}$ as in the centred potential scenario, and distinguish two relevant cases: $p=2$ and $p>2$. In each case we will consider different initial mass ratios $R_*$ [c.f.~Eq.~\ref{eq:massratio}], in correspondence with different values of the initial resonance parameter $q_*$. \vspace{0.3cm}

\begin{figure}
    \centering
    \includegraphics[width=0.48\textwidth]{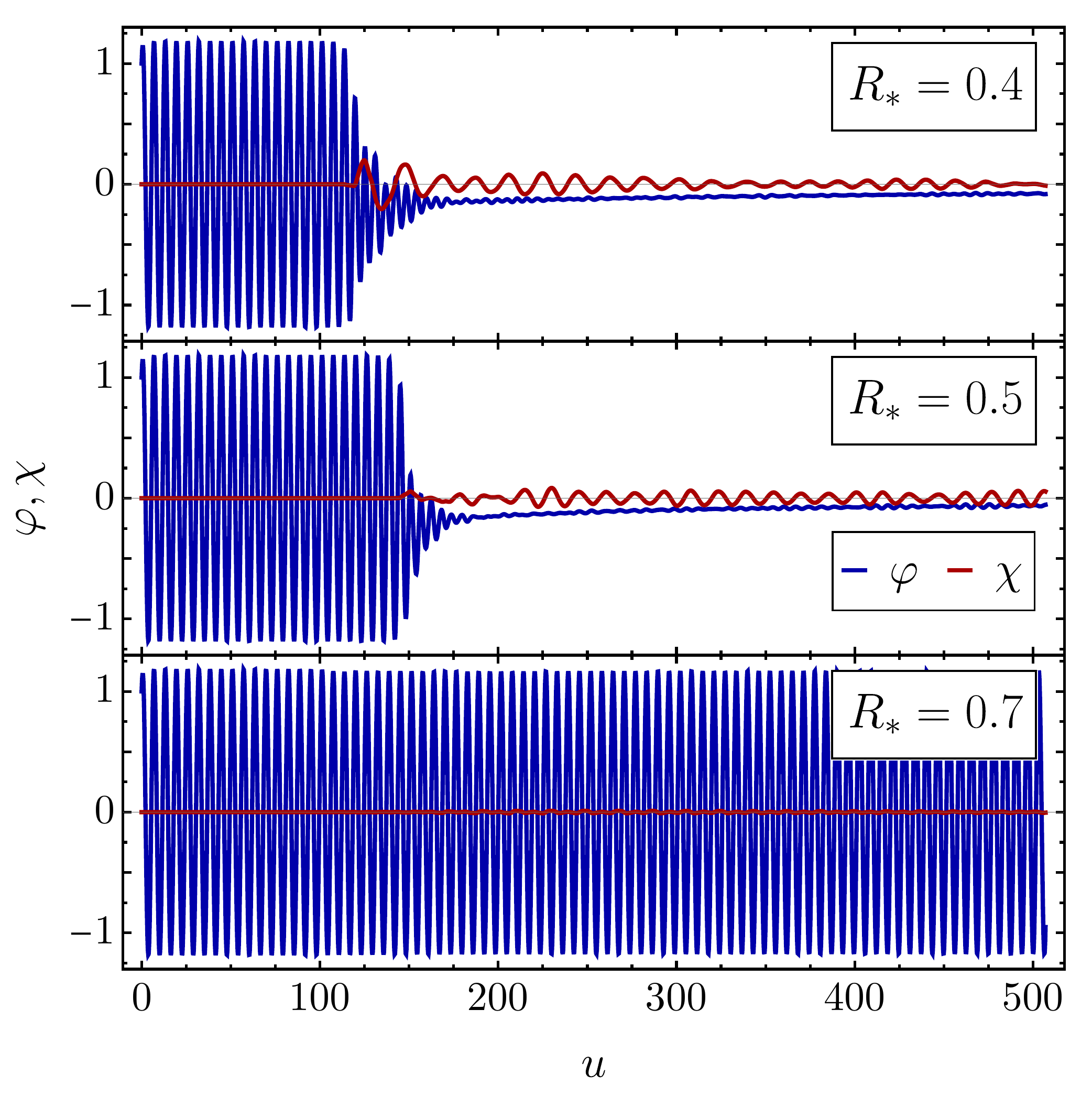}    
    \caption{[$v>0$] Evolution of the volume-averaged amplitudes of the inflaton $\varphi$ (blue) and the daughter field $\chi$ (red) for $p=2$, $M=10 m_{\rm pl}$,  $v=10^{-2}m_\mathrm{pl}$, and the mass ratios $R_*=0.4$ (top), 0.5 (middle) and 0.7 (bottom).}  \label{fig:meansmm040507}
\end{figure}

\textbf{i) \boldsymbol{$p=2$}:}  In Fig.~\ref{fig:meansmm040507} we show the evolution of the volume-averaged amplitudes of the inflaton and daughter fields for the mass ratios $R_*=0.4$, 0.5, and 0.7. As expected, the energy budget of the system is initially dominated by the oscillations of the homogeneous inflaton in all cases. However, the later evolution of the amplitude of both fields strongly depends on the choice of $R_*$. For $R_*=$ 0.4 and 0.5, the homogeneous inflaton mode decays quickly at the onset of backreaction at $u_{\rm br}\simeq 100 - 140$, but for $R_*=0.7$ the homogeneous oscillatory regime survives and remains till the end of the simulation. This indicates that the excitation of the daughter field is significantly stronger in the first two cases than in the third.

\begin{figure*} \centering
        \includegraphics[width=0.45\textwidth]{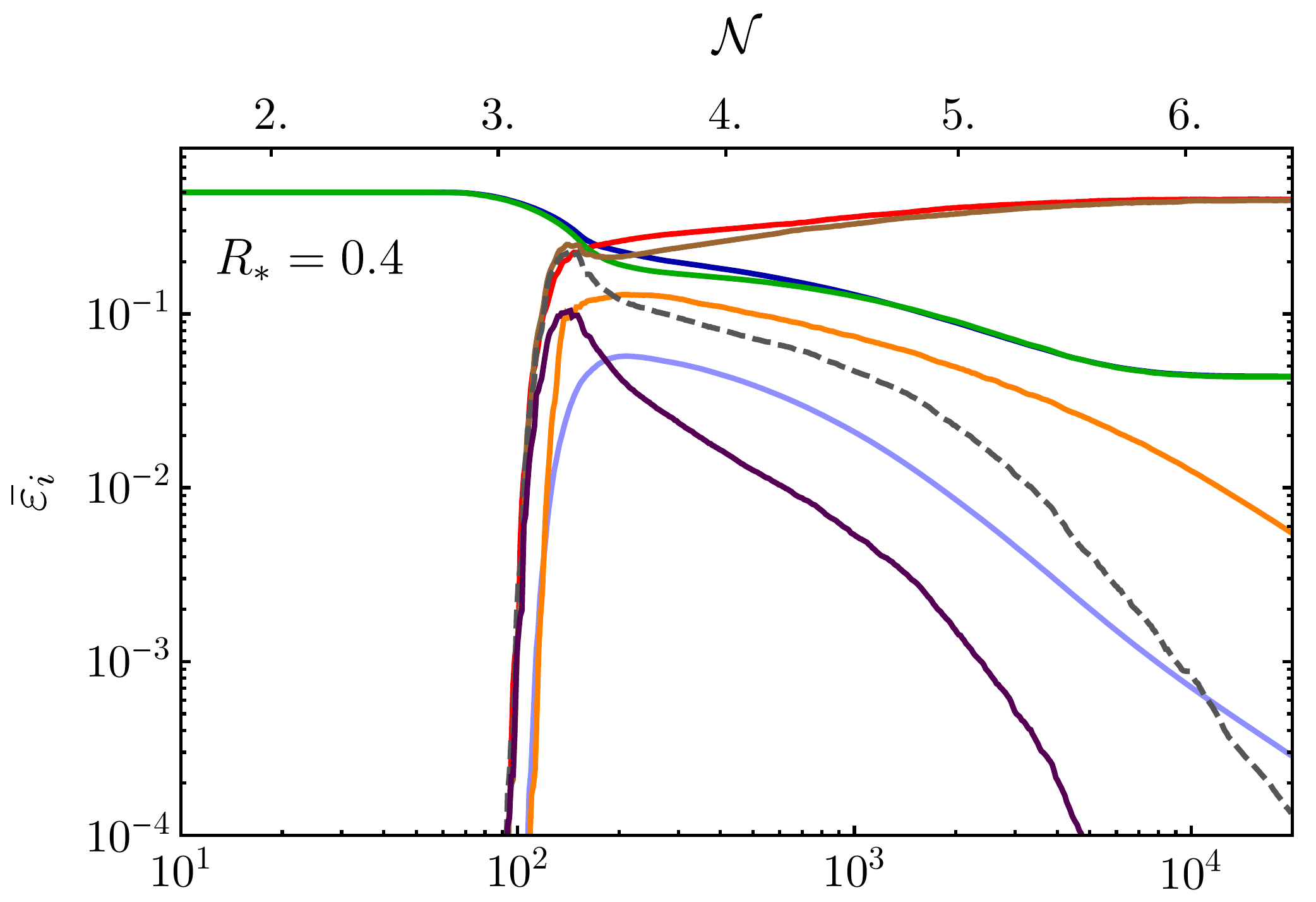}  \hspace{0.2cm}
        \includegraphics[width=0.45\textwidth]{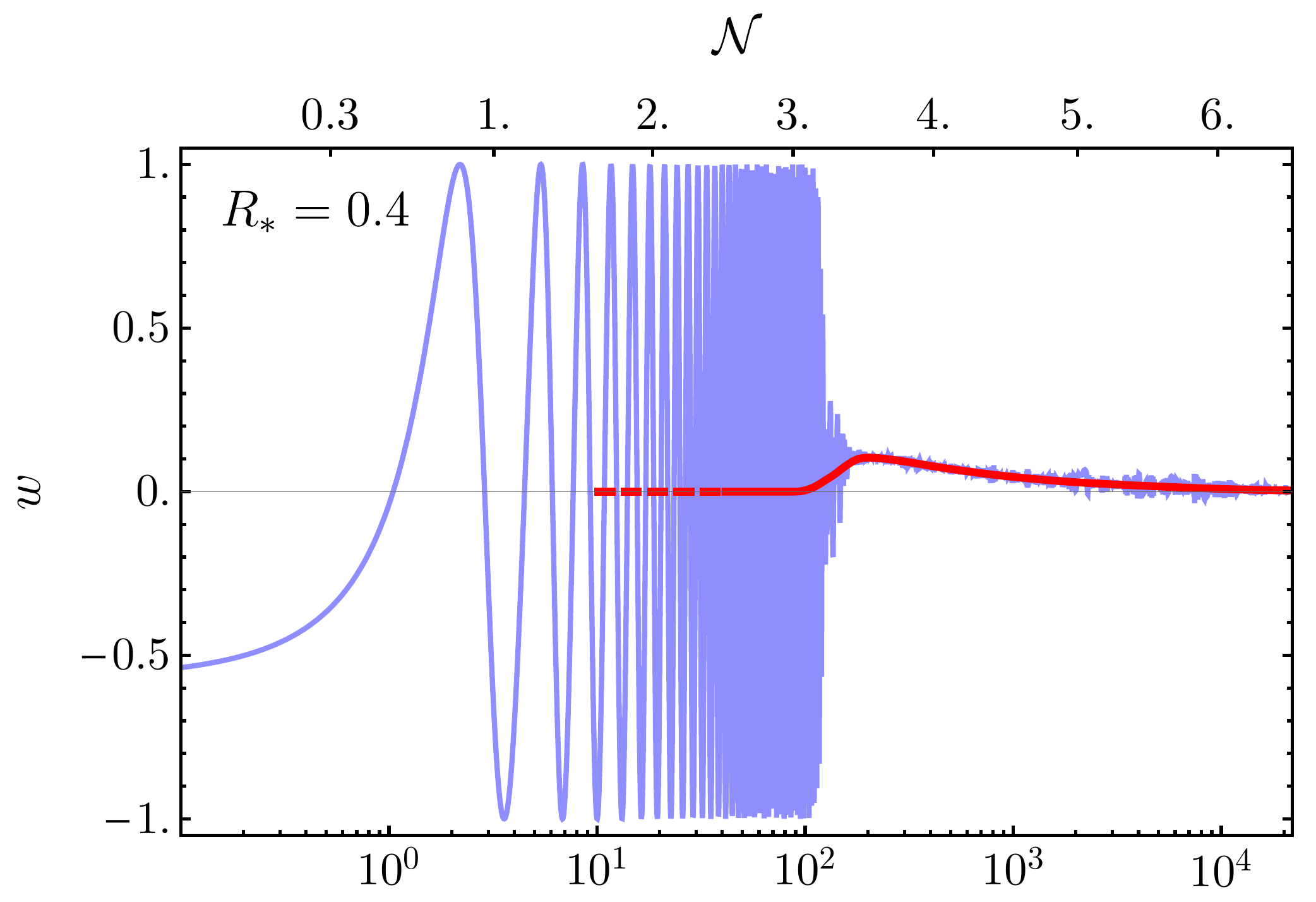}  \\
        \includegraphics[width=0.45\textwidth]{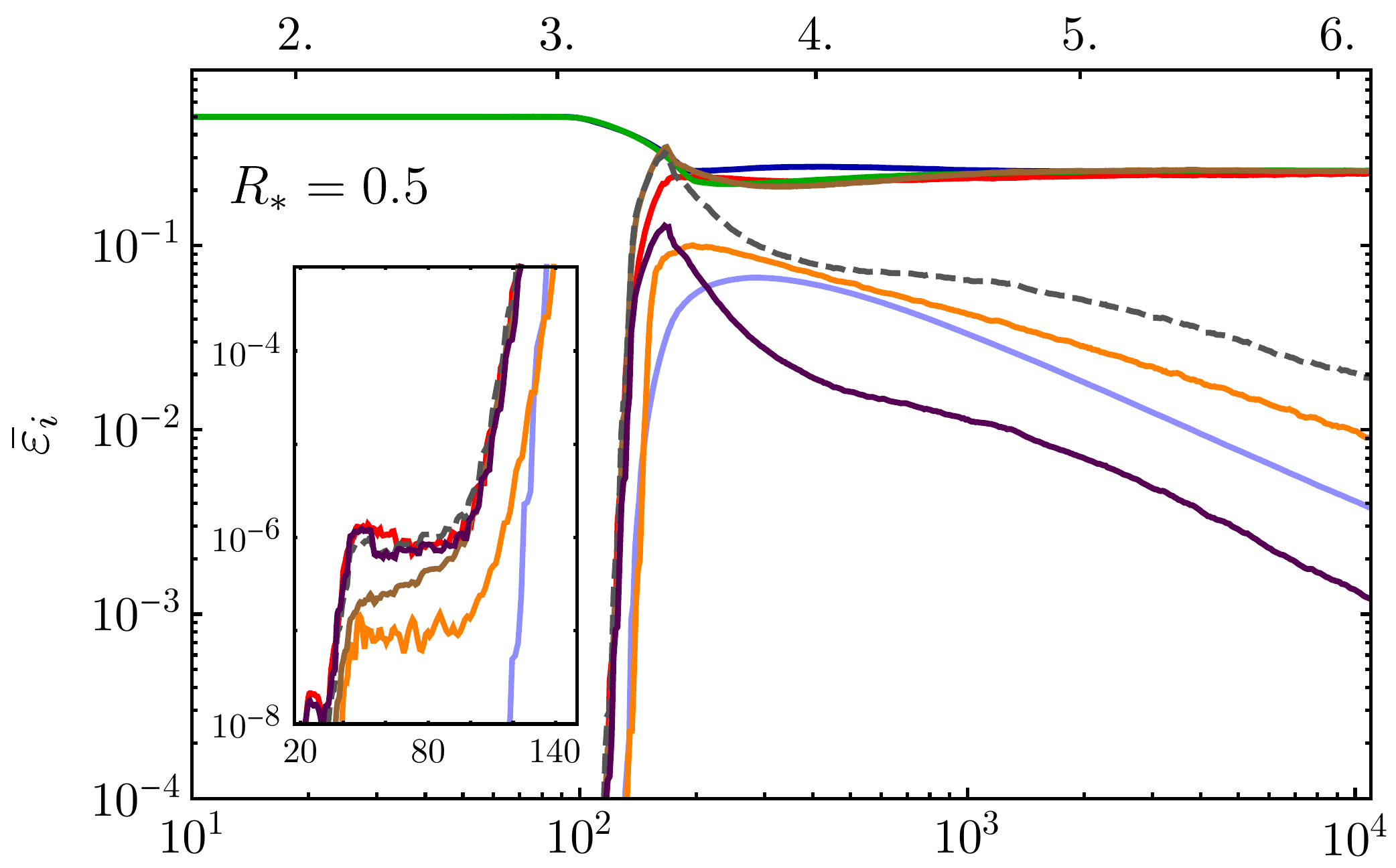}   \hspace{0.2cm}
        \includegraphics[width=0.45\textwidth]{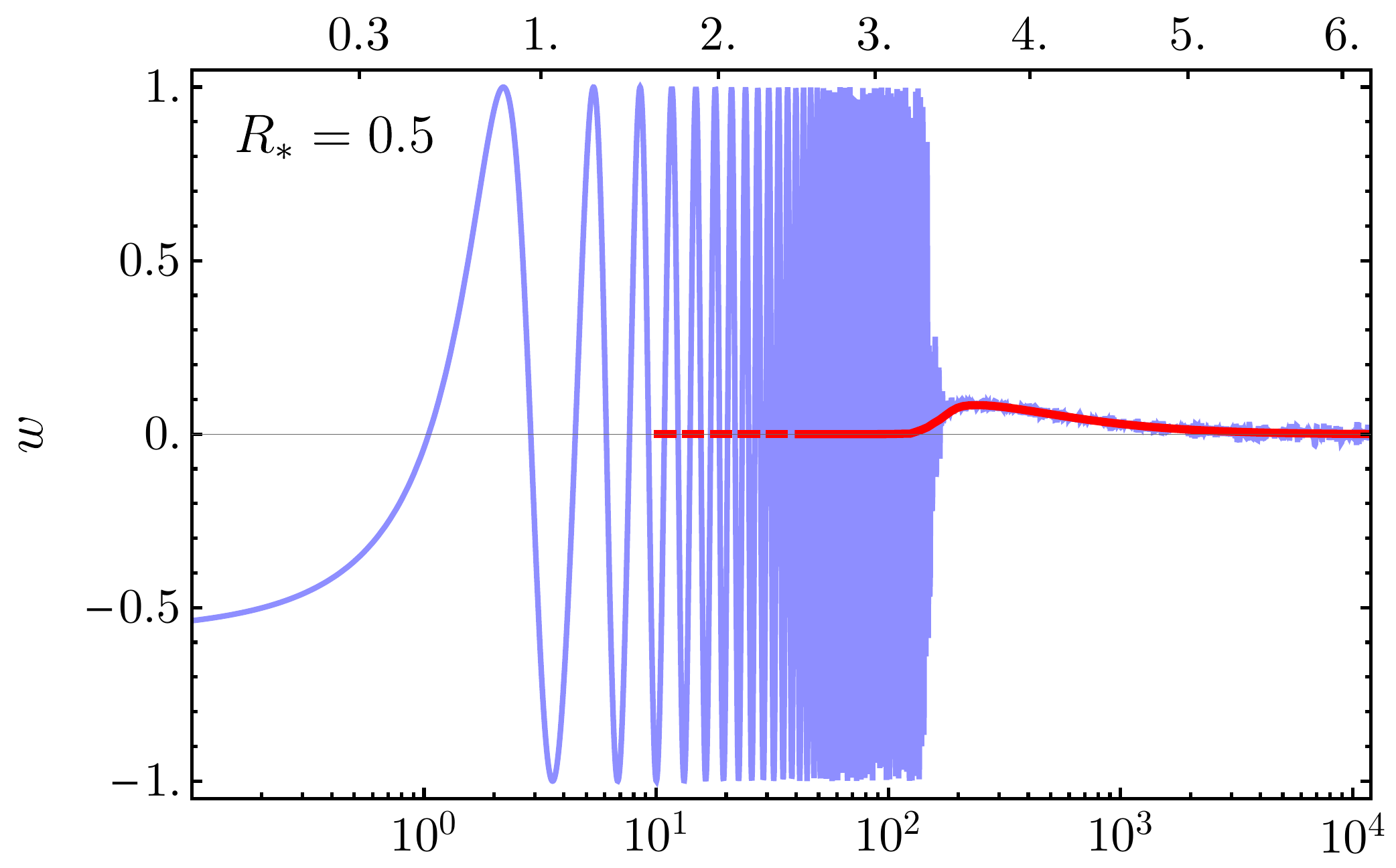}  \\ 
        \includegraphics[width=0.45\textwidth]{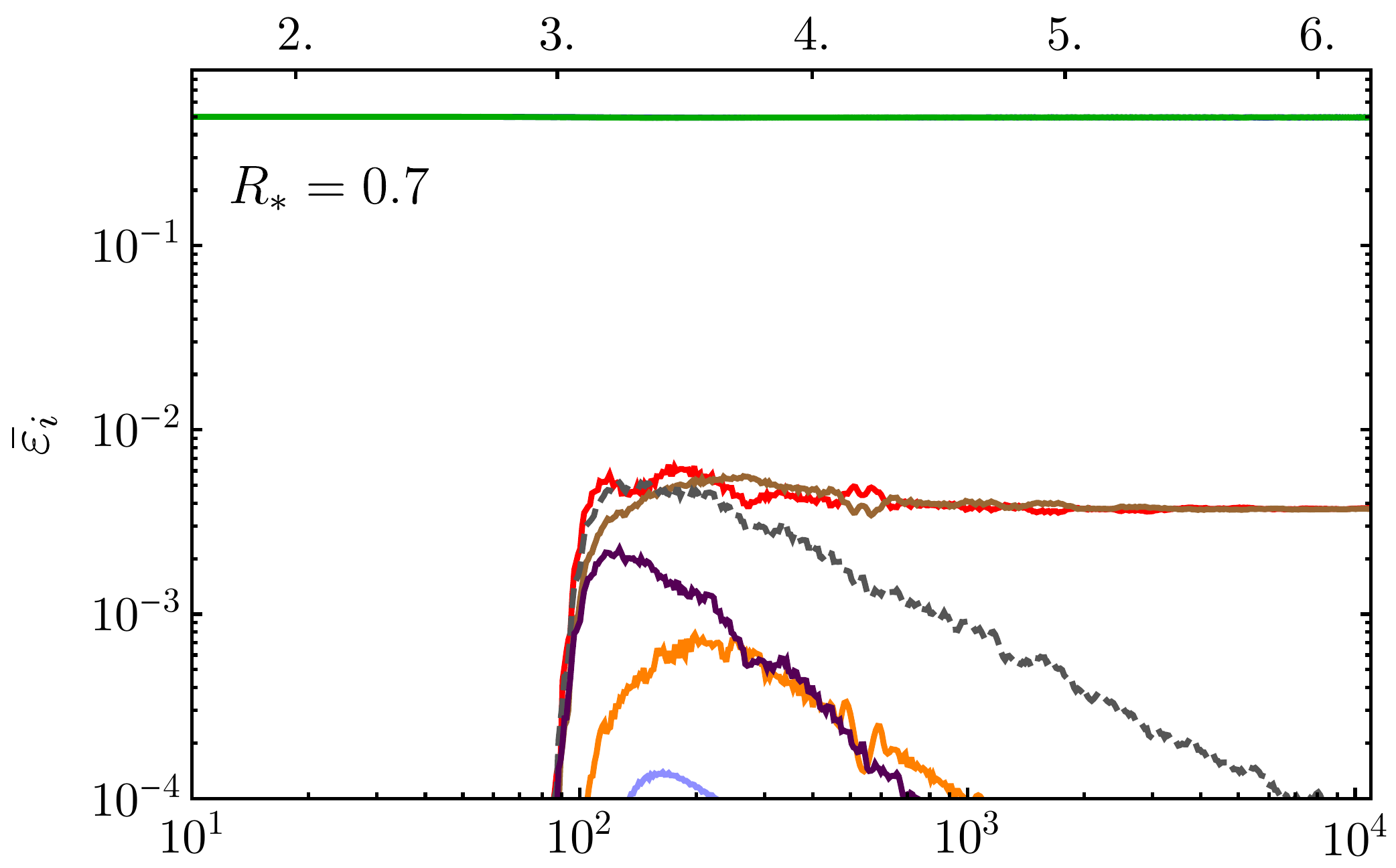}   \hspace{0.2cm}
        \includegraphics[width=0.45\textwidth]{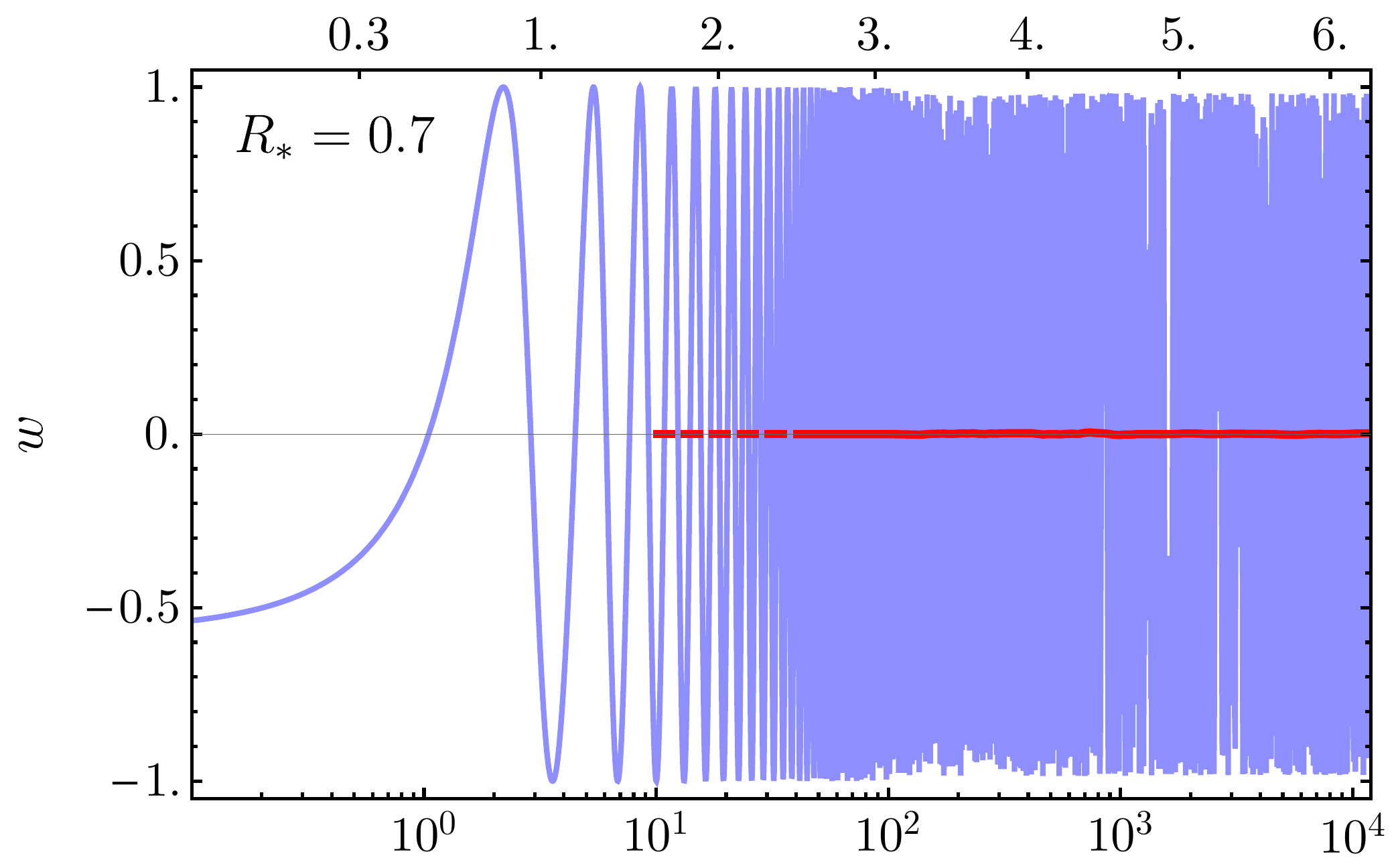} \\
        \includegraphics[width=0.45\textwidth]{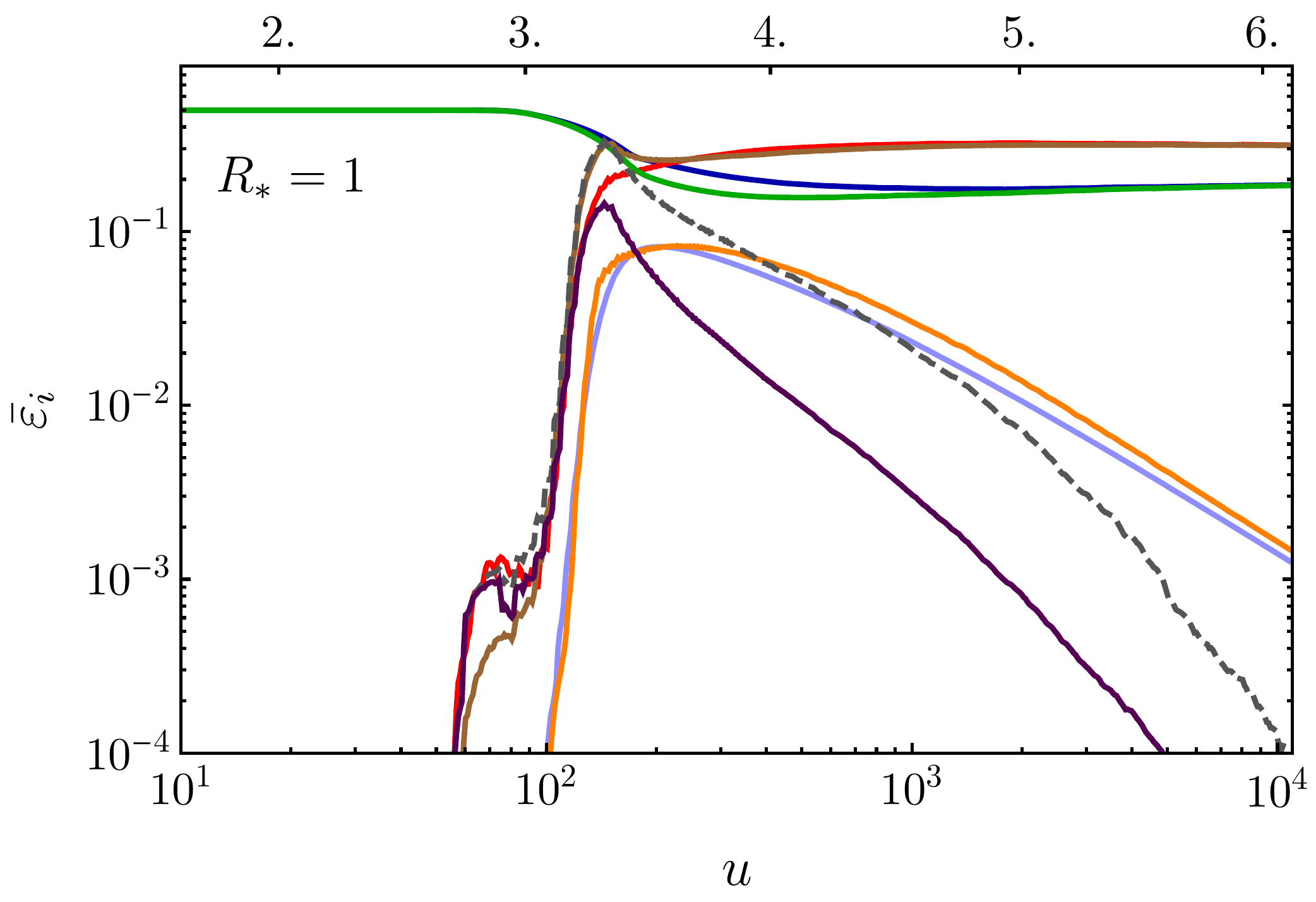} \hspace{0.2cm}
        \includegraphics[width=0.45\textwidth]{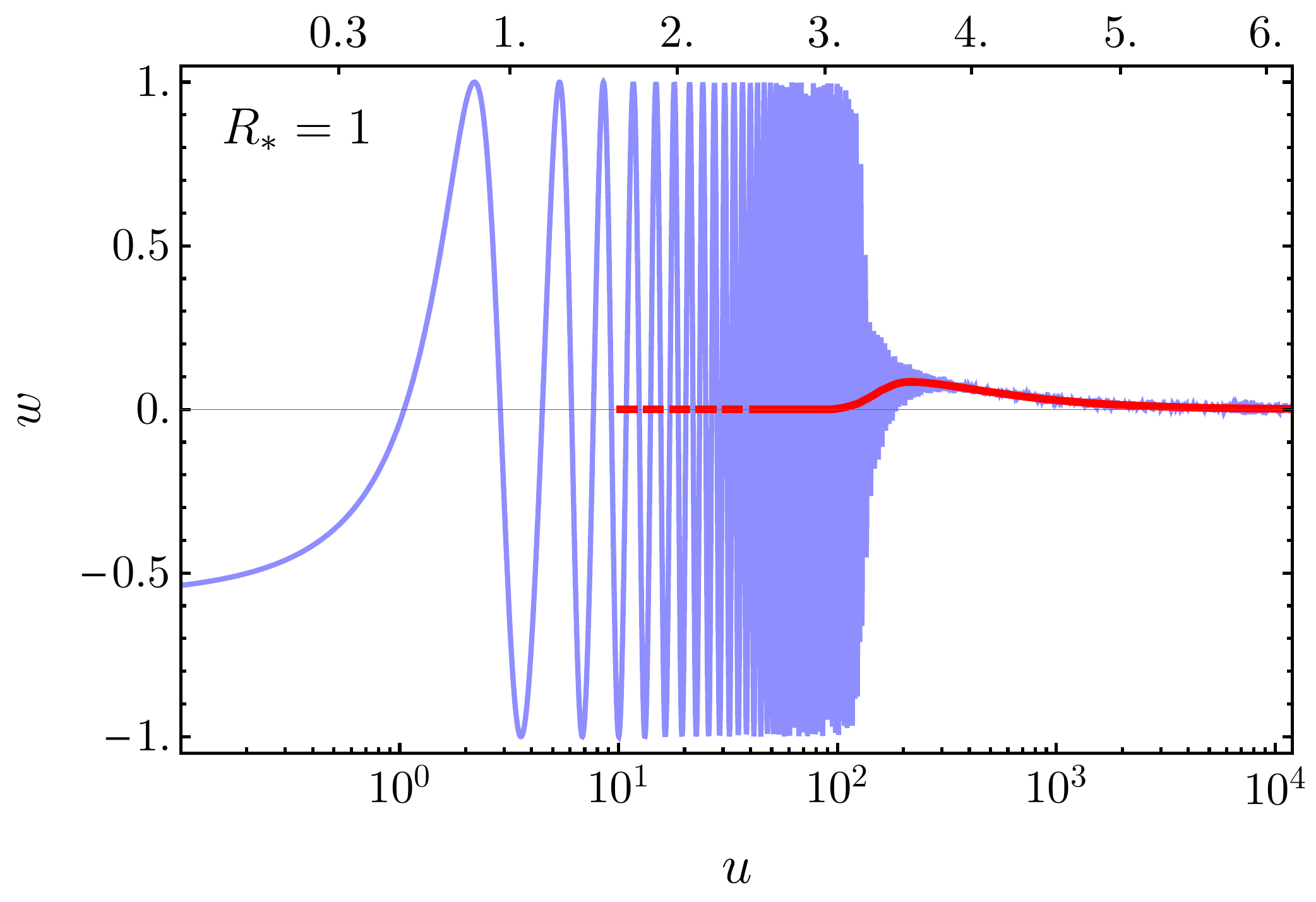} \\\vspace{0.4cm}
    \includegraphics[width=0.5\textwidth]{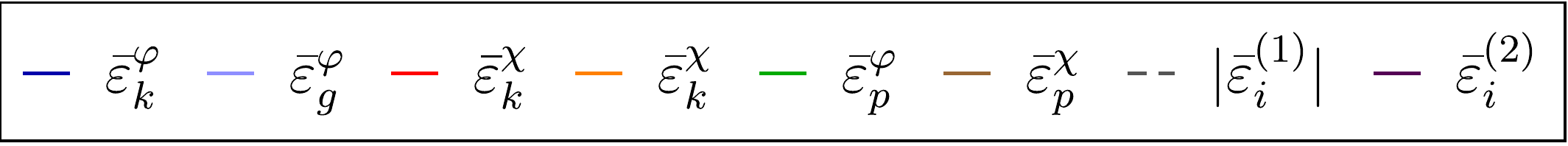} 
    \hspace{0.2cm}
    \includegraphics[width=0.12\textwidth]{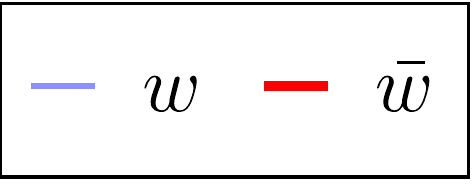}
    \vspace{0.4cm}             

    \caption{[$v>0$] Left panels: Evolution of the averaged energy ratios $\bar{\varepsilon}_i$ [see Eqs.~(\ref{eq:energyshift1})-(\ref{eq:energyshift4})] for $p=2$, $M=10m_{\rm pl}$, $v=10^{-2}m_\mathrm{pl}$, and four different mass ratios: $R_* = 0.4$, 0.5, 0.7, and 1. Note: the interaction energy ratio is negative, so we depict $|\bar{\varepsilon}_i^{(1)}|$. Right panels: Evolution of the instantaneous and oscillation-averaged equation of state (blue and red respectively) for the same model parameters.} \label{fig:p2-En_vev}
\end{figure*}

We plot now the corresponding evolution of the energy distribution in order to better understand the above behavior. In Fig.~\ref{fig:p2-En_vev} we show, for the same choices of mass ratios together with $R_* = 1$, the post-inflationary evolution of the averaged energy ratios $\bar{\varepsilon}_i$ and equation of state. The energy budget is initially dominated by the kinetic and potential energies of the inflaton, with $\bar{\varepsilon}_k^{\varphi} = \bar{\varepsilon}_p^{\varphi} \simeq 0.5$ in agreement with the equipartition identity (\ref{eq:Virial1}). The remaining energy ratios grow exponentially due to the parametric resonance effect analyzed in Sec.~\ref{sec:Lin:DisplacedPotential}. In particular, the stochastic behaviour of the system, where intervals of strong excitation and no excitation alternate can be observed in the inset of the $R_* = 0.5$ case. For the mass ratios $R_* = 0.4$, $0.5$, and $1$, the resonant excitation of the daughter field modes takes long enough for backreaction effects to become important, and for the inflaton homogeneous regime to break down at the backreaction time $u_{\rm br}$. On the contrary, for $R_*=0.7$ these ratios stop growing and saturate at $\bar{\varepsilon}_i \sim 10^{-2}$, before becoming of the same order of magnitude than $\bar{\varepsilon}_k^{\varphi}$ and $\bar{\varepsilon}_p^{\varphi}$. Therefore, it is clear that the strength and duration of the daughter field resonance depends very sensitively on $R_*$. This can be understood in light of the stability charts depicted in the top panels of Fig.~\ref{fig:shifted-floquet}. First, parametric resonance of the daughter field is significantly stronger for the critical mass ratios $R_* \simeq 0.5, 1, 1.5 \dots$, etc, see Eq.~(\ref{eq:MassRatioSpikes}). Second, the \textit{effective} vacuum amplitude $\tilde v \equiv v_* a^{3/2}$ grows in time, so the resonance eventually terminates for all values of $R_*$, which can be seen by following the different arrows in the Floquet chart. This happens at later times when $R_*$ is close to the critical values, which explains why backreaction effects break apart the inflaton homogeneous mode for $R_* = 0.4$, $0.5$ and $1$, but not for $R_* = 0.7$.

The different evolutions of the energy distribution are also reflected in the corresponding behaviour of the equation of state. For $R_* = 0.4$, 0.5, and 1, the averaged equation of state enters a transitory phase during which it deviates from the value $\bar{w}_{\rm hom} = 0$ and attains a maximum $\bar{w}=\bar{w}_{\rm max} < 1/3$ around $u = u_{\rm max} \gtrsim u_{\rm br}$. On the contrary, for $R_* = 0.7$ it remains approximately constant at $\bar{w} = \bar{w}_{\rm hom} = 0$.

Let us now analyze the evolution of the system at late times. After parametric resonance ends, no more field fluctuations are produced, so the gradient energy ratios stop growing and start diluting as radiation, such that $\bar \varepsilon_{g}^{\varphi}$, $\bar \varepsilon_{g}^{\chi} \ll 1$ at late times. A similar decrease is observed in the interaction energy ratios. Due to this, the equation of state for $R_*=0.4, 0.5, 1$ relaxes from the local maximum attained at $\bar{w}= \bar{w}_{\rm max}$, back to $\bar{w} \rightarrow \bar{w}_{\rm hom} = 0$ at late times. Only the kinetic and potential energy ratios of \textit{both} the inflaton and the daughter field remain and become constants at late times, obeying $\bar{\varepsilon}_{k}^{\varphi} \simeq \bar{\varepsilon}_{p}^{\varphi}$ and $\bar{\varepsilon}_{k}^{\chi} \simeq \bar{\varepsilon}_{p}^{\chi}$, in accordance with the equipartition identities (\ref{eq:Virial1}) and (\ref{eq:Virial2}). In particular, we end in a situation in which \textit{both} the inflaton and the daughter field become oscillating homogeneous condensates, with their oscillation periods being $T_\varphi\approx2\pi$ and $T_\chi\approx T_\varphi/R_*$ respectively. The final amount of energy transferred to the daughter field depends roughly on the strength of the resonance, i.e.~on $R_*$. In particular, for $R_* = 0.4$ and $1$ we end up with more energy in the daughter field than in the inflaton, for $R_* = 0.5$ we end with the same amount of energy in both fields, whereas almost all the energy remains in the inflaton for $R_* = 0.7$. This is in sharp contrast with the results for the centred potential, for which there is at the end no significant transfer of energy to the daughter field, no matter the value of the coupling. Remarkably, let us note that the significant transfer of energy to the daughter field for mass ratios $R_*>0.5$, as we see here, is forbidden for perturbative decays. 

Finally, we briefly discuss how the system evolves for vacuum expectation values smaller and larger than our canonical choice $v=10^{-2} m_{\rm pl}$. We have carried out lattice simulations for $v=10^{-3} m_{\rm pl}$ and $v=10^{-1} m_{\rm pl}$, considering in each case several mass ratios. The smaller the value of $v$, the closer the post-inflationary dynamics of the system is to the centred potential case ($v=0$). For $v=10^{-3}m_{\mathrm{pl}}$ we have observed that, for all considered mass ratios including $R_* = 0.7$, the initial linear stage of field excitations takes long enough for backreaction effects to fragment the initial homogeneous configuration of the inflaton, as expected in centered potentials. On the other hand, for our larger choice $v=10^{-1}m_{\mathrm{pl}}$, we find that a non-negligible excitation of the daughter field can only be seen for mass ratios very close to $R_*\simeq 1/2$. For these mass ratios, the daughter field may still carry a relevant part of the total energy density of the universe at late times. However, we note that even in those cases, only a very small amount of gradient energy is produced during parametric resonance, so the averaged equation of state stays roughly at $\bar{w}\simeq 0$ during the entire preheating process.  \vspace{0.3cm}

\begin{figure*} \centering
    \includegraphics[width=0.43\textwidth]{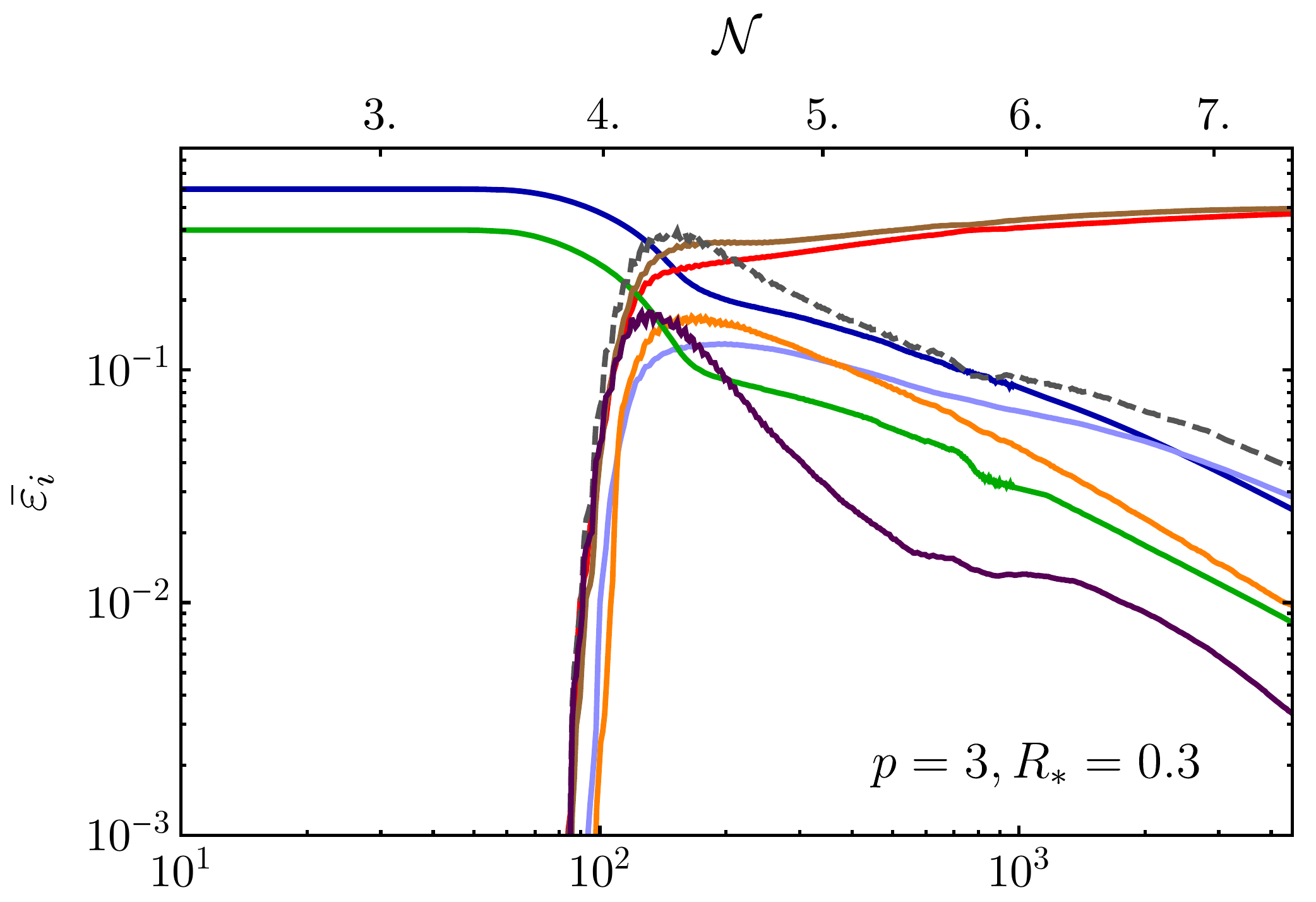}  \hspace{0.2cm}
    \includegraphics[width=0.43\textwidth]{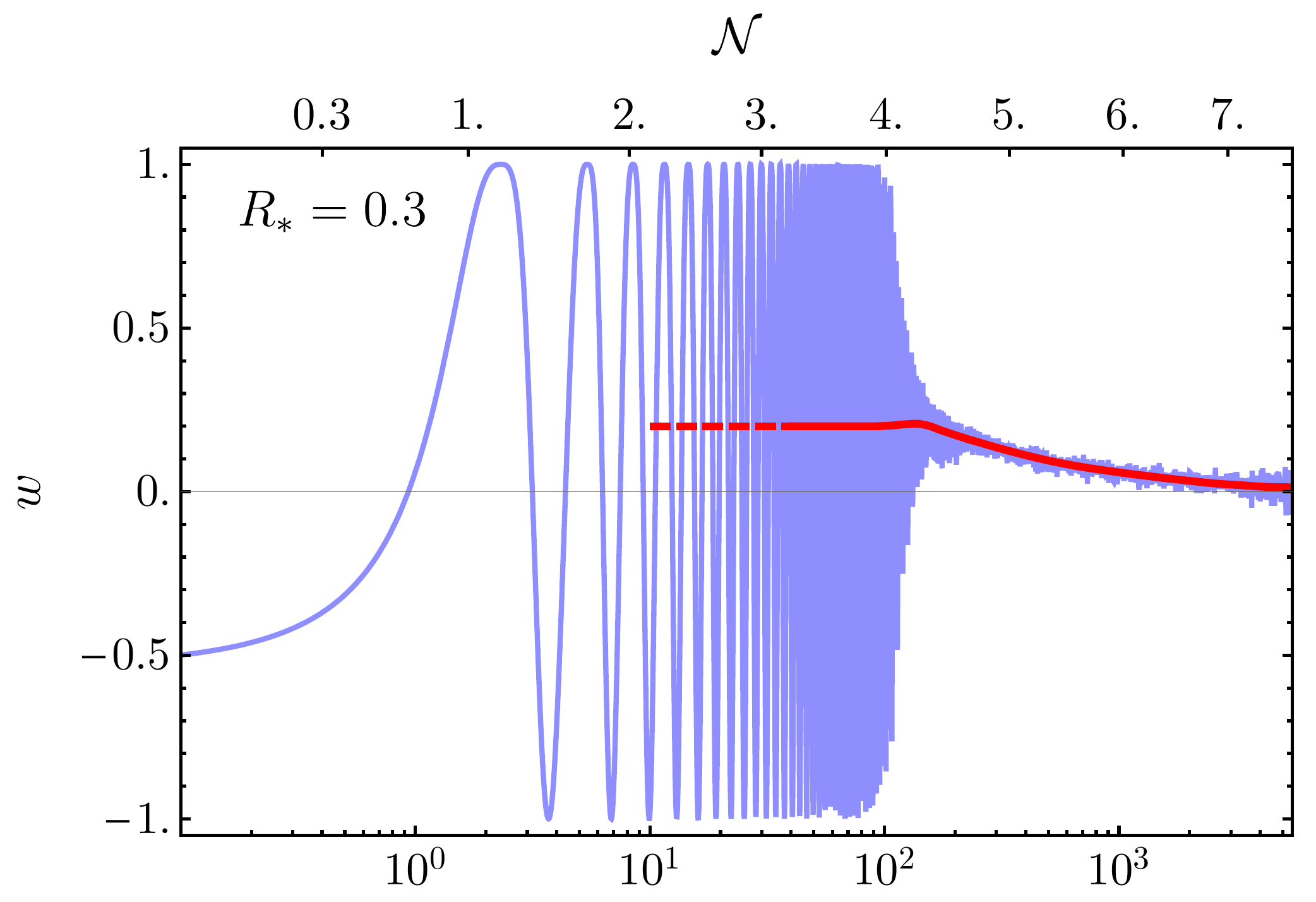}  \\ 
    \includegraphics[width=0.43\textwidth]{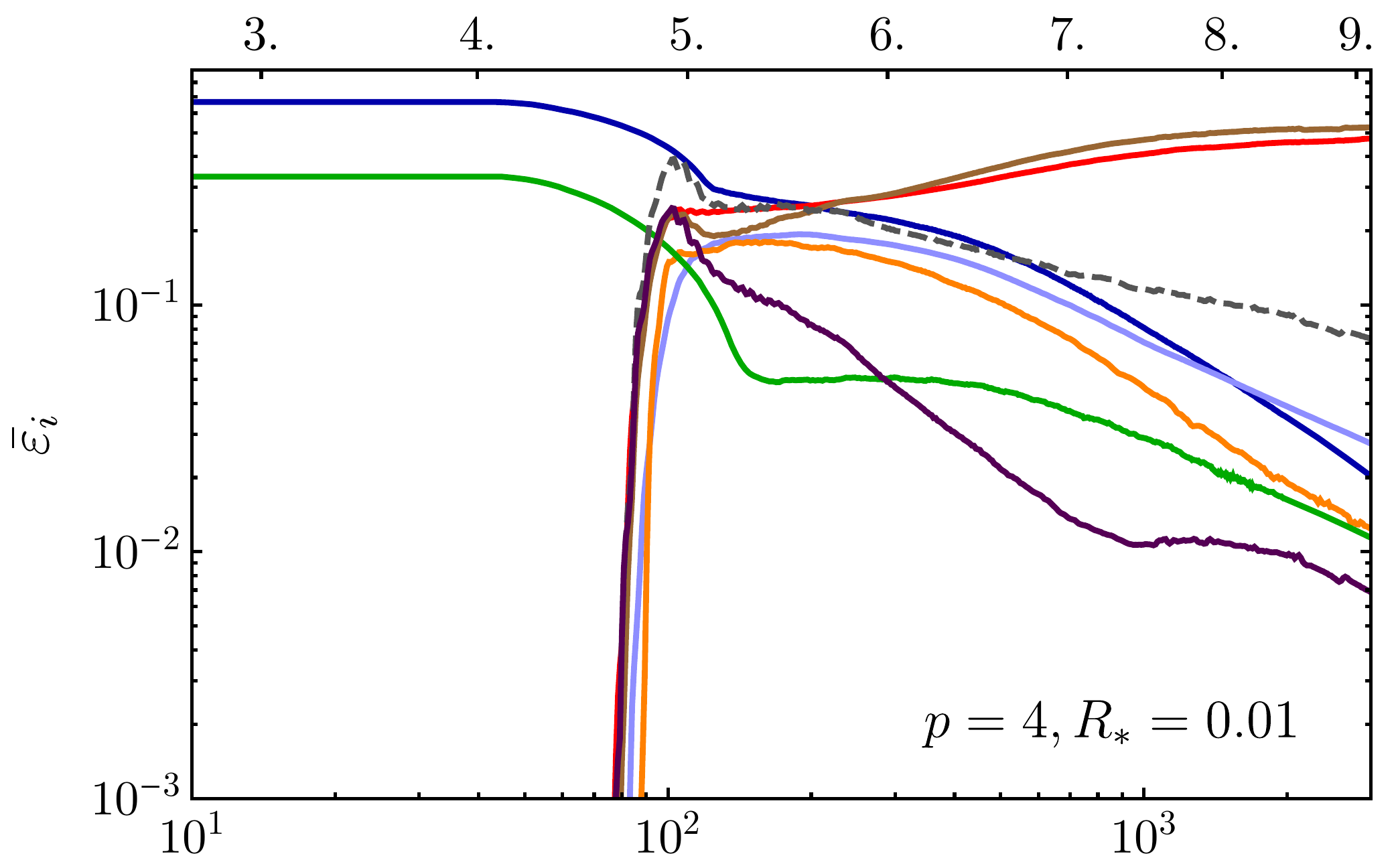} \hspace{0.2cm}
    \includegraphics[width=0.43\textwidth]{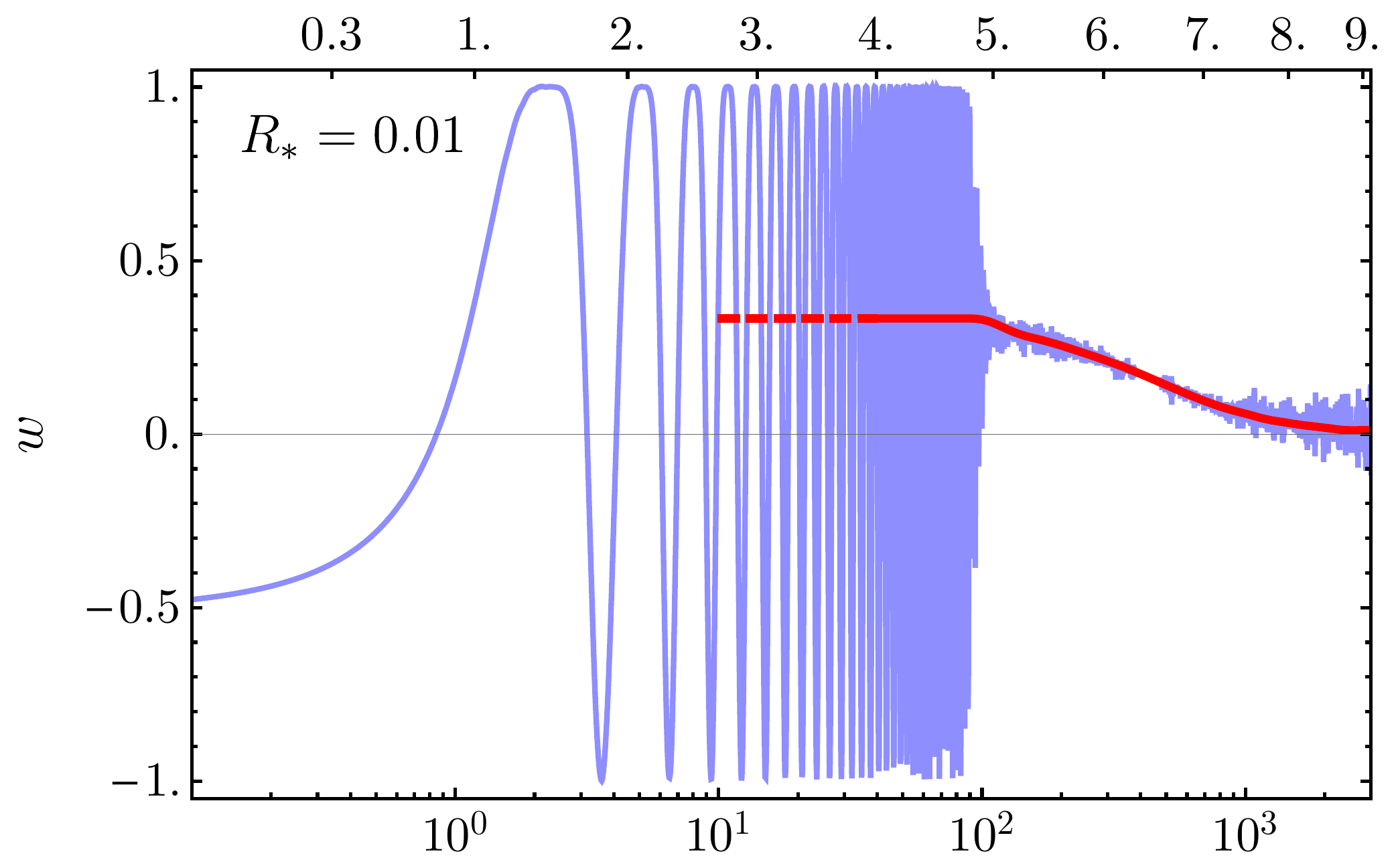}  \\
    \includegraphics[width=0.43\textwidth]{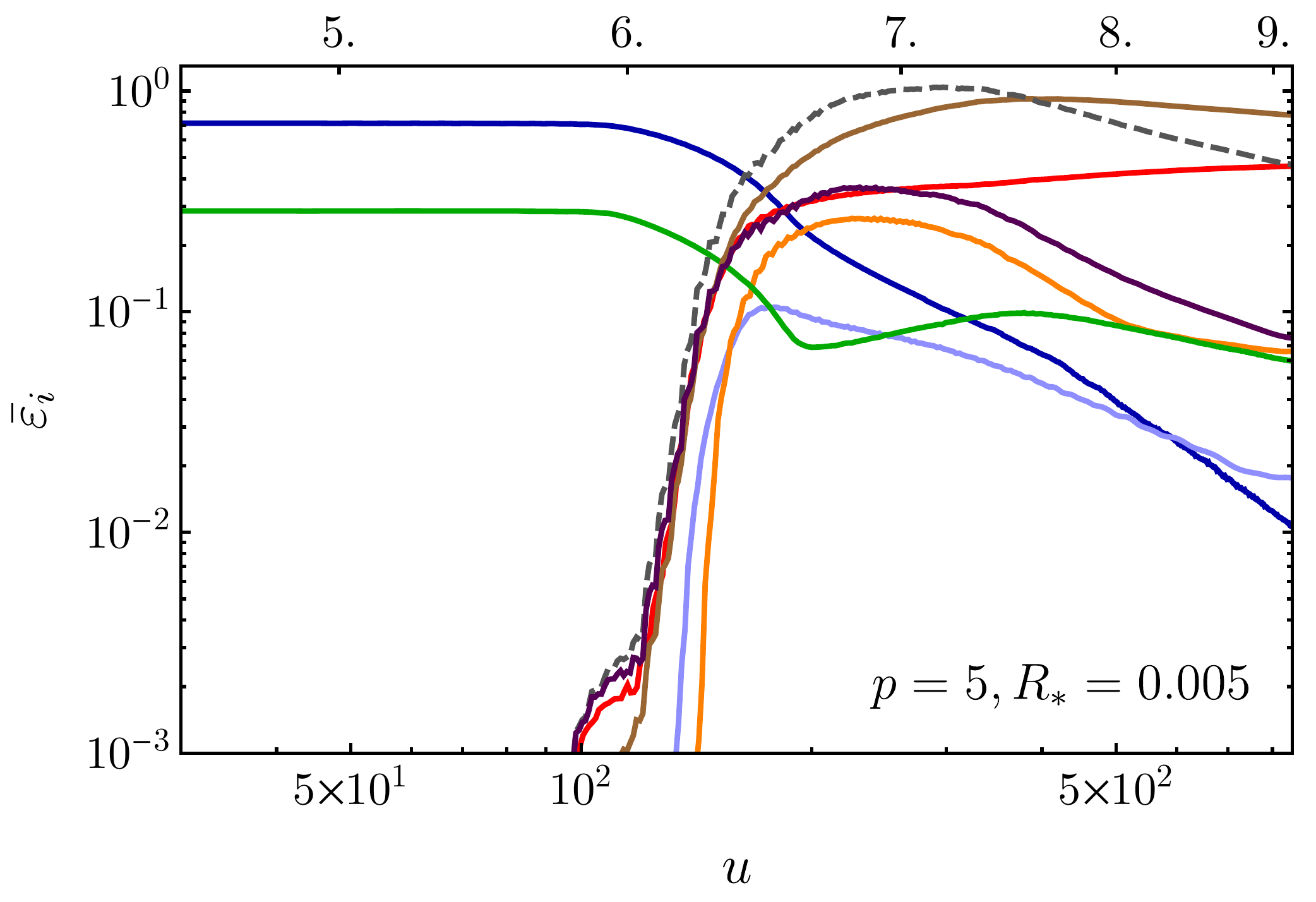} \hspace{0.2cm}
    \includegraphics[width=0.43\textwidth]{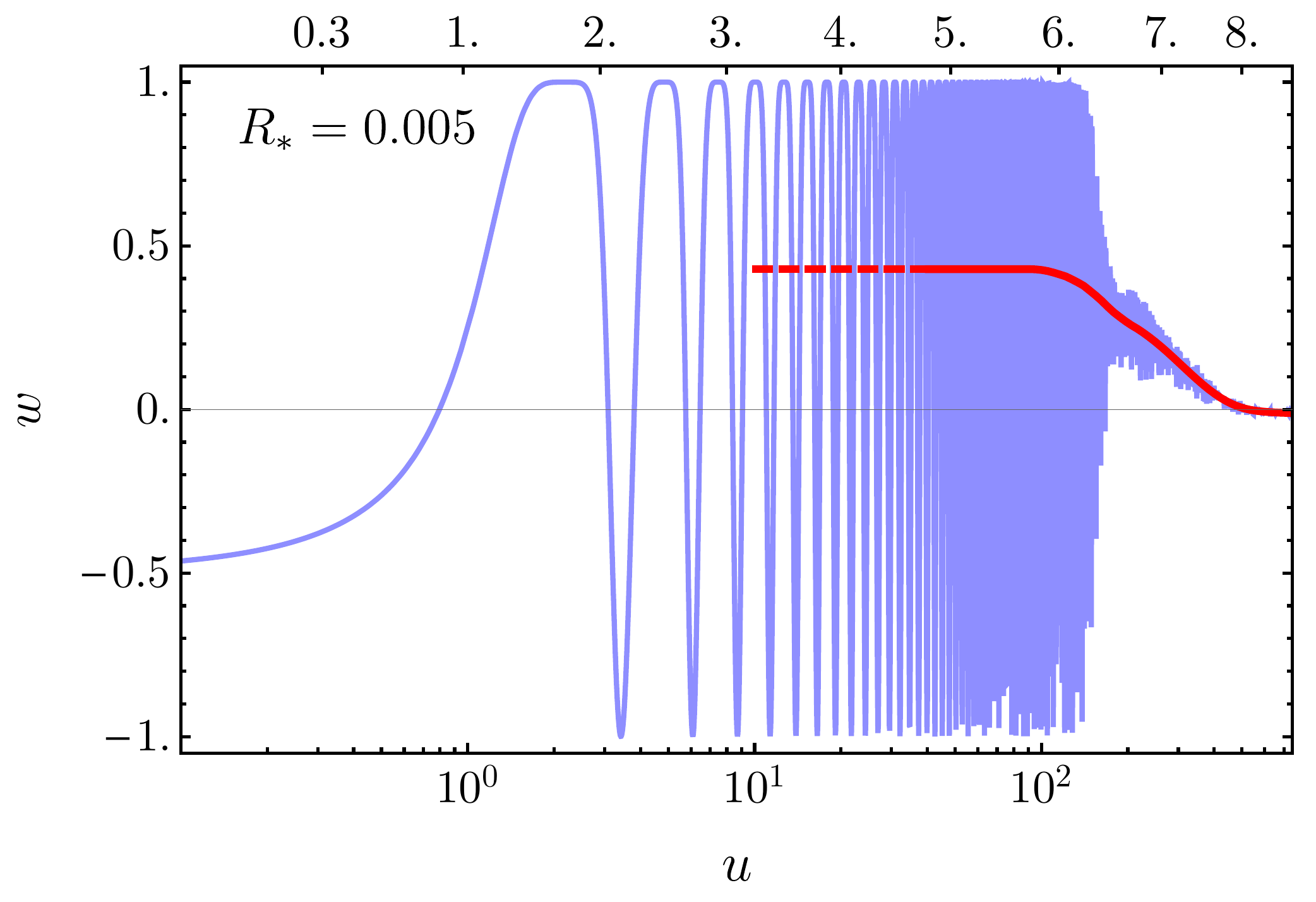}   \\ \vspace{0.4cm}    
    \includegraphics[width=0.5\textwidth]{en_vev_legend.pdf} 
    \hspace{0.2cm}
    \includegraphics[width=0.12\textwidth]{eos_legend.pdf}
    \vspace{0.4cm}         
    \caption{[$v>0$] Evolution of the averaged energy ratios (left panels) and equation of state $w$ (right panels) for $M=10m_{\rm pl}$, $v=10^{-2}m_\mathrm{pl}$, and different values of $p$ and $R_*$: $p=3$ and $R_*=0.3$ (top), $p=4$ and $R_*=0.01$ (middle), and $p=5$ and $R_*=0.005$ (bottom). The right panels show the equation of state $w$ in blue and its oscillation average $\bar{w}$ in red. Note: the energy ratio associated to the trilinear interaction is depicted in terms of its absolute value $|\bar{\varepsilon}_{\rm i}^{1}|$.}  \label{fig:eneosp345}
\end{figure*} 

\textbf{ii) \boldsymbol{$p>2$}:} In Fig.~\ref{fig:eneosp345} we present results from lattice simulations for $p=3, 4$ and $5$, each case for a particular choice of $R_*$. In the left panels we have depicted the evolution of the averaged energy ratios. Initially, all the energy is stored in the homogeneous inflaton condensate, which approximately satisfies the equipartition identity $ \bar{\varepsilon}_{\rm k}^\varphi  \simeq (p/2) \bar{\varepsilon}_{\rm p}^\varphi $. The kinetic and gradient energies of the daughter field get excited via parametric resonance, and when they become sizable, they induce the decay of the inflaton condensate due to backreaction effects at the time scale $u_{\rm br} \sim 80-120$. As analyzed in Section \ref{sec:Lin:DisplacedPotential}, parametric resonance eventually ends, so the gradient energies stop growing and start decreasing. At very late times, we end in a situation in which the daughter field dominates the entire energy budget and shows a homogeneous configuration. Note that the equipartition identities Eq.~(\ref{eq:Virial1}) and (\ref{eq:Virial2}) are approximately preserved at all times during preheating, so at late times the daughter field kinetic and potential energies approximately obey the virial identity $ \bar{\varepsilon}_{\rm k}^\chi \simeq  \bar{\varepsilon}_{\rm p}^\chi$. This result is independent on the choice of $p$ as long as $p>2$.

In the right panels of Fig.~\ref{fig:eneosp345} we show the evolution of the equation of state $w$ (blue) and its average $\bar{w}$ (red) for three choices of $R_*$. During the initial homogeneous phase we have $\bar w = \bar{w}_{\rm hom} = (p-2)/(p+2)$ as expected. With the later growth of the gradient energies due to parametric resonance, $\bar{w}$ is pushed slightly towards the radiation-domination result $\bar{w}\rightarrow1/3$ for $p=3$ and $5$ (for $p=4$, it stays at $\bar{w}=\bar{w}_{\rm hom} = 1/3$). This result is similar to the centred case potential. Once parametric resonance terminates, the gradient energy ratios start to dilute and the massive daughter field eventually dominates the energy budget, so that the averaged equation of state evolves towards $\bar{w}\rightarrow 0$ at late times.

\section{Determination of inflationary constraints} \label{Sec:CMB} 

Inflation generates an almost scale-invariant spectrum of scalar fluctuations, which can be parametrized as 
\be \Delta_{\mathcal{R}}^2(k)= A_s\left(\frac{k}{k_\mathrm{CMB}}\right)^{n_s-1} \ , \ee
where $A_s$ and $n_s$ are the scalar amplitude and spectral tilt at a particular pivot scale $k_\mathrm{CMB}$. We denote the inflaton amplitude when the pivot scale crosses the Hubble scale as $\phi_k$. In terms of slow roll parameters, $A_s$ and $n_s$ are determined as a function of the inflationary potential as
\bea A_s &=& \frac{V(\phi_k)}{24\pi^2\epsilon_{_V} (\phi_k)m_\mathrm{pl}^4} , \label{eq:As} \\
n_s &=& 1-6 \epsilon_{_V}(\phi_k)+2\eta_{_V}(\phi_k) \ . \label{eq:ns} \eea
Observations of CMB anisotropies~\cite{Aghanim:2018eyx} have constrained
the scalar amplitude to be $\rm{ln} (10^{10} A_s) = 3.043 \pm 0.014$ at $k_\mathrm{CMB} =0.05\,\rm{Mpc}^{-1}$, which we adopt from now on as our fiducial choice of pivot scale. We can then use Eq.~(\ref{eq:As}) to determine the value of $\phi_k$ when the pivot scale crossed the Hubble radius. Furthermore, inflation also generates a nearly scale-invariant spectrum of tensor perturbations with amplitude $A_t$ and tensor tilt $n_t$. The tensor-to-scalar ratio $r \equiv A_t / A_s$ can be computed as
\be r = 16 \epsilon_{_V} (\phi_k)\ . \label{eq:tensor-to-scalar} \ee  

Given an inflationary model, it is interesting to obtain predictions for $n_s$ and $r$ to compare them with experimental constraints. However, these predictions depend on the number of e-folds of expansion from the moment at which the pivot scale $k_\mathrm{CMB}$ crossed the Hubble scale till the end of inflation. This quantity, which we denote as $N_k$, can be written as
\be N_k \equiv \mathrm{ln} \frac{a_\mathrm{end}}{a_k} \simeq \frac{1}{m_\mathrm{pl}^2}\int_{\phi_k}^{\phi_\mathrm{end}}\frac{V}{V_{,\phi}}\rvert d\phi \rvert\ , \label{eq:Nk1} \ee
where $\phi_{\rm end}$ is the field amplitude at the end of inflation. To determine $N_k$ exactly one needs to know the evolution of the equation of state $w$ from the end of inflation until the onset of a radiation-dominated (RD) stage.

A more convenient expression for $N_k$ can be obtained by comparing the pivot scale $k_\mathrm{CMB}$ to the present Hubble radius $1/(a_0H_0)$, and then considering the expansion of the Universe from the time of horizon crossing (when $k_\mathrm{CMB}=a_k H_k$) till today. We parametrize the expansion history as a series of four stages characterised by different equations of state. The first stage corresponds to the inflationary expansion, from the time of horizon crossing till the end of inflation. The second stage goes from the end of inflation till the onset of RD (which does not necessarily coincide with the moment of thermalization of the relativistic species). The subsequent third and fourth stages go, respectively, from the onset of RD to the onset of matter domination (MD), and from the latter moment to today, including the present smooth transition into a dark energy (DE) dominated universe. We can write the ratio $k_\mathrm{CMB} /(a_0H_0)$ as follows,
\be \frac{k_\mathrm{CMB}}{a_0H_0}=\frac{a_kH_k}{a_0H_0}= e^{-N_k}\frac{a_\mathrm{end}}{a_\mathrm{rd}}\frac{a_\mathrm{rd}}{a_\mathrm{eq}}\frac{a_\mathrm{eq} }{a_0}  \frac{H_\mathrm{eq}}{H_0}\frac{H_k}{H_\mathrm{eq}} \ , \label{kcmba0H0}\ee
where the different labels indicate at which times the quantities must be evaluated: $a_\mathrm{end}$, $a_{\rm rd}$, and $a_\mathrm{eq}$ denote, respectively, the scale factor at the end of inflation, onset of RD, and when the matter-radiation equality holds. We note that for inflationary potentials of the form $V(\phi) \propto \phi^4$, the Universe enters RD almost immediately after the end of inflation. In such a case we have $a_{\rm end} \simeq a_{\rm rd}$.

We can write the above expression in a more useful form by considering the following relations, 
\bea
H_k^2 &\simeq& \frac{V_k}{3m_\mathrm{pl}^2} \ , \hspace{0.4cm} V_k \equiv V(\phi_k) \ , \\
\frac{ a_\mathrm{rd} }{ a_\mathrm{eq} } &=& \left(\frac{\rho_\mathrm{rd}}{\rho_\mathrm{eq}} \right)^{-1/4}\ , \\
\frac{a_0}{a_\mathrm{eq}} &=& (1+z_\mathrm{eq})\simeq3387\ , \\
\rho_\mathrm{eq} &=& 6\Omega_{m,0}m_\mathrm{pl}^2H_0^2(1+z_\mathrm{eq})^3 \ ,
\eea
where in the second expression we have neglected changes in the number of relativistic degrees of freedom. Using the constraints of $H_0= (67.66 \pm 0.42) \mathrm{km}\,\mathrm{s}^{-1}\mathrm{Mpc}^{-1}$ and $\Omega_{m,0}\simeq0.311 \pm 0.006$~\cite{Aghanim:2018eyx}, we obtain the following expression
\be N_k\simeq 61.5 + \frac{1}{4}\mathrm{ln} \left( \frac{V_k^2}{m_\mathrm{pl}^4\rho_\mathrm{rd}} \right) - \Delta N_{\rm end}^{\rm rd}\ , \label{Nk1} \ee
where $\Delta N_{\rm end}^{\rm rd} \equiv {\rm ln} (a_{\rm rd} / a_{\rm end})$ is the number of e-folds between the end of inflation and the onset of the RD stage, which depends on the evolution of the equation of state during this period. Under reasonable assumptions for the reheating stage, $N_k$ is estimated to be around $N_k \simeq 50 - 60$~\cite{Liddle:2003as}. However, we are now in position to use our in-depth analysis of the post-inflationary evolution of the scale factor and energy densities to exactly compute $N_k$, and hence $n_s$ and $r$. Note that in Appendix \ref{App:Nbr}, we have rewritten Eq.~(\ref{Nk1}) by decomposing the expansion history between the end of inflation and radiation domination in different sub-stages. This allows to discuss in more detail the role that the interaction term plays in the determination of $N_k$, and hence on the inflationary observables $n_s$ and $r$.

\subsection*{Determination of \texorpdfstring{$N_k$}{Nk}, \texorpdfstring{$n_s$}{ns} and \texorpdfstring{$r$}{r}}

\begin{table}
  \begin{center}     \def\arraystretch{1.3}
    \begin{tabular}{|p{0.7cm}|c|c|c|c|}
    \hline
    \centering $\mathbf{ N_k}$ & \multicolumn{2}{ c| }{$M = 2m_{\rm pl}$} & \multicolumn{2}{ c| }{ $M = 10m_{\rm pl}$} \\
    \hline
    \centering \textbf{$p$} &  $q_*=0$ & $q_*\geq q_*^{\rm (min)}$ &  $q_*=0$  & $q_*\geq q_*^{\rm (min)}$\\
    \hline
    \centering \textbf{$3$} & $55.2$ & $55.5$ &$56.7$  & $56.9$\\
    \hline
    \centering \textbf{$4$} & $55.8$ & $55.8$ & $57.4$  & $57.4$\\
    \hline
    \centering \textbf{$5$} & $56.9$ & $56.1$&   $58.1$& $57.8$\\
    \hline
    \centering \textbf{$6$} & $57.1$ &  $56.5$ &$58.9$  & $58.3$\\
 \hline
    \end{tabular}
  \end{center}
  \caption{Values of $N_k$ computed with lattice simulations of the centred potential scenario for different choices of $p$ and $M$, when i) there is no interaction between the inflaton and the daughter field, so $q_* = 0$, and ii) the inflaton is coupled to the daughter field with $q_*\geq q_*^{\rm (min)}$ [see Eq.~(\ref{eq:qstarM})].\vspace{-0.1cm}}
  \label{tab:Nk}
\end{table}

We will now use our lattice results in order to compute exact predictions for $N_k$ in the case of the $\alpha$-attractor model (\ref{eq:inflaton-potential}),  and then constrain the CMB observables $n_s$ and $r$. Given an inflationary model, we can simulate the post-inflationary dynamics on the lattice and observe when a RD stage is achieved in the simulations. The values of $\rho_{\rm rd}$ and $\Delta N_{\rm end}^{\rm rd}$ can be extracted from the numerical simulations and be used to compute $N_k$ exactly via Eq.~(\ref{Nk1}). This only applies to the centred potential scenario for $p>2$, as for a centred potential with $p = 2$ or for a displaced potential with arbitrary $p$, the universe always ends in a transient MD stage within our simulation time scales.

An important challenge with the procedure just mentioned is that, in order to perform the lattice simulations, the value of $\Lambda$ must be fixed beforehand, even though it depends on $N_k$ via Eq.~(\ref{eq:LabelNk}). We circumvent this issue by applying an iterative procedure as follows. First, an initial simulation is carried out with the parameter $\Lambda^{(1)} = \Lambda (N_k^{(1)})$ corresponding to $N_k^{(1)} = 60$. By extracting $\rho_{\rm rd}$ and $\Delta N_{\rm end}^{\rm rd}$ from the simulation, we can then use Eq.~(\ref{Nk1}) to compute a better approximation $N_k^{(2)}$. A new simulation with $\Lambda = \Lambda^{(2)} \equiv \Lambda(N_k^{(2)})$ is now carried out, the outcome of which is used to improve further the estimation of $N_k$. The iterative cycle is repeated several times, until the value of $N_k$ only changes marginally in each iteration. We observe that after four to five cycles, $N_k$ changes only by a factor $\mathcal{O}(10^{-2})$. 

The following results for $N_k$ are obtained by the numerical routine presented above. We considered power-law coefficients and mass parameters in the ranges $p=3-6$ and $M/m_{\rm pl}=2-10$, such that the inflaton oscillates only in the positive curved region of the potential (see Eq.~\ref{eq:inflection-potential}). In each case we distinguish two situations: first, when the inflaton is not coupled to a daughter field (i.e.~$q_* = 0$) and the homogeneous regime breaks down due to inflaton self-resonance; second, when a quadratic-quadratic $g^2\phi^2X^2$ interaction is present with $q_* \gtrsim q_*^{\rm (min)}$ [see Eq.~(\ref{eq:qstarM})], and the homogeneity of the inflaton breaks down due to parametric resonance of the daughter field. In the later case, for sake of being specific, simulations have been performed for $g\simeq10^{-4}$, as for other reasonable couplings the e-folding needed to achieve RD are rather similar as long as $q_* \gtrsim q_*^{\rm (max)}$, see Fig.~\ref{fig:PvsN}. \vspace{0.3cm}

\textbf{Results for $N_k$:} Table \ref{tab:Nk} shows values of $N_k$ computed for the power-law coefficients $p=3$, 4, 5, 6, for the mass parameters $M=2 m_{\rm pl}$ and $10 m_{\rm pl}$. In these cases, $N_k$ lies roughly in the range $N_k\approx 55-59$. We note that uncertainties due to experimental bounds of observational data, such as $H_0$ and $\Omega_{m,0}$~\cite{Aghanim:2018eyx}, lead to uncertainties in $N_k$ of $\mathcal{O}(10^{-2})$. The biggest uncertainty comes therefore from the variation of $M$: as this scale decreases, the transition between the monomial and flat regimes of the potential happens at smaller field amplitudes, so the value of $N_k$ changes correspondingly.

\begin{figure}
    \centering
    \includegraphics[width=0.45\textwidth]{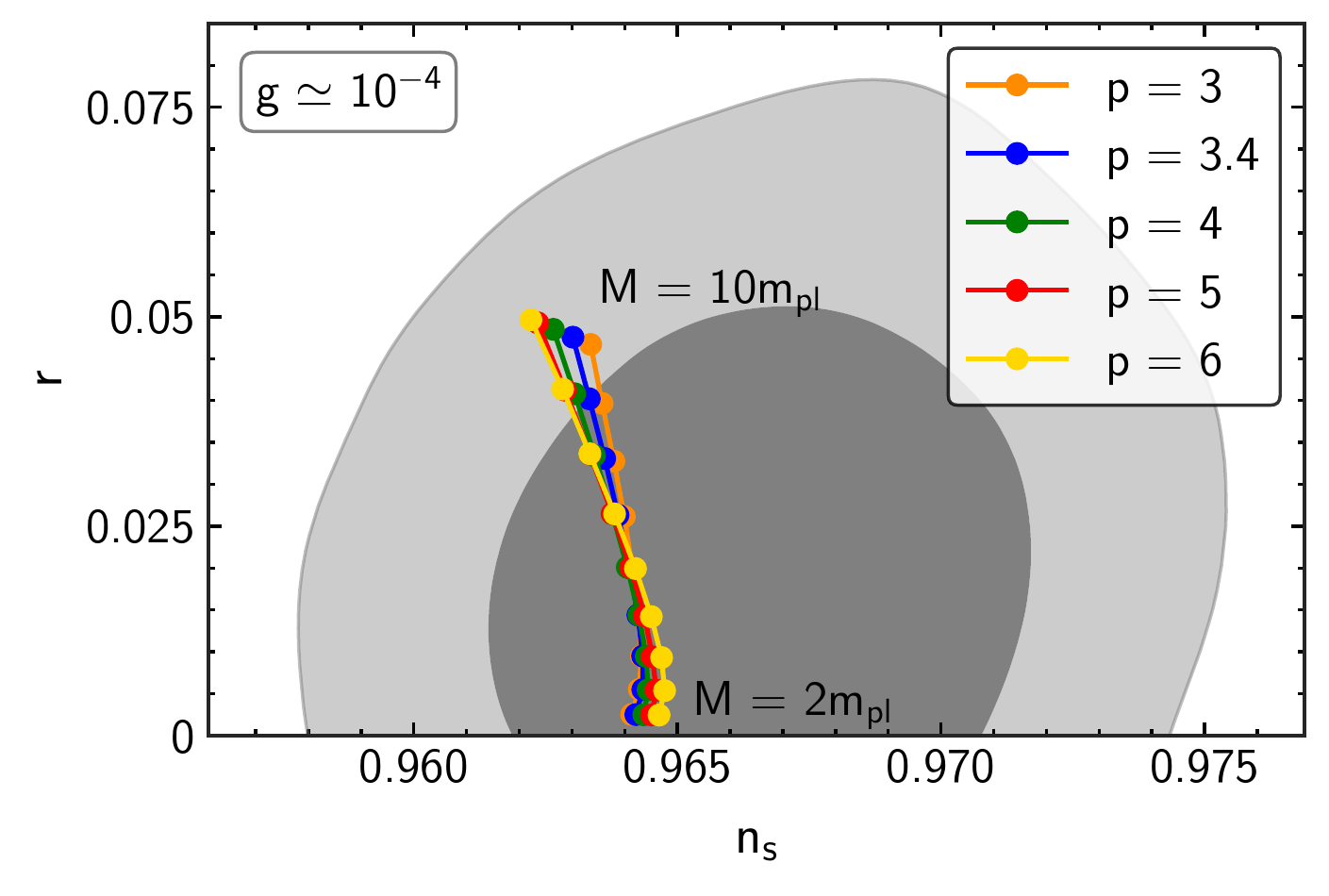}
    \includegraphics[width=0.45\textwidth]{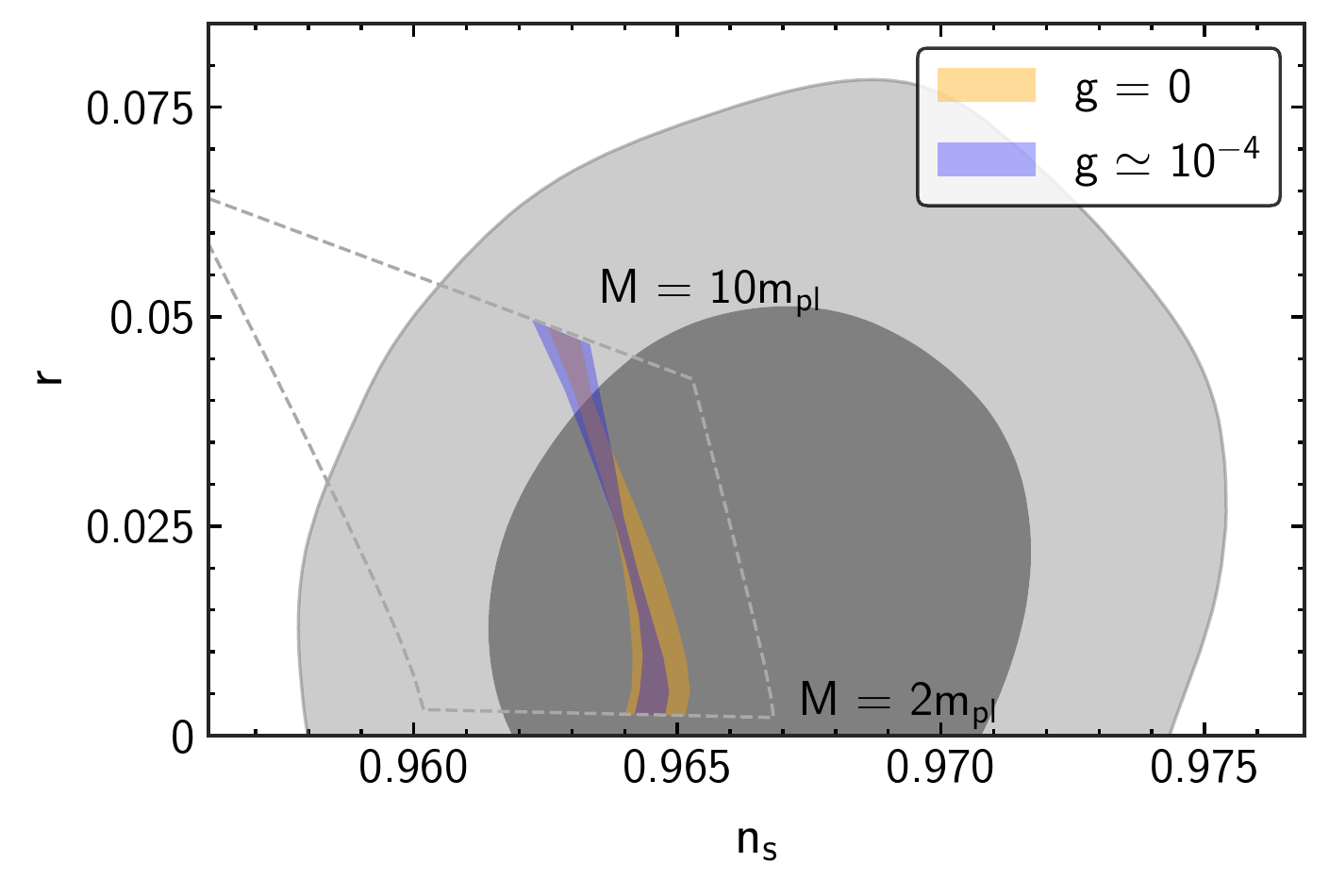}

    \caption{[$v=0$] Top: Predictions for $n_s$ and $r$ for potential (\ref{eq:inflaton-potential}), in the presence of an interaction between the inflaton and a daughter field with $g\simeq10^{-4}$. We have considered potential parameters $M/m_\mathrm{pl}=2-10$ and $p=3,3.4,4,5$ and $6$. Bottom: Comparison of the predicted values of $n_s$ and $r$ in the presence and absence of quadratic-quadratic interaction. The yellow area shows, for $g = 0$, the range of values of $n_s$ and $r$ predicted for the potential parameters just indicated. The blue area shows the same but for $g\simeq10^{-4}$. Finally, the area enclosed by dashed lines shows the predictions for $N_k \in [50,60] $. The dark and light gray areas indicate the $68\%$ and $95\%$ CL regions for $n_s$ and $r$ respectively. }  \label{fig:CMBpred}
\end{figure}

Note that for $p=4$, the equation of state after the first oscillation is already very close to radiation-domination ($w_{\rm hom} = 1/3$ for $p=4$, c.f.~\ref{eq:EoSoscillations}), so for a fixed $M$, the values of $N_k$ are the same for all choices of $q_*$. On the other hand, $N_k$ for $p<4$ is smaller with respect to the $p=4$ case, because all the $\Delta N_i^j$-terms in Eq.~(\ref{eq:e-folds-approx}) become negative in this case. Similarly, $N_k$ becomes larger for $p>4$, as the different $\Delta N_i^j$-terms in Eq.~(\ref{eq:e-folds-approx}) become positive. Moreover, note that for a given choice of $M$ and $p$, the value of $N_k$ for $q_*>q_*^{(m)}$ only changes by less than an e-fold with respect to the $q_* = 0$ case (for a more detailed discussion, see Appendix \ref{App:Nbr}).

Let us briefly comment here on the centred potential case with $p=2$. Independently of the specific preheating dynamics, the system eventually returns to MD, so $N_k$ cannot be determined without adding further ingredients in the scenario. In order to reheat the universe and arrive into  a RD stage, further ingredients should be added to the theory, such as e.g.~perturbative decay channels.\footnote{Note that the shifted potential provides such a perturbative decay channel, as it generates a trilinear coupling between the inflaton and the daughter field. However, this does not solve the problem because $X$ becomes a massive field.} In any case, the stage of parametric resonance induces a temporary deviation of the equation of state from $\bar w=0$ towards $\bar w = \bar w_{\rm max} < 1/3$, which makes $\delta N_k$ slightly change with respect to the case without interaction. Although $\delta N_k$ becomes larger for increasingly larger values of $q_*$, we obtain $\delta N_k<1$ for all values of $q_*$ considered in this work.  \vspace{0.3cm}

\textbf{Predictions for $n_s$ and $r$}: With the numbers for $N_k$, we can now compute predictions for the CMB observables using Eqs.~(\ref{eq:ns}) and (\ref{eq:tensor-to-scalar}) (for $p>2$). Our results are summarized in Fig.~\ref{fig:CMBpred}. The top panel shows results in the $n_s$-$r$ plane for $g\simeq10^{-4}$ and several power-law coefficients $p$. Each dot indicates an integer value of the mass scale between $M/m_{\rm pl}=2-10$ in $\Delta M/m_{\rm pl} = 1$ intervals. The tensor-to-scalar ratio is mainly controlled by the mass scale $M$ and falls in the range $r\simeq0.0025-0.05$, though much smaller values can be achieved for $M\ll m_{\rm pl}$ (which we have not considered here because the inflaton potential can no longer be approximated by a monomial potential during preheating). On the other hand, $n_s$ decreases for all values of $p$ as we increase $M$ from $M \gtrsim 3m_{\rm pl}$, though $n_s$ increases slightly when we move from $M =2 m_{\rm pl}$ to $M \simeq 3 m_{\rm pl}$.

A direct comparison between the results in the absence and presence of interaction is shown in the lower panel. There we also show the constraints corresponding to the approximate range $N_{\rm k}=50-60$. We note that an accurate investigation of the preheating phase strongly reduces the uncertainty. The space of possible spectral tilts ranges for the considered parameters, ranges from $n_s\simeq0.9622$ (for $p=6$, $M=10m_{\rm pl}$ and $q_* \gtrsim q_*^{\rm (max)}$) to $n_s\simeq0.9652$ (for $p=6$, $M=3m_{\rm pl}$ and $g=0$).

\section{Summary and discussion}  \label{Sec:Conclusion}

In this work we have characterized the evolution of the energy distribution and equation of state after inflation, by using a combination of analytical techniques and lattice simulations.  As a proxy for preheating, we have considered two-field scenarios where the inflaton $\Phi$ has a monomial potential around a minimum at some scale $\Phi = v$ after inflation, so that $V(|\phi|) \propto |\phi|^p$ ($p\geq2$) with $\phi \equiv \Phi - v $. We have also considered that the inflaton is coupled to a daughter field $X$ via a quadratic-quadratic interaction $g^2 \Phi^2 X^2$. We have examined two situations: 1) a minimum at $v = 0$ so that $V(\Phi) \propto |\Phi|^p$ (\textit{centred} potentials), and 2) a minimum at $v > 0$ so that $V(\Phi) \propto |\Phi - v|^p$ (\textit{displaced} potentials).

\subsection{Summary of results}

Due to the length of the paper, we provide here a short `guide' to the most important results, with references to the corresponding equations and figures:

\begin{itemize}[align=left,leftmargin=*]
    \item In Section \ref{Sec:Inflation} we have studied the properties of the considered potentials. In particular, in Section \ref{sec:2A} we have parametrized the oscillatory properties of the homogeneous inflaton condensate during the initial stage after inflation. This configuration gives rise to the equation of state $\bar{w}_{\rm hom} = (p-2)/(p+2)$.
    
    \item In Section \ref{Sec:LinearAn} we have studied analytically the resonant excitation processes of the inflaton and daughter fields triggered by the oscillations of the homogeneous inflaton, which are valid during the initial linear regime before backreaction effects become relevant. In particular:
    
    \begin{itemize}[align=left,leftmargin=*]
        \item[\tiny $\blacksquare$] In Section \ref{sec:3A} we have studied the process of inflaton self-resonance, which exists for $p>2$. The corresponding Floquet diagram is depicted in Fig.~\ref{fig:FloquetDiagrams}. Using its properties, we have computed an analytical estimate of the backreaction time as a function of $p$ in Eq.~(\ref{eq:ubr-selfr}), depicted in Fig.~\ref{fig:PvszbrNbr}.
        
        \item[\tiny $\blacksquare$]  In Section \ref{sec:3B} we have studied the process of parametric resonance of the daughter field for the centred potential scenario. The corresponding Floquet diagram is depicted in Fig.~\ref{fig:AllFloquetDiagrams}, and an analytical estimate for the backreaction time has been computed in Eq.~(\ref{eq:zBR}), and depicted in Fig.~\ref{fig:Analytic-ParRes}.
        
        \item[\tiny $\blacksquare$] In Section \ref{sec:Lin:DisplacedPotential} we have studied the process of parametric resonance of the daughter field in the displaced potential scenario. The corresponding Floquet charts for different model parameters, which summarize the most important information from our analysis, are depicted in  Fig.~\ref{fig:shifted-floquet}.
        
    \end{itemize}
    
    \item In Section \ref{Sec:LatResults} we have presented results from our lattice simulations, which go beyond the linearized analysis and capture the later non-linear dynamics. We have considered both the centred and displaced potential scenarios:
    
    \begin{itemize}[align=left,leftmargin=*]
        \item[\tiny $\blacksquare$] In Section \ref{Sec:LatResults1} we have presented our lattice results for the centred potential scenario. Depending on the choice of $p$, we can distinguish three different regimes: 
        
        \begin{itemize}[align=left,leftmargin=*]
            \item For $p=2$ there is no inflaton self-resonance. Moreover, even if parametric resonance of the daughter field is initially broad, it will always become narrow at late times due to the expansion of the universe. Therefore, even if both fields can get excited and develop fluctuations at initial times, at late times the inflaton gets homogeneous again and dominates the energy budget. Similarly, the equation of state deviates temporarily from $\bar{w} = \bar{w}_{\rm hom} = 0$ to $\bar{w} = \bar{w}_{\rm max} < 1/3$, but goes back to $\bar{w} \rightarrow \bar{w}_{\rm hom} = 0$ at late times. This behaviour can be seen in Figs.~\ref{p2-EosEn}, \ref{fig:p2qParam} and \ref{p2-Fits}, which illustrate results from different simulations with different coupling strengths.
            
            \item  For $2 < p < 4$, parametric resonance of the daughter field is also narrow at late times, but inflaton self-resonance is always active (unlike the $p=2$ case). Due to this, the inflaton also dominates the energy budget of the universe at late times, but in this case it fragments completely, with its energy is equally distributed between its kinetic and gradient components. Correspondingly, the equation of state goes from $\bar w = \bar{w}_{\rm hom} = (p-2)/(p+2)$ at initial times, to $ \bar{w} \rightarrow 1/3$ at late times. Results from specific lattice simulations have been depicted in Figs.~\ref{p23-EosEnergy} and \ref{fig:Energiesp3}.
            
            \item For $p \geq 4$, if parametric resonance is initially broad, it will remain as such at later times. As the inflaton self-resonance is also active, we end up in a situation in which the energy is equally distributed ($\sim$ 50 $\%$ - 50 $\%$) between the inflaton and the daughter field, and in each case, shared equally between the corresponding kinetic and gradient components. Consequently, the equation of state jumps also from $\bar w = \bar{w}_{\rm hom} = (p-2)/(p+2)$ at initial times, to $\bar w \rightarrow 1/3$ at late times. Results from lattice simulations of this scenario have been depicted in Fig.~\ref{fig:Energiespg4}.
            
            \item Finally, note that all these results have been summarized in Table \ref{Tab:ResultsCentred}, which indicates the final values attained by the equation of state and energy ratios for different choices of $p$ and $q_*$. Also, the various evolutions of the equation of state for different $p$ are shown in Fig.~\ref{Fig:EoSvsP}.
            
        \end{itemize}
        
        Note also that a spectral analysis of the centred potential scenario is presented in Appendix \ref{app:Spectra}.
        
        \item[\tiny $\blacksquare$] In Section \ref{Sec:LatResults2} we have presented results from lattice simulations for the displaced potential scenario. Depending on $p$ we have identified two different scenarios, and in each case we have  simulated the dynamics for different `mass ratios' $R_*$ (defined in Eq.~\ref{eq:massratio}):
        
        \begin{itemize}[align=left,leftmargin=*]
            \item For $p=2$, our results are summarized in Fig.~\ref{fig:p2-En_vev}. We have observed that the amount of energy transferred to the daughter field via parametric resonance depends significantly on the value of $R_*$, and unlike in the centred case, we can transfer more than 50\% for some specific choices of $R_*$ and $v$. However, parametric resonance becomes weak at late times due to the expansion of the universe, so the contribution of the gradient energies of both fields to the energy budget eventually becomes negligible. Therefore, at late times both the inflaton and daughter fields are completely homogeneous and oscillate around the minimum of the potential with different oscillation periods. Similarly, the equation of state deviates initially from $\bar w = \bar w_{\rm hom} = 0$ to $\bar w = \bar w_{\rm max} < 1/3$, but recovers $\bar w \rightarrow 0$ at late times.
            
            \item For $p>2$, the dynamics is very similar to the $p=2$ case, but now the inflaton is massless at late times (while the daughter field is still massive). Therefore, the energy budget at late times is dominated, in all cases, by the oscillating homogeneous mode of the daughter field. This can be observed in Fig.~\ref{fig:eneosp345}.
        \end{itemize}
        
    \end{itemize}
   
            \item In Section \ref{Sec:CMB}, we have used our information on the equation of state to determine the number of e-folds between the time of horizon crossing and the end of inflation exactly, see Table \ref{tab:Nk} and Fig.~\ref{fig:PvsN}. Note that this is only possible for the centred potential with $p>2$, as these are the only cases where the equation of state achieves an RD stage at late times. With this information, we have been able to predict exactly the values of $n_s$ and $r$, which are given in Fig.~\ref{fig:CMBpred}.
    
\end{itemize}

We remark that for the cases where the system stabilises with equation of state $\bar w \rightarrow 0$, the ultimately required transition to the RD stage has to happen at a later time, e.g.\ via perturbative decays. This requires some slight extensions of the scenarios considered here by e.g.\ small mass terms of the respective fields and additional interactions. For situations where the RD stage is reached but the inflaton field still carries some substantial  fraction of energy, such slight extensions (with negligible effects during the phase of our simulation) can also lead to the transfer of energy to secondary light fields via perturbative decays, while staying within a RD universe. In these situations the transition to RD can be considered as already completed in the here simulated phase of reheating. \vspace{-0.3cm}

\subsection{Future work}   

In this paper, we have focused on the cases when the inflationary potential can be approximately described as a monomial around the minimum ($V(|\phi|) \propto |\phi|^p$ during the stages following inflation, with $p\geq2$ and $\phi \equiv \Phi - v$). We have restricted our analysis to the situation where the dominant interaction between the inflaton and a daughter field is described by a quadratic-quadratic coupling $g^2\Phi^2 X^2$. 

We envisage our present project (formed by letter \cite{Antusch:2020iyq} and this work) as a first step towards a more complete characterization of the post-inflationary stage in general - hence the  `\textit{Part I}' in the title. In particular, this work will be followed by \textit{Part II}, in which we study the energy distribution and equation of state when the inflaton is coupled to \textit{multiple} daughter fields instead of one, and where these can also have quartic self-interactions. In such type of setups, we will show that one can transfer far more than 50\% of the energy to the daughter field sector, while simultaneously achieving a RD stage.

Another interesting case of study will be to include trilinear interactions between the inflaton and the daughter field (similar to the case studied in \cite{Dufaux:2006ee}). This set-up has been partially considered in the displaced potential scenario, as the non-zero $v$ can be mapped, after a convenient field redefinition, to the case of a massive daughter field with a specific trilinear interaction, see Eq.~(\ref{eq:shifted-inflaton-potential1}). However, both the coupling strength and the mass depend explicitly on the one parameter $v$ here, while ideally it would be interesting to study separately the role of a trilinear coupling during the post-inflationary dynamics for arbitrary values.

Other possible extensions of this work might consist in considering: 1) Inflaton potentials with shapes different than monomial. For example, if we consider inflaton potentials with regions flatter than quadratic at largest field values of the inflaton oscillations, the formation of oscillons is expected. Oscillons can form  via self-resonance effects~\cite{Amin:2011hj} or tachyonic oscillations~\cite{Antusch:2015nla}, typically pushing the equation of state towards $\bar w = 0$ during their lifetime \cite{Gleiser:2014ipa,Lozanov:2016hid,Lozanov:2017hjm}. 2) Different mechanisms of parametric excitation during the linear regime, such as tachyonic preheating \cite{Felder:2000hj,Felder:2001kt,GarciaBellido:2002aj,Copeland:2002ku}. 3) Scenarios where non-minimal couplings to gravity or non-minimal kinetic terms are considered, like e.g.~in~\cite{DeCross:2015uza,DeCross:2016fdz,DeCross:2016cbs,Child:2013ria,Krajewski:2018moi,Iarygina:2018kee}. Two-field models with non-minimal gravitation couplings have been studied in detail with lattice simulations in \cite{Nguyen:2019kbm,vandeVis:2020qcp}, showing that for quartic potentials and quadratic-quadratic interactions RD can be achieved in $\lesssim 3$ e-folds after inflation for generic coupling choices. Scenarios of geometric preheating where spectator fields are non-minimally coupled to gravity represent also interesting cases to be considered~\cite{Bassett:1997az,Tsujikawa:1999jh,Fu:2019qqe,Figueroa:2021iwm}.  \newline

\textbf{Acknowledgements:} DGF (ORCID 0000-0002-4005-8915) is supported by a Ram\'on y Cajal contract with Ref.~RYC-2017-23493. FT has been supported by the Research Fund for Junior Researchers of the U.~Basel. SA, KM and FT acknowledge support by the Swiss National Science Foundation (project number 200020/175502). This work is also supported by project PROMETEO/2021/083 from Generalitat Valenciana, and by project PID2020-113644GB-I00 from Ministerio de Ciencia e Innovaci\'on.

\appendix

\section{Details on the T-Model}\label{App:TanhPotential}

Here we lay out some details of the inflationary potential (\ref{eq:inflaton-potential}), in particular expressions and constraints concerning the phase of inflation. Inflation terminates when the first slow-roll condition breaks, at the field amplitude $\phi_*$ given by Eq.~(\ref{eq:end-of-inflation}). The number of e-folds $N_k$ between the moment when the pivot scale $k_{\rm CMB}$ leaves the horizon and the end of inflation is given by
\be N_k \equiv \mathrm{ln} \frac{a_*}{a_k} \simeq \frac{1}{m_\mathrm{pl}^2}\int_{\phi_k}^{\phi_*}\frac{V}{\partial_\phi V}\rvert d\phi \rvert\ . \ee
With this expression, we can determine the field value $\phi_k$,
\be
\phi_k=\frac{1}{2}M{\rm arccosh}\left(\frac{\mathcal{I}}{pN_k}+\mathcal{J}\right),
\ee
where $\mathcal{I}=4p^2N_k^2 m_{\rm pl}^2/M^2$ and $\mathcal{J}^2=1+2p^2 m_{\rm pl}^2/M^2$. By evaluating the two slow-roll parameters at $\phi_k$, we can give expressions for the spectral index $n_s=1-6 \epsilon_{_V}(\phi_k)+2\eta_{_V}(\phi_k)$ and tensor-to-scalar ratio $r= 16\eta_{_V}(\phi_k)$,
\bea n_s &=& 1-\frac{p^2+2p\mathcal{J}+2\mathcal{I}/N_k}{p^2/2+2pN_k\mathcal{J}+\mathcal{I}} ,\\
r &=& \frac{8p^2}{p^2/2+2pN_k\mathcal{J}+\mathcal{I}} \ . \eea
In the limit $M\rightarrow \infty$ we have $\mathcal{I}\rightarrow 0$ and $\mathcal{J}\rightarrow 1$, and we recover the expressions of the chaotic scenario for $n_s$ and $r$. Moreover, we can constrain $\Lambda^4$ from the scalar amplitude $A_s$ via $V(\phi_k)=24\pi^2\epsilon_V(\phi_k)A_s m_{\rm pl}^4$. We then obtain the following relation,
\be
\Lambda^4=\frac{3\pi^2 A_s M^2 m_{\rm pl}^2}{N_k^2}f(p,M,N_k) \ ,
\ee
with
\be
f(p,M,N_k)=\frac{p \mathcal{I}^\frac{2-p}{2}(\mathcal{I}+2pN_k\mathcal{J}+pN_k)^p}{{(p^2/2+2pN_k\mathcal{J}+\mathcal{I})^\frac{2+p}{2}}}.\ee
In the limit $M\rightarrow 0$ this function goes to $f(p,M,N_k)\rightarrow 1$ and we recover the expression for the chaotic inflation scenario, see Eq.~(\ref{eq:mu}).

\section{Oscillation-averaged expressions for the inflaton homogeneous amplitude}\label{App:PeriodAverages}

We compute here the oscillation-averaged quantities $\langle {\varphi}'^2 \rangle$, $\langle \varphi^2 \rangle$, $\langle |{\varphi}|^p \rangle$ and $\langle |{\varphi}|^{p-2} \rangle$, where ${\varphi}$ is the homogeneous component of the inflaton amplitude in natural variables, used in the linearized analysis of Sections \ref{Sec:Inflation} and \ref{Sec:LinearAn}. The EOM of $\varphi$ is given in Eq.~(\ref{eq:inflaton-homeq}). As explained in the bulk text, its solution can be approximately written, under the approximation $\Delta = 0$, as ${\varphi} \simeq \cos [2 \pi u / (T_{{\varphi}} \omega_*) ]$ (c.f.~\ref{eq:Infl-sol}), where $T_{{\varphi}}$ is the oscillation period given in Eq.~(\ref{eq:Inflaton-period}).

Let us start with $\langle {\varphi}'^2 \rangle$. The oscillation-averaged quantity can be written as four times the integration over a quarter-period, which goes from ${\varphi} = 0$ (the minimum of the oscillation) to ${\varphi} = 1$ (the maximum). The computation proceeds as follows,
\bea 
\langle {\varphi}'^2 \rangle &=& \frac{4}{T_{\rm {\varphi}}} \int_{u({\varphi}=0)}^{u({\varphi}=1)}\, {\varphi}'^2(u) du = \frac{4}{T_{\rm {\varphi}}} \int_{{\varphi}=0}^{{\varphi}=1} {\varphi}'\, d {\varphi}  \nonumber \\
&=& \frac{4}{T_{\rm {\varphi}}}\sqrt{\frac{2}{p}} \int_{{\varphi}=0}^{{\varphi}=1} \sqrt{1-|{\varphi }|^p}\, d {\varphi } =\frac{2}{2+p} \ , \label{eq:dotVarphi2}
\eea   
where in the second line we have used that the energy $E_{{\varphi}} \equiv \frac{1}{2} {\varphi}'^2 + \frac{1}{p} |{\varphi}|^p = 1/p$ is conserved during the oscillation.  Similarly, $\langle \varphi^2 \rangle$  can be computed as
\bea 
\langle {\varphi}^2 \rangle &=& \frac{4}{T_{\rm {\varphi}}} \int_{u({\varphi}=0)}^{u({\varphi}=1)}\, {\varphi}^2(u) du =   \label{eq:Varphi2}  \\
&=& \frac{\sqrt{8p}}{T_{\rm {\varphi}}}  \int_{{\varphi}=0}^{{\varphi}=1}  \frac{ \varphi ^2}{\sqrt{1-|{\varphi}|^p}}d {\varphi} = \frac{ \Gamma [\frac{3}{p}]\Gamma [\frac{p+2}{2p}]}{\Gamma [\frac{1}{p}] \Gamma [\frac{p+6}{2p}]} \ , \nonumber
\eea   
Also, $\langle |{\varphi}|^p \rangle$  can be computed as
\bea
\langle | {\varphi}|^p \rangle &=& \frac{4}{T_{\rm {\varphi}}}  \int_{u({\varphi}=0)}^{u({\varphi}=1)}  | {\varphi}(u) |^p du  \label{eq:Varphip} \\
&=& \frac{\sqrt{8p}}{T_{\rm {\varphi}}}  \int_{{\varphi}=0}^{{\varphi}=1}  \frac{| {\varphi} |^p}{\sqrt{1-|{\varphi}|^p}}d {\varphi} = \frac{2}{2+p} \ . \nonumber \eea
Finally, $ \langle| {\varphi}|^{p-2} \rangle$ is given by
\bea \langle | {\varphi}|^{p-2} \rangle &=& \frac{4}{T_{\rm {\varphi}}}  \int_{u({\varphi}=0)}^{u({\varphi}=1)}  | {\varphi}(u) |^{p-2}du \nonumber \\
&=& \frac{\sqrt{8p}}{T_{\rm {\varphi}}}  \int_{{\varphi}=0}^{{\varphi}=1}  \frac{| {\varphi} |^{p-2}}{\sqrt{1-|{\varphi }|^p}}d {\varphi} \nonumber \\
&=& \frac{\Gamma[\frac{p-1}{p}]  \Gamma[\frac{p+2}{2p}]}{\Gamma[\frac{1}{p}] \Gamma[\frac{3p-2}{2p}]}\ . \label{eq:dotVarphipm2}\eea

\section{Spectral Analysis}\label{app:Spectra}

\begin{figure*} \centering
    \includegraphics[width=0.46\textwidth]{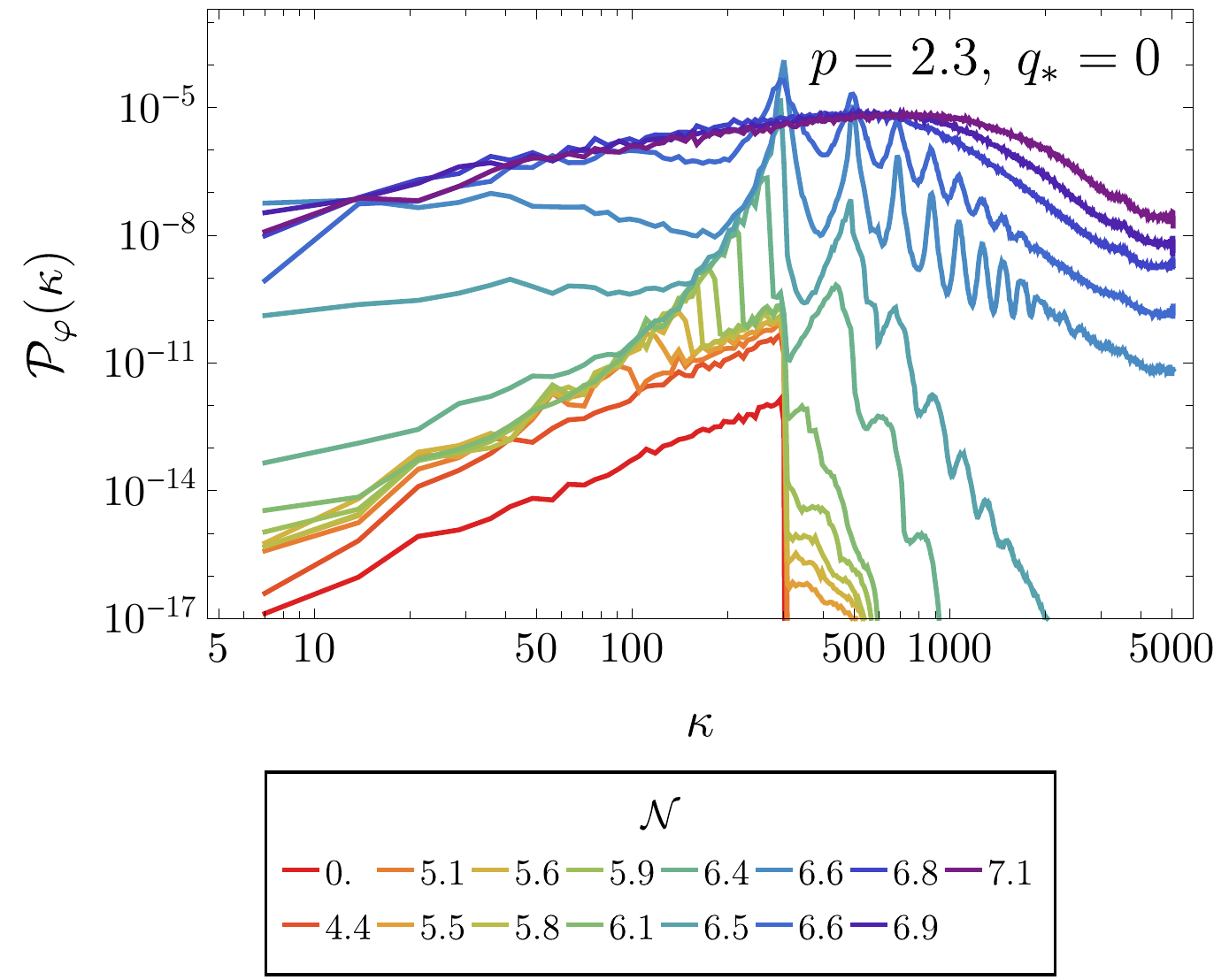} \,
    \includegraphics[width=0.46\textwidth]{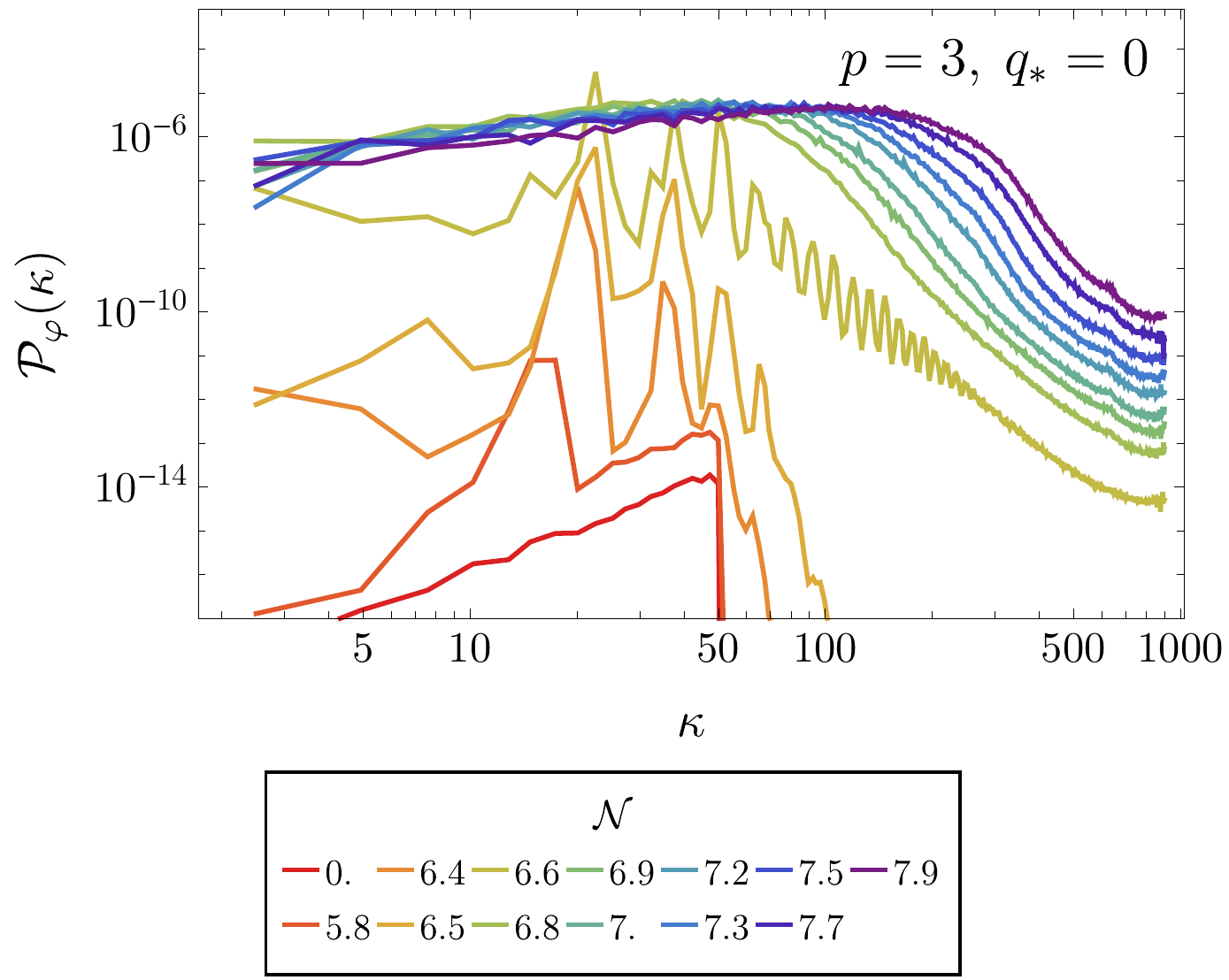}  \\ 
    \includegraphics[height=6.5cm]{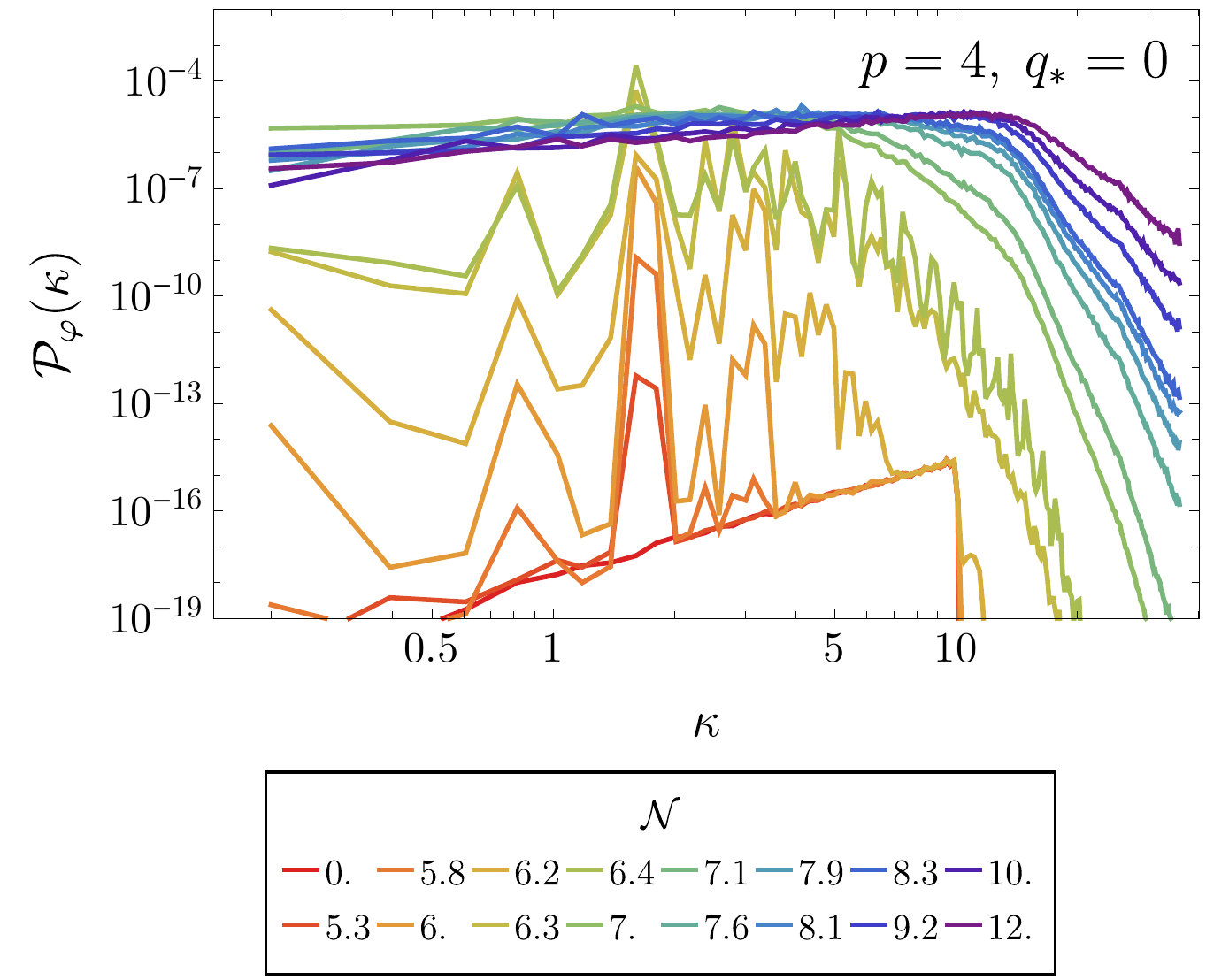} \,
    \includegraphics[height=6.5cm]{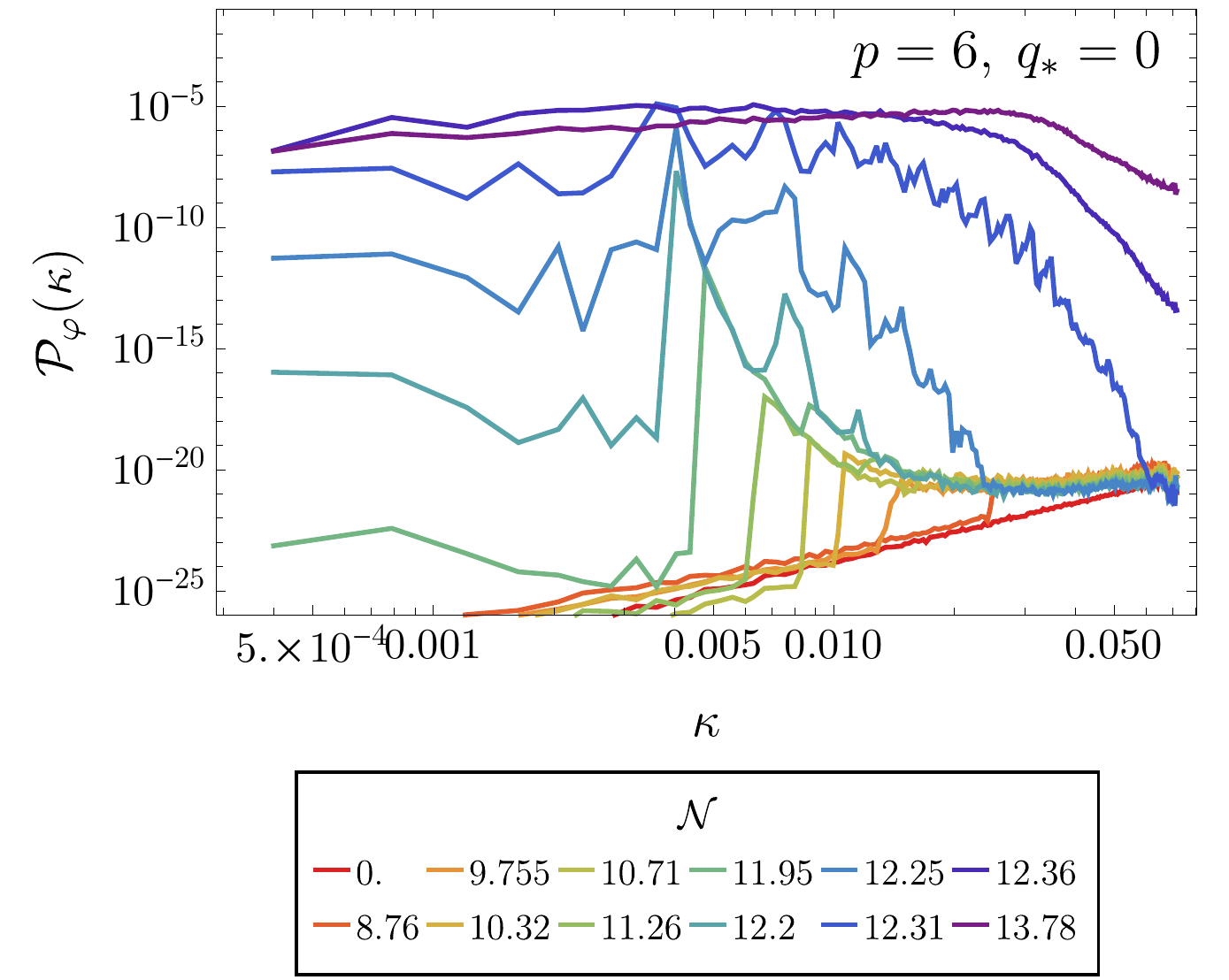} 
    \caption{[$v=0$] Spectra of the inflaton for the case $q_*=0$ and the power-law coefficients $p=2.3$, $3$, $4$ and $6$, as a function of natural comoving momentum $\kappa \equiv k / \omega_*$. Each line shows the spectrum at different times, from red (early times) to purple (late times). We indicate the corresponding number of e-folds after inflation for each simulation. }  \label{Fig:Spectra-NoInteraction}
\end{figure*}

\begin{figure*} \centering
    \centering
     \includegraphics[height=4.9cm]{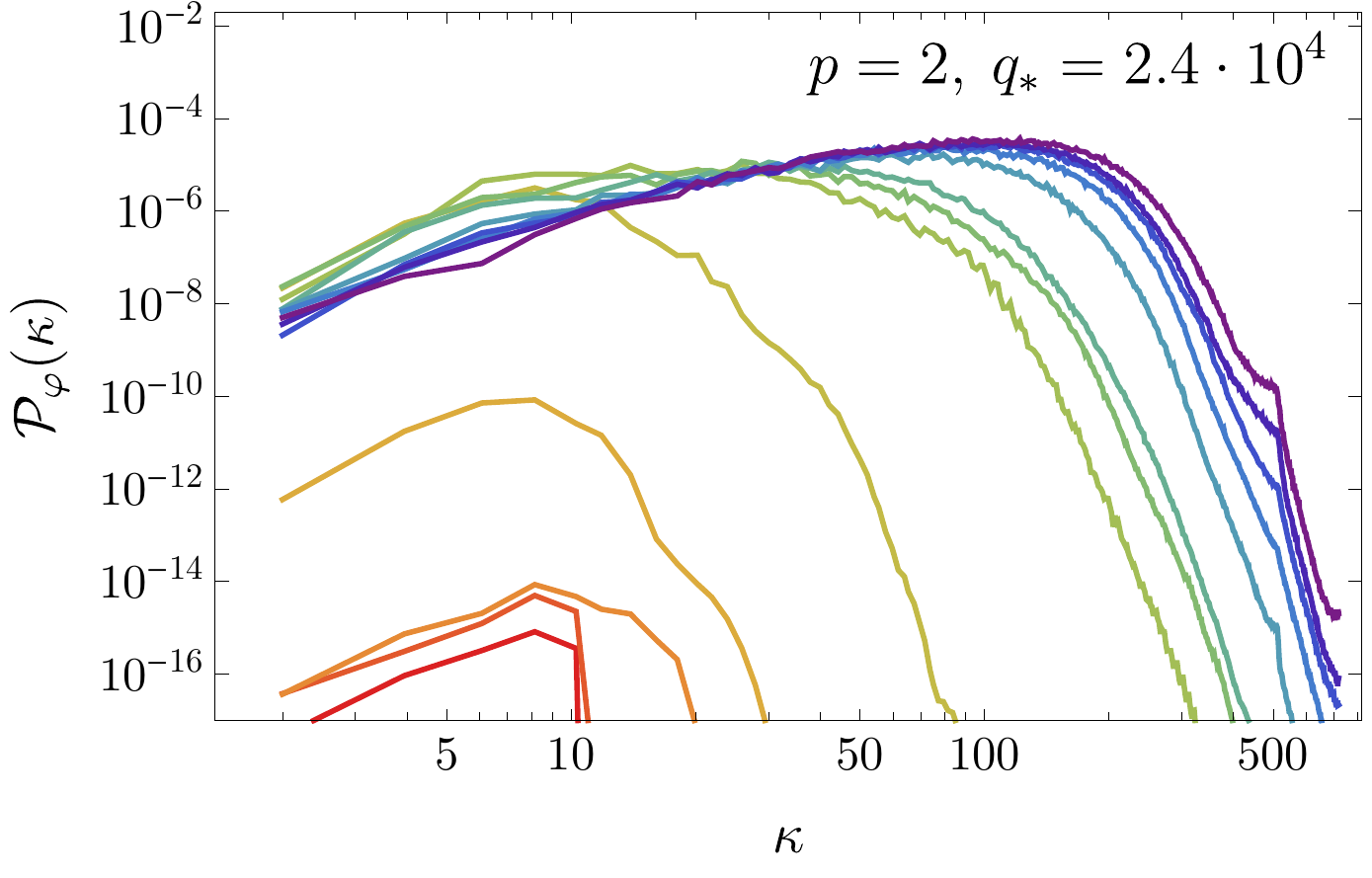} \,\,\,
    \includegraphics[height=4.9cm]{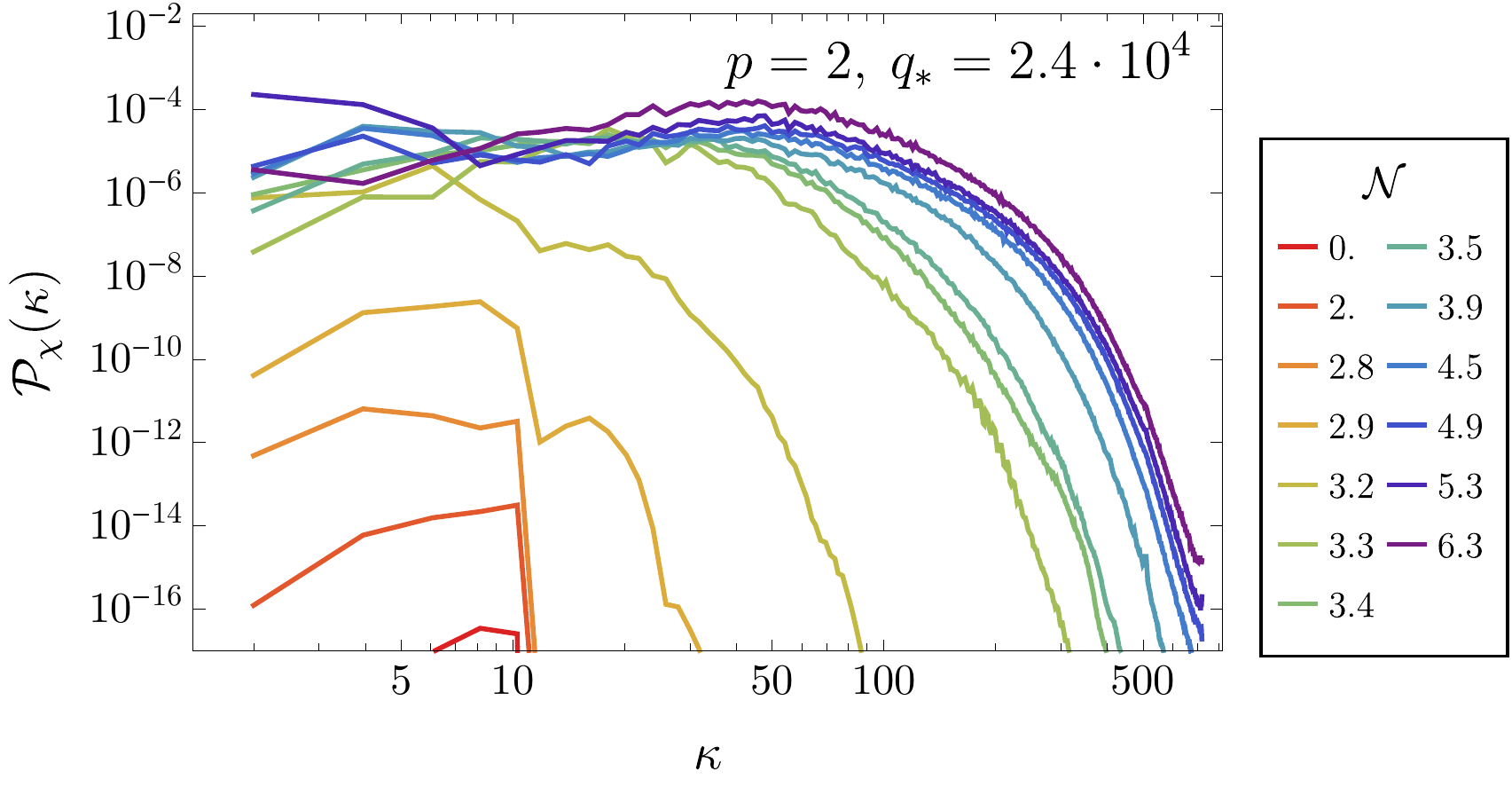} 
    \includegraphics[height=4.9cm]{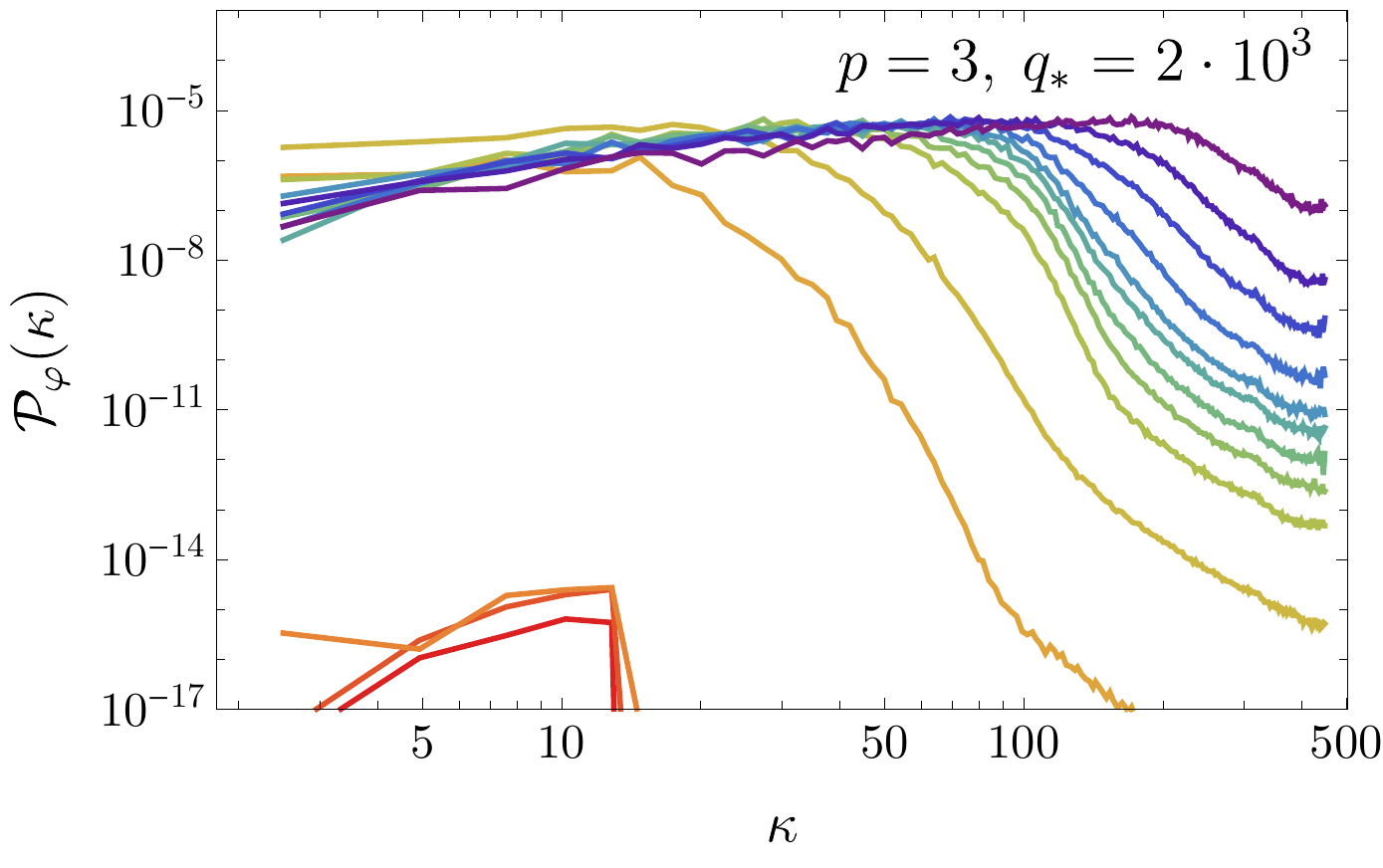} \,\,\,
    \includegraphics[height=4.9cm]{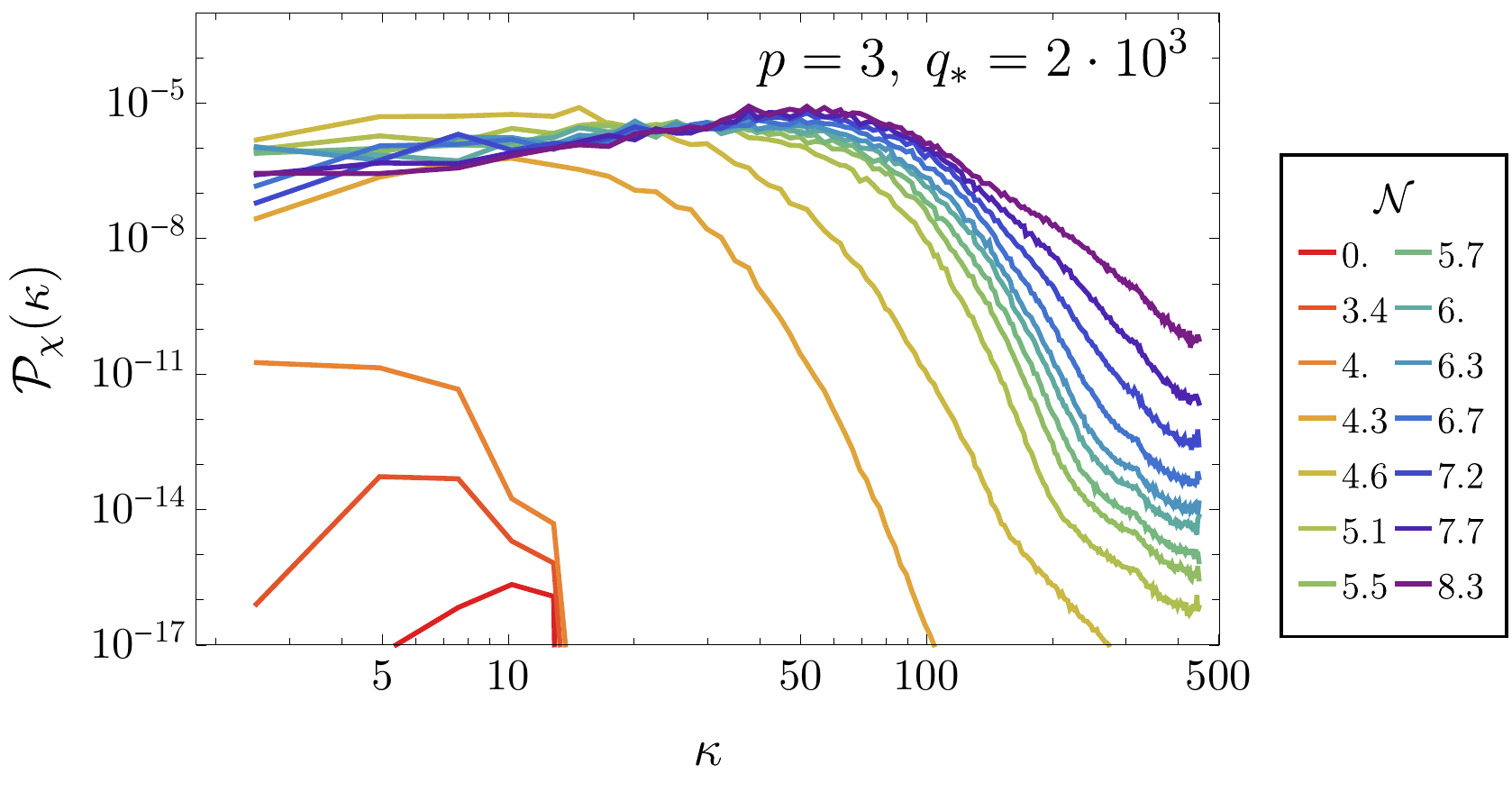} 
    \includegraphics[height=4.9cm]{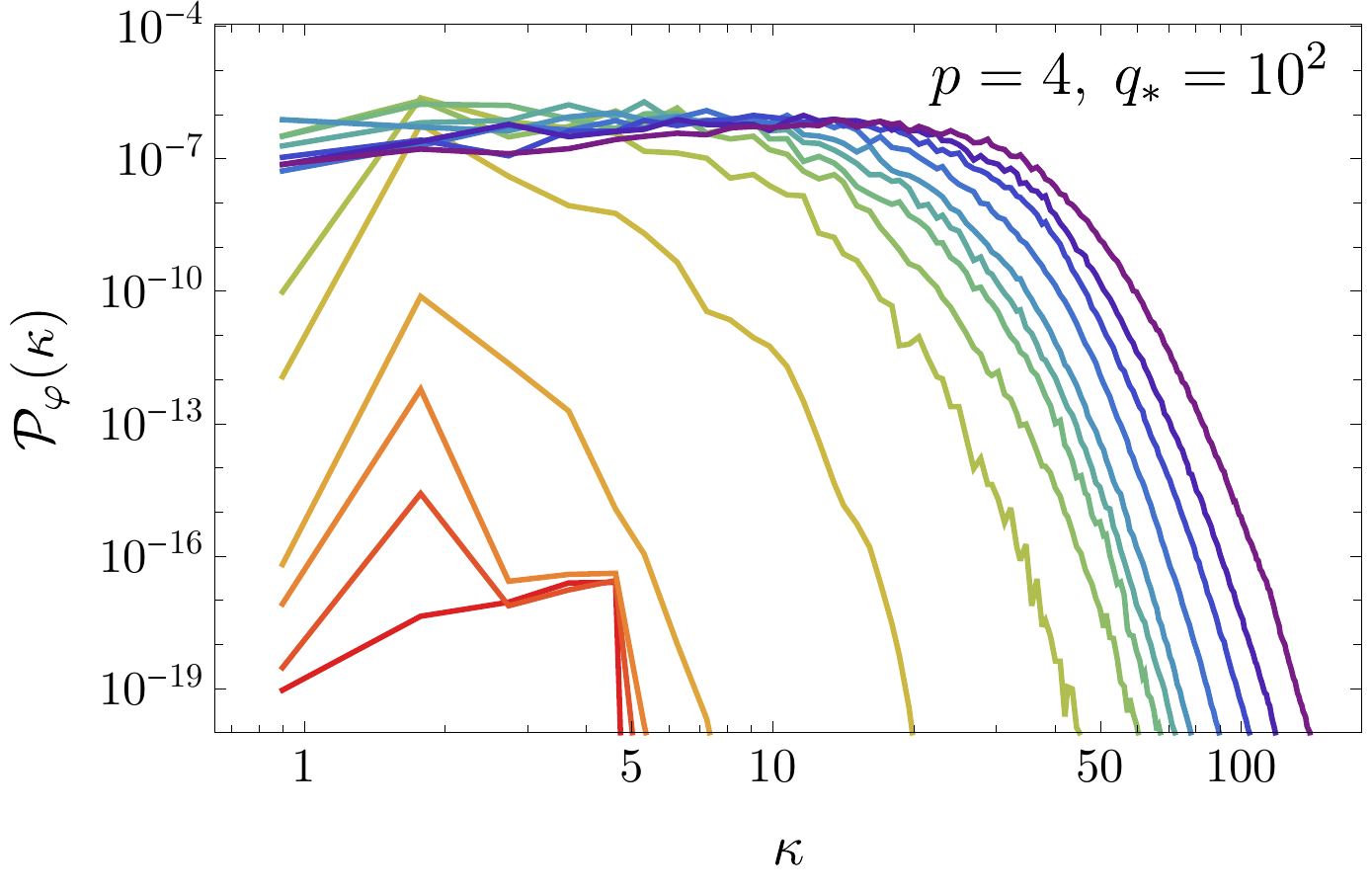} \,\,\,
    \includegraphics[height=4.9cm]{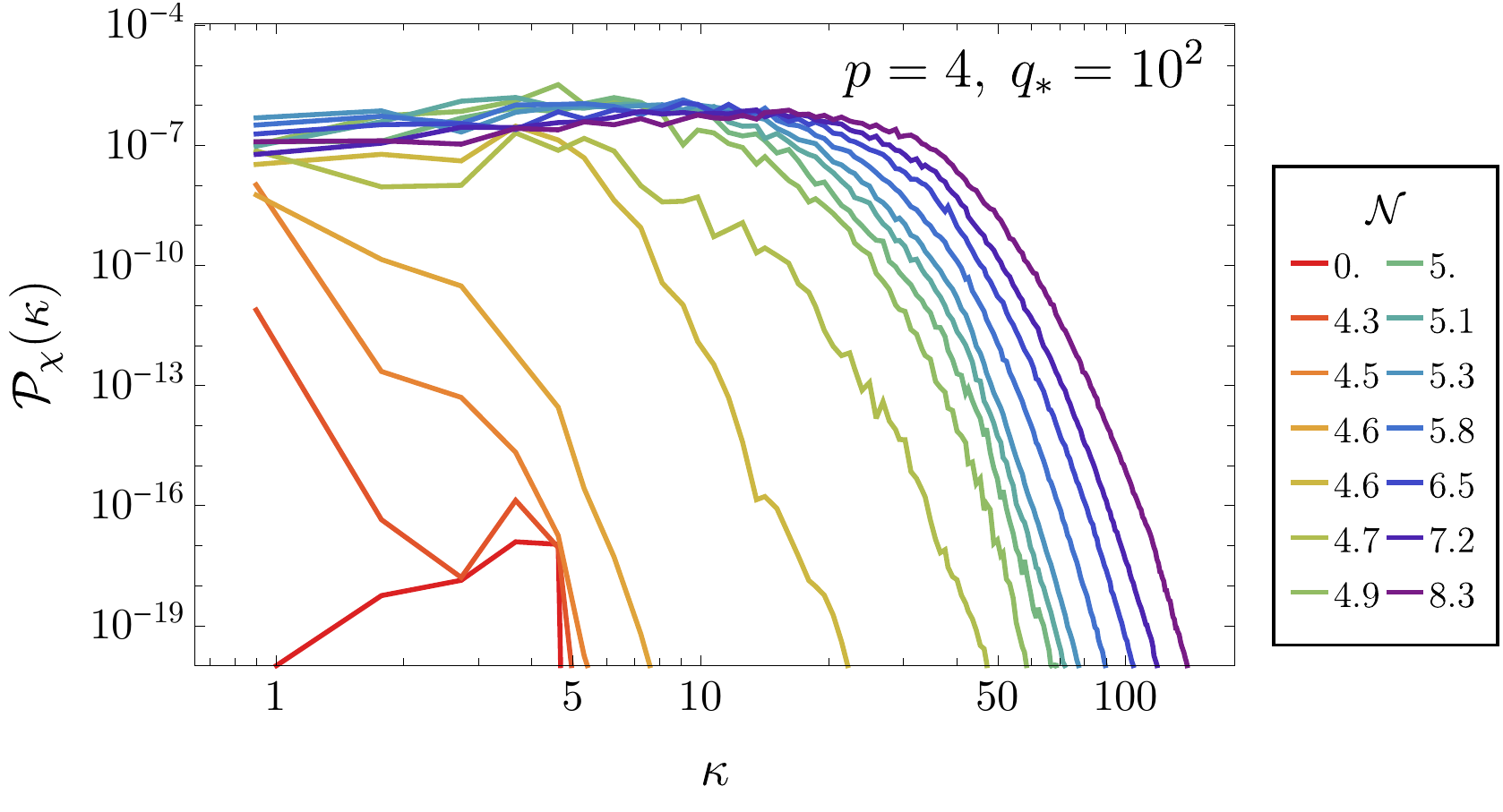} 
    \includegraphics[height=4.9cm]{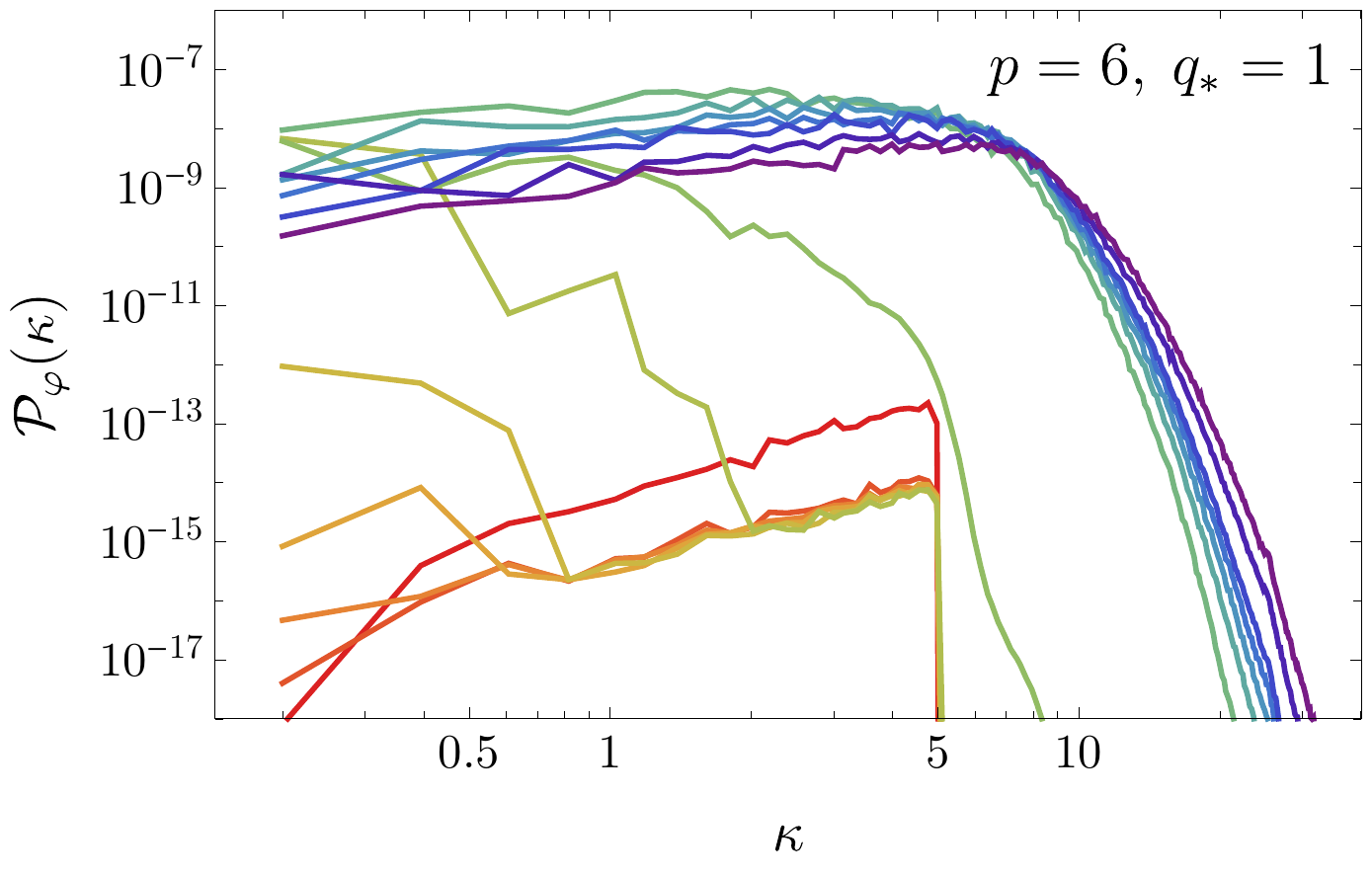} \,\,\,
    \includegraphics[height=4.9cm]{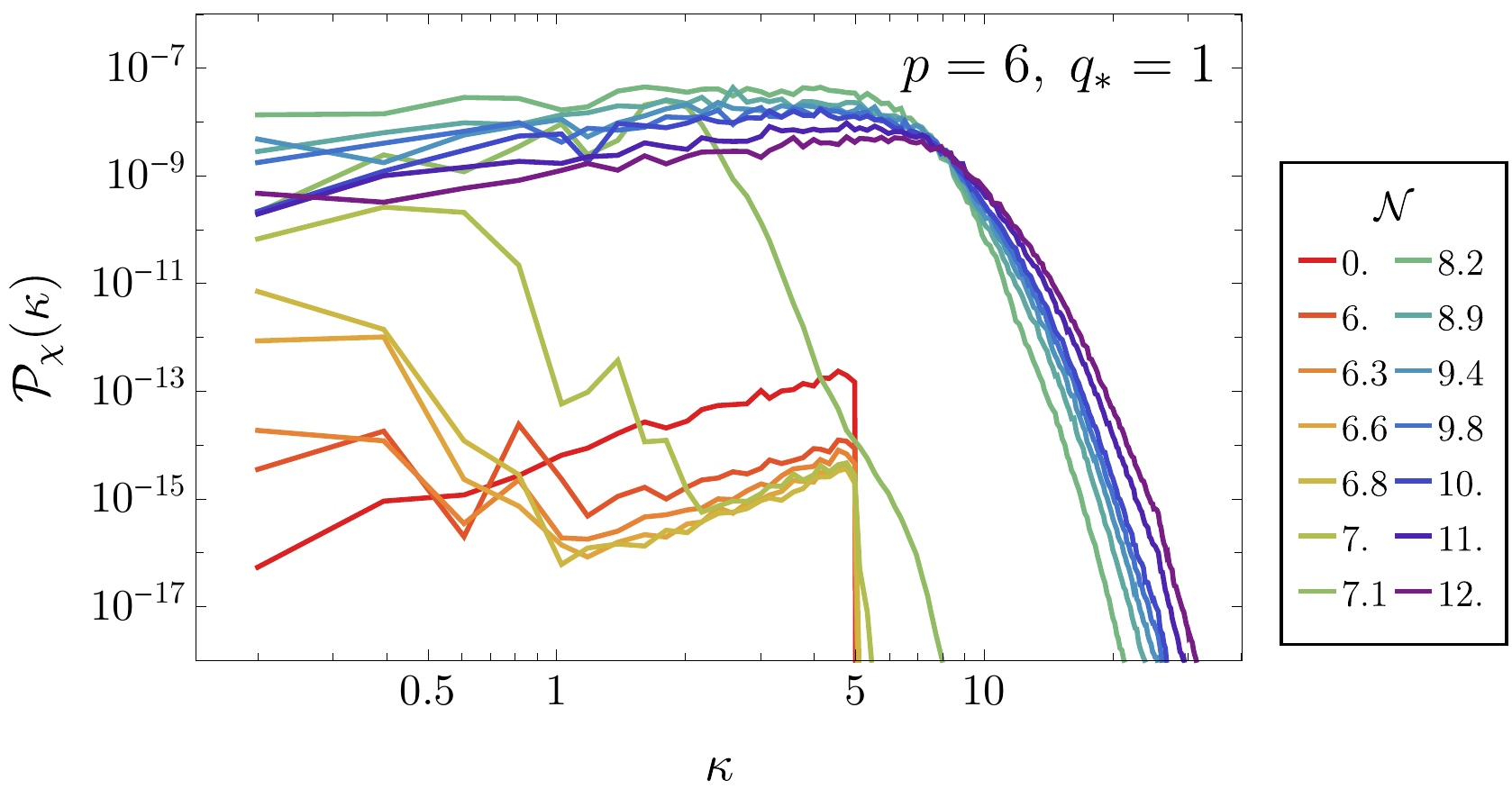} 
\caption{[$v=0$] Spectra of the inflaton and daughter fields for $p=2,3,4,6$, as a function of natural comoving momentum $\kappa \equiv k / \omega_*$. Each line shows the spectrum at different times, from red (early times) to purple (late times). We indicate the corresponding number of e-folds after inflation for each simulation. }  \label{Fig:Spectra-WithInteraction}
\end{figure*}

Here, as a supplement to the results of Section \ref{Sec:LatResults}, we briefly discuss how the spectra of the inflaton and daughter field evolve in momentum space in the case of the centred potential. The power spectra of both fields, $\mathcal{P}_{\varphi} (k)$ and $\mathcal{P}_{\chi} (k)$, can be defined as 
\bea \langle \varphi^2 \rangle &=& \int d \log k \, \mathcal{P}_{\varphi} (k) \ , \\
\langle \chi^2 \rangle &=& \int d \log k \, \mathcal{P}_{\chi} (k) \ . \eea
We consider two scenarios: when $q_* = 0$ and only the inflaton gets excited via self-resonance (see Fig.~\ref{Fig:Spectra-NoInteraction}), and when $q_*>q_*^{\rm (min)}$ and the daughter field also gets excited via parametric resonance (see Fig.~\ref{Fig:Spectra-WithInteraction}). In each case we consider different choices of $p$.  \vspace{0.3cm}

\textbf{i) \boldsymbol{$q_* = 0$}:}  In Fig.~\ref{Fig:Spectra-NoInteraction} we have depicted the time-evolution of the inflaton spectra for $p=2.3$, $3$, $4$ and $6$  (note that there is no excitation for $p=2$). In all cases, the main growth of the spectrum during the initial linear regime takes place in very narrow bands of fixed \textit{resonance} momenta, given by the Floquet chart of Fig.~\ref{fig:FloquetDiagrams}. The natural resonance momenta is defined in terms of comoving momenta as $\tilde \kappa (a) \equiv  \kappa a^{-\frac{(8-2p)}{p+2}}$ with $\kappa = k / \omega_*$ (c.f.~\ref{eq:resonant-momentum}), so the position of the bands in comoving momenta changes as the universe expands: they move to the infrared for $p<4$, to the ultraviolet for $p>4$, and remain constant for $p=4$. For example, the main resonance bands for $p=2.3$ and $3$ are emplaced, according to the chart, at the constant values $\tilde \kappa \simeq 1.7$ and $1.5$ respectively, so in terms of comoving momenta they move to the ultraviolet as $\kappa \equiv \tilde \kappa (a) a^{\frac{8-2p}{p+2}} \sim \tilde \kappa a^{0.79}$, $\tilde \kappa a^{0.4}$ respectively. Note that, as the propagation towards the ultraviolet for $p=3$ is slower than for $p=2.3$, the simulation for the first case has only required a lattice with $N^2 = 512^2$ points, while the second one has required instead $N^2 = 1024^2$ points. For $p=4$, the growth of the field modes during the linear stage takes place at constant comoving momenta $\kappa$ for $p=4$, while for $p=6$ it takes place at red-shifting comoving momenta $\kappa \equiv \tilde \kappa (a) a^{\frac{8-2p}{p+2}} \sim \tilde \kappa a^{-1/2}$.

In all cases, approximately $u{\rm br} \approx 6 - 12$ e-folds after the end of inflation, backreaction takes place, and the sharp peaks in the spectrum get washed out. During the subsequent non-linear stage, the whole spectrum slowly propagates towards the ultraviolet, in a rescattering process induced by the inflaton self-resonance, which is also reflected in the growth of gradient energy seen in Fig.~\ref{p23-EosEnergy}. \vspace{0.3cm}

\textbf{ii) \boldsymbol{$q_* > q_*^{\rm (min)}$}:} In Fig.~\ref{Fig:Spectra-WithInteraction} we have plotted the evolution of the inflaton and daughter field spectra for $p=2$, $3$, $4$ and $6$. For $p=2$, we observe that the spectra of both fields evolve in very similar ways, showing that the interaction term is very efficient in coupling the evolution of both fields. During the initial linear regime, the spectrum of the daughter field gets populated in an infrared band $0 \leq \kappa \leq \kappa_{+}$, as expected from our linealized analysis. However, the inflaton spectrum also grows within this band, which is purely an effect of the daughter field modes backreacting onto the inflaton (as there is no inflaton self-resonance). After backrection time, we enter the non-linear regime, in which both spectra start propagating towards the ultraviolet due to scattering processes, eventually saturating at a certain momentum scale.

In the $p=3$ case, parametric resonance of the daughter field is the dominant resonance process during the linear regime, so the growth of the spectra takes place mainly within a wide infrared band. Therefore, the sharp narrow peaks induced by the inflaton self-resonance cannot be observer here. Eventually, backreaction takes place $\sim 4$ e-folds after the end of inflation, and during the following non-linear regime, both spectra slowly propagate to the ultraviolet. We clearly see that the inflaton spectrum broadens towards much larger momenta than the daughter field. At this stage, the parametric resonance exciting the daughter field has become narrow, $\tilde q < 1$, so the only relevant effect is the excitation of the inflaton via self-resonance. This is constantly stimulating the different inflaton modes \textit{even after backreaction has destroyed the inflaton homogeneous mode}. This explains why the inflaton spectrum moves to the ultraviolet much faster than the daughter field, and eventually makes the inflaton gradient energy dominate over the daughter (kinetic and gradient) energies, as observed in Fig.~\ref{p23-EosEnergy}.

Finally, the evolution for the $p=4$ and $p=6$ cases is similar to the $2<p<4$ scenario. First, the initial growth of both spectra takes place within a broad resonance band $0 < \kappa < \kappa_+$ due to parametric resonance. Then, after backreaction time, the peaks get washed out, and the spectra start propagating to the ultraviolet. The evolution of both field spectra is very similar because the resonance parameter $\tilde q$ stays constant or grows, thus enables efficient exchange of energy via scattering processes.

\section{Dependence of \texorpdfstring{$N_k$}{Nk} on the quadratic-quadratic interaction}\label{App:Nbr}

In order to illustrate the influence of the transition phase from the end of inflation to radiation domination on $N_k$, Eq.~(\ref{Nk1}) can be developed further, by including explicitly the time scale at which backreaction effects break the homogeneous inflaton condensate. We can decompose the expansion stage between the end of inflation and the onset of radiation domination as $a_{\rm end} / a_{\rm rd} = (a_{\rm end} / a_{\rm hom} )(a_{\rm hom} / a_{\rm br} ) (a_{\rm br} / a_{\rm rd} )$, where $a_{\rm hom}$ denotes the scale factor when the equation of state becomes $\bar{w}_{\rm hom} = (p-2)/(p+2)$ (see \ref{eq:EoSoscillations}), and $a_{\rm br}$ the scale factor at the backreaction time. Similarly, we can write the corresponding number of e-folds as $\Delta N_{\rm end}^{\rm rd} = \Delta N_{\rm end}^{\rm hom} + \Delta N_{\rm hom}^{\rm br} + \Delta N_{\rm br}^{\rm rd}$, where $\Delta N^j_i \equiv \mathrm{ln}(a_j/a_i)$. We can then write $\rho_{\rm rd}$ in terms of $\rho_{\rm end}$ as
\bea \rho_{\rm rd} &=& \frac{ \rho_{\rm rd}}{\rho_{\rm br} } \frac{ \rho_{\rm br}}{\rho_{\rm hom} }\frac{ \rho_{\rm hom}}{\rho_{\rm end} } \rho_{\rm end} \\
&=& \left( \frac{a_{\rm rd}}{a_{\rm br}}\right)^{- 3 (1 + \bar{w}_{\rm br}^{\rm rd} )  } \left( \frac{a_{\rm br}}{a_{\rm hom}}\right)^{- 3 (1 + \bar{w}_{\rm hom} ) } \nonumber\\ 
&\times&\left( \frac{a_{\rm hom}}{a_{\rm end}}\right)^{- 3 (1 + \bar{w}_{\rm end}^{\rm hom})  }\rho_{\rm end} \nonumber  \eea
where $\bar w_i^j$ is the averaged equation of state  over number of e-folds between $a_i$ and $a_j$,
\be \bar w_i^j \equiv \frac{1}{\Delta N_i^j }\int_{N_i}^{N_j}w(N')dN'\  .\ee
We can then write Eq.~(\ref{Nk1}) as follows,
\bea N_k\approx 61.5 &+& \frac{1}{4}\mathrm{ln}\frac{V_k^2}{m_\mathrm{pl}^4\rho_\mathrm{end}} + \frac{3 \bar w_{\rm end}^{\rm hom} - 1}{4} \Delta N_\mathrm{end}^{\rm hom} \nonumber \\
&+&\frac{p-4}{2p+4}\Delta N_\mathrm{hom}^{\rm br} + \frac{3 \bar w_{\rm br}^{\rm rd} - 1}{4} \Delta N_\mathrm{br}^{\rm rd} \ . \label{eq:e-folds-approx} \eea

A similar expression has been derived in \cite{Lozanov:2017hjm}, although $\Delta N_\mathrm{end}^{\rm hom}$ has been omitted and the explicit dependence on $\Delta N_\mathrm{br}^{\rm rd}$ approximated by an instant jump to radiation domination. The terms in the first line of Eq.~(\ref{eq:e-folds-approx}) depend on observational data and the shape of the inflaton potential. The quantities in the second line are directly affected by the preheating dynamics. While for $p=2$ the time of radiation domination is unknown without further assumption (thus $\Delta N_{\rm br}^{\rm rd}$ cannot be determined), for $p>2$ we can estimate how the coupling to the daughter field affects the value of $N_k$. With this aim, we approximate the fast transition from $\bar{w}_{\rm hom}$ to $\bar{w} = 1/3$ with $\Delta N_{\rm br}^{\rm rd}\approx0$ (which is well justified for cases $p \geq 3$, see e.g.~Fig.~\ref{Fig:EoSvsP}), such that only the term including $\Delta N_{\rm hom}^{\rm br}$ is left in the second line. For $p=2-6$ we have $\frac{p-4}{2p+4} \sim (-0.2) - 0.1$, while the values of $\Delta N_{\rm hom}^{\rm br}$ are depicted in Fig.~\ref{fig:PvsN} for $M=10m_{\rm pl}$ and different interaction strengths. The difference in number of e-folds until backreaction between the two resonance cases is $(\Delta N_{\rm hom}^{\rm br})^{(\rm sr)} - (\Delta N_{\rm hom}^{\rm br})^{(\rm pr)} \simeq 2 - 6$, which leads to a change in $N_k$ of less than an e-fold.\pagebreak

\bibliography{References.bib,ReferencesExtra.bib}

\end{document}